 \documentclass[final,3p,times,compress]{elsarticle}

\usepackage{amssymb}
\usepackage{lipsum}
\usepackage{color}
\usepackage{amsthm}
\usepackage{braket}
\usepackage{mathrsfs}
\usepackage{nccmath}
\usepackage{bm}
\usepackage{graphics}
\usepackage{multirow}
\usepackage[colorlinks=true]{hyperref}
\usepackage{comment}
\usepackage{empheq}

\journal{Annals of Physics}

\begin{document}

\begin{frontmatter}



\title{Criteria for Feasible Monte Carlo Stochastic Simulations of\\Bosonic Markovian Open Quantum Dynamics}


\author[first]{Toma~Yoneya}
\author[second]{Kazuya~Fujimoto}
\author[first,third]{Yuki~Kawaguchi}
\affiliation[first]{organization={Department of Applied Physics, Nagoya University},
            city={Nagoya},
            state={464-8603},
            country={Japan}}

\affiliation[second]{organization={Department of Physics, Institute of Science Tokyo},
            city={Tokyo},
            state={152-8551},
            country={Japan}}

\affiliation[third]{organization={Research Center for Crystalline Materials Engineering, Nagoya University},
            city={Nagoya},
            state={464-8603},
            country={Japan}}
\begin{abstract}
The Monte Carlo sampling of the stochastic differential equations based on the quasiprobability distribution function,
such as the Glauber--Sudarshan P,
Wigner,
and Husimi Q functions provides a powerful framework for investigating bosonic open quantum many-body dynamics described by the Gorini--Kossakowski--Sudarshan--Lindblad (GKSL) equation,
while considering the effects of quantum fluctuations beyond the mean-field approximation.
However,
the stochastic Monte Carlo simulation is possible only when the corresponding Fokker--Planck equation has a positive-semidefinite diffusion matrix,
and the general conditions for the diffusion matrix to be positive semidefinite have remained unclear.
In this work,
starting from the path integral formulation,
we first derive the sufficient conditions under which the diffusion matrix is positive semidefinite for an arbitrary Hamiltonian,
jump operators,
and choice of quasiprobability distribution functions.
We also analytically derive the corresponding stochastic differential equations to be solved.
We then investigate the dynamics of the GKSL equation in the thermodynamic limit and show that,
depending on the form of the jump operators,
the mean-field approximation may fail to describe the dynamics accurately,
making stochastic Monte Carlo simulations indispensable.
Furthermore, 
we derive the sufficient conditions under which the higher-order quantum fluctuation terms beyond the Fokker--Planck description vanish identically,
even when the jump operators contain quadratic terms.
Under these conditions,
whenever the corresponding stochastic differential equations can be derived,
the stochastic Monte Carlo simulation reproduces the exact dynamics.
These results clarify the conditions under which the stochastic Monte Carlo simulations are both feasible and necessary for accurately describing the dynamics governed by the GKSL equation in phase space.

\end{abstract}




\end{frontmatter}

\tableofcontents



\section{\label{introduction}Introduction}
The phase-space formulation of quantum mechanics \cite{Hillery,Lee} provides a powerful framework for investigating the bosonic quantum many-body dynamics while considering the effects of quantum fluctuations,
and has been widely applied to the study of open quantum many-body systems \cite{Gardiner,Milburn,Carmichael1,Carmichael2} described by the Gorini--Kossakowski--Sudarshan--Lindblad (GKSL) equation \cite{Gorini,Lindblad}.
In this formulation,
the density operator is represented by a quasiprobability distribution function,
such as the Glauber-Sudarshan P \cite{Glauber,Sudarshan},
Wigner \cite{Wigner},
and Husimi Q \cite{Husimi,GlauberQ} functions,
and the GKSL equation is mapped onto a partial differential equation for the quasiprobability distribution function that generally contains higher-order derivative terms.
By neglecting derivative terms beyond second order,
the GKSL equation is approximated by the Fokker--Planck equation,
whose dynamics can be investigated through a Monte Carlo simulation of the corresponding stochastic differential equations \cite{Gardiner,Milburn,Carmichael1,Carmichael2}.

However, depending on details of the Hamiltonian, jump operators,
and choice of the quasiprobability distribution function,
the Fokker--Planck equation does not always reduce to the stochastic differential equations because the diffusion matrix is not necessarily positive semidefinite \cite{Gardiner,Milburn,Carmichael1,Carmichael2}.
For the Wigner function,
where the corresponding approximation is known as the truncated Wigner approximation (TWA),
the diffusion matrix depends only on the details of the jump operators \cite{Gardiner,Milburn,Carmichael1,Carmichael2,Huber,Plimak,Yoneya2025,Yoneya2026}.
Thus, for an isolated system,
the Fokker--Planck equation involves no diffusion terms,
and the dynamics can therefore be simulated simply by the Monte Carlo sampling of the classical equations of motion.
Owing to this tractability,
the TWA has been widely used for isolated systems \cite{Steel,Alice,Blakie,Polkovnikov2010} and 
has been extended to describe the many-body dynamics of spins \cite{Schachenmayer,Davidson2015,Wurtz,Zhu} and fermions \cite{Davidson2017}.
The validity of the TWA,
including these generalizations,
has been investigated through comparisons with experiments \cite{Orioli,Fersterer,Takasu,Nagao2021,Christopher2022,Christopher2023,Nagao2024}.
For open quantum systems,
the TWA has been applied to investigate the dynamics of the dissipative Bose--Hubbard model \cite{kordas2015,Vicentini,PRXQuantumDeuar2021},
cavity systems \cite{Iacopo2005,Dagvadorj,KeBler2020,Seibold},
and dissipative spin systems \cite{Huber,Singh,Mink,Huber2022}.
On the other hand,
for the Glauber--Sudarshan P and Husimi Q functions,
in isolated systems,
the diffusion matrix has the particle--hole symmetry \cite{Yoneya2026},
which implies that the diffusion matrix of the Fokker--Planck equation always has at least one negative eigenvalue.
As a consequence,
the Monte Carlo simulation is generally unfeasible except in non-interacting systems.
However,
in open quantum systems,
the couplings with environments can render the diffusion matrix positive semidefinite even when the Hamiltonian involves many-body interactions \cite{Gardiner,Milburn,Carmichael1,Carmichael2}.

In Refs.~\cite{Yoneya2025,Yoneya2026},
based on the path-integral representation of the GKSL equation,
we formulated the stochastic Monte Carlo method for the Glauber--Sudarshan P,
Wigner,
and Husimi Q functions for an arbitrary Hamiltonian and jump operators.
We also derived an analytical expression for the diffusion matrix of the corresponding Fokker--Planck equation, whose positive semidefiniteness allows the dynamics to be represented by an associated stochastic differential equation. For jump operators that do not couple different degrees of freedom, we analytically diagonalized the diffusion matrix and obtained explicit conditions for its positive semidefiniteness, as well as an analytical expression for the stochastic term. For more general jump operators, however, our previous formulation requires numerical diagonalization of the diffusion matrix at every time step to check its positive semidefiniteness and construct the stochastic term, which incurs a high computational cost. Extending both the positivity criteria and the analytical construction of the stochastic term to more general jump operators remains an open problem. Then, the following questions naturally arise:
\begin{itemize}
    \item Under what conditions does the diffusion matrix become positive semidefinite for more general jump operators?
    \item When is the stochastic Monte Carlo simulation necessary to accurately describe the dynamics?
    \item How do the higher-order quantum fluctuations beyond the Fokker--Planck equation affect the dynamics?
\end{itemize}
To the best of our knowledge,
the positive semidefiniteness of the diffusion matrix has been investigated on a case-by-case basis for specific models \cite{Yoneya2025,Yoneya2026,Huber,kordas2015,Vicentini,PRXQuantumDeuar2021,Iacopo2005,Dagvadorj,KeBler2020,Seibold,Singh,Mink,Huber2022},
and the general framework for constructing the positive-semidefinite diffusion matrix is still lacking.
Regarding the third question,
in isolated systems,
the stochastic method that accounts for the effects of the higher-order quantum fluctuations has been developed from the path-integral formalism \cite{Polkovnikov2010,Polkovnikov2003,Plimak_2001}.

In this work,
we address the three questions raised above from the path-integral representation of the GKSL equation in phase space.
From the diffusion matrix derived in Refs.~\cite{Yoneya2025,Yoneya2026}, 
we first derive sufficient conditions under which the diffusion matrix is positive semidefinite.
The resulting conditions can be applied to a broad class of systems with Hamiltonians and jump operators that involve higher-order interaction terms and couple different degrees of freedom.
In particular,
for quadratic jump operators,
we show that the diffusion matrix for the Wigner function becomes positive semidefinite whenever the infinite-temperature state is a steady-state solution of the GKSL equation.
We also analytically derive the corresponding stochastic differential equations to be solved and verify their validity by numerically simulating relaxation dynamics.

In order to tackle the second question,
we investigate the dynamics of the GKSL equation in the thermodynamic limit.
In isolated systems,
it is well known that,
by appropriately scaling the Hamiltonian parameters so that the thermodynamic limit is well defined,
the mean-field approximation provides an accurate description of the dynamics in this limit \cite{Hepp,Spohn,Yaffe}.
However,
in open quantum systems,
we show that,
depending on the details of the jump operators,
the mean-field approximation may fail to describe the dynamics,
making the stochastic Monte Carlo simulation necessary for an accurate description.
Although it is nontrivial whether the corresponding stochastic differential equations can be derived in the thermodynamic limit,
we show that,
for the Wigner function,
the diffusion matrix becomes positive semidefinite in this limit independently of the details of the Hamiltonian and jump operators.
In benchmark calculations,
we numerically demonstrate the breakdown of the mean-field approximation.

Regarding the third question,
we derive sufficient conditions under which the effects of higher-order quantum fluctuations do not affect the dynamics.
In the presence of quadratic jump operators,
the GKSL equation is generally mapped onto a partial differential equation involving up to fourth-order derivatives.
However,
depending on the form of the jump operators and the choice of the quasiprobability distribution function, we show that the third- and fourth-order derivative terms vanish.
As a result,
the GKSL equation reduces exactly to the Fokker–Planck equation in phase space,
and the stochastic Monte Carlo simulation can reproduce the exact dynamics whenever the corresponding stochastic differential equations can be derived.

We summarize the conditions obtained in this work in Tab.~\ref{tab:brief summary}:
The diffusion matrix is positive semidefinite,
the mean-field approximation breaks down,
and the effects of the higher-order quantum fluctuation vanish.
These results clarify both the feasibility and the necessity of the stochastic Monte Carlo simulation for the GKSL equation in phase space.

This paper is organized as follows.
In Sec.~\ref{sec:Target of this paper},
we introduce the GKSL equation and the systems under consideration.
In Sec.~\ref{sec:Review of functional representation of Markovian open quantum systems in the phase space},
we briefly review the phase-space mapping and $s$-ordered quasiprobability distribution function,
which provides a unified description of the Glauber--Sudarshan P,
Wigner,
and Husimi Q functions by tuning the real parameter $s$.
The review of the path-integral formulation of the stochastic Monte Carlo simulation is also given in the same section.
In Sec.~\ref{sec:Feasibility of the second-order approximation using the stochastic differential equation},
we derive the sufficient conditions under which the diffusion matrix becomes positive semidefinite,
together with the corresponding stochastic differential equations.
In Sec.~\ref{sec:Equation of motion in the high occupancy limit},
we introduce the high-occupancy limit as an analogue of the thermodynamic limit,
in which each degree of freedom,
corresponding to spatial coordinates and/or internal degrees of freedom,
is occupied by a large number of bosons.
We then show the conditions on the jump operators that lead to the breakdown of the mean-field approximation in this limit.
In the same section,
we show that,
for the Wigner function,
the diffusion matrix becomes positive semidefinite regardless of the Hamiltonian and jump operators in the high-occupancy limit.
The sufficient conditions under which the effects of the higher-order quantum fluctuations vanish are given in Sec.~\ref{sec:Higher order of quantum fluctuations}.
We show some benchmark calculations for two- and three-site models in Sec.~\ref{sec:Benchmark calculations}.
Summary and conclusions are given in Sec.~\ref{sec:Summary and conclusions}.

\section{\label{sec:Target of this paper}Target of this paper}


\subsection{\label{subsec:Gorini--Kossakowski--Sudarshan--Lindblad equation}Gorini--Kossakowski--Sudarshan--Lindblad equation}
In this work, we consider an open quantum system interacting with environments following a Markovian dynamics described by the GKSL equation \cite{Gorini,Lindblad}:
\begin{align}
    \label{eq:def of GKSL equation}
    \frac{d\hat{\rho}(t)}{dt} = -\frac{i}{\hbar}\left[\hat{H},\hat{\rho}(t)\right]_- + \sum_{k=1}^{k_{\rm max}}\gamma_k\left(\hat{L}_k\hat{\rho}(t)\hat{L}^{\dagger}_k - \frac{1}{2}\left[\hat{L}^{\dagger}_k\hat{L}_k,\hat{\rho}(t)\right]_+\right),
\end{align}
where $\hat{\rho}(t)$ is a density operator of the system we focus on and $[\dots]_{\mp}$ denote the commutator $(-)$ and anti-commutator $(+)$.
The first term of the right-hand side of the GKSL equation describes unitary dynamics generated by the system's Hamiltonian $\hat{H}$, and the second term describes non-unitary dynamics, where the jump operator $\hat{L}_k$ characterizes the interaction between the system and the environment, $\gamma_k$ represents the corresponding strength, and the subscript $k$ labels the different system-environment couplings,
with $k_{\rm max}$ denoting the total number of couplings.


\subsection{Setup}
We consider a bosonic system with total $M$ degrees of freedom,
which we identify by using subscripts $m,n,p,q \in \{1,2,\dots,M\}$,
where the degrees of freedom can correspond to spatial coordinates and/or internal degrees of freedom.
The Hamiltonian $\hat{H}$ and the jump operators $\hat{L}_k$ for $\forall k$ are composed of bosonic creation and annihilation operators $\hat{a}_m$ and $\hat{a}^{\dagger}_m$, which satisfy the commutation relation $[\hat{a}_m,\hat{a}^{\dagger}_n]_- = \delta_{mn}$.
Here, $\hat{H}$ and $\hat{L}_k$ for $\forall k$ can include higher-body interactions and couple different degrees of freedom unless otherwise specified.


\section{\label{sec:Review of functional representation of Markovian open quantum systems in the phase space}Review of functional representation of Markovian open quantum systems in phase space}
In the phase-space method, a bosonic operator is mapped into a $c$-number function, and the density operator is expressed as a quasiprobability distribution function.
Here, the way of the mapping is not unique, and the most general and comprehensive description has been established in Refs.~\cite{Agarwal1,Agarwal2,Agarwal3}, where the quasiprobability distribution function generally takes complex values depending on the mapping.
In this work, we utilize the phase-space mapping that leads to a real-valued quasiprobability distribution function \cite{Cahill1,Cahill2}.
This condition is necessary for performing the numerical calculation using a classical computer. 
Using the phase-space mapping, we have formulated the path-integral representation of the GKSL equation \cite{Yoneya2025,Yoneya2026}, where the Lagrangian involves the classical and quantum fields which respectively characterizes the classical motion and quantum fluctuations.
Here, the second-order perturbative expansion of the Lagrangian with respect to the quantum fields leads us to obtain the stochastic differential equation to be solved.
Below, we first introduce the phase-space mapping and the resulting quasiprobability distribution function with focusing on the relation between the mapping and the operator ordering in Sec.~\ref{sec:Mapping to phase space and quasiprobability distribution functions}.
In Sec.~\ref{subsec:Path-integral representation}, we introduce the path-integral representation based on the phase-space mapping.
By expanding the action with respect to the quantum fields order by order,
we obtain equations of motion at each order of quantum fluctuations in Sec.~\ref{subsec:Equations of motion in the phase space}.


\subsection{\label{sec:Mapping to phase space and quasiprobability distribution functions}Mapping to phase space and quasiprobability distribution functions}
An arbitrary bosonic operator $\hat{A}$ is mapped into a $c$-number function in phase space, $\hat{A}\mapsto A_s(\vec{\alpha},\vec{\alpha}^*)$, via
\begin{gather}
    \label{eq:s-parametrized mapping of A}
    A_{s}(\vec{\alpha},\vec{\alpha}^*) = \int\frac{d^2\vec{\eta}}{\pi^M}\chi_A(\vec{\eta},s)e^{\vec{\alpha}^*\cdot\vec{\eta} - \vec{\alpha}\cdot\vec{\eta}^*},\\
    \label{eq:definition of the characteristic function}
    \chi_A(\vec{\eta},s) = {\rm Tr}\left[\hat{A}\hat{D}^{\dagger}(\vec{\eta},-s)\right],
\end{gather}
where $\vec{\alpha}=(\alpha_1, \alpha_2, \dots, \alpha_M)^{\rm T}$ with ${\rm T}$ being the transposition and $\alpha_m = \alpha^{\rm re}_m + i\alpha_m^{\rm im}$ ($\alpha^{\rm re}_m,\alpha^{\rm im}_m \in \mathbb{R}$) for $\forall m$, $\vec{\eta}=(\eta_1, \eta_2, \dots, \eta_M)^{\rm T}$, $\int d^2\vec{\eta} = \prod_{m=1}^M\int d^2\eta_m = \prod_{m}\int_{-\infty}^\infty d\eta_m^{\rm re} \int_{-\infty}^\infty d\eta_m^{\rm im}$ with $\eta_m=\eta_m^{\rm re} + i\eta_m^{\rm im}\in\mathbb{C}$ ($\eta_m^{\rm re}, \eta_m^{\rm im}\in\mathbb{R}$) for $\forall m$, $\cdot$ indicates the inner product, and $\chi_{A}(\vec{\eta},s)$ is the characteristic function.
Here, $\hat{D}(\vec{\eta},s)$ is given by $\hat{D}(\vec{\eta},s) = \bigotimes_{m=1}^M\hat{D}(\eta_m,s)$,
where $\hat{D}(\eta_m,s)$ is defined by using the displacement operator $\hat{D}(\eta_m) =e^{\eta_m\hat{a}_m^\dagger - \eta_m^*\hat{a}_m}$ as
\begin{align}
    \label{eq:s-ordered displacememt operator}
    \hat{D}(\eta_m,s) = \hat{D}(\eta_m)e^{s|\eta_m|^2/2}.
\end{align}
In Eq.~\eqref{eq:s-parametrized mapping of A}, the parameter $s$ takes $-1\leq s \leq 1$ and characterizes the ordering of bosonic creation and operators \cite{Cahill1,Cahill2}.
We can show that $A_s^*(\vec{\alpha},\vec{\alpha}^*)$ is the phase-space representation of $\hat{A}^{\dagger}$, i.e., $A_s^*(\vec{\alpha},\vec{\alpha}^*) = [\hat{A}^{\dagger}]_s(\vec{\alpha},\vec{\alpha}^*)$, by taking the complex conjugate of Eq.~\eqref{eq:s-parametrized mapping of A} and replacing $\vec{\eta}$ with $-\vec{\eta}$.
Subsequently,
we can also show that $A_s^*(\vec{\alpha},\vec{\alpha}^*) = A_s(\vec{\alpha},\vec{\alpha}^*) \in \mathbb{R}$ if $\hat{A} = \hat{A}^{\dagger}$.
We refer to $A_s(\vec{\alpha},\vec{\alpha}^*)$ as the $s$-ordered phase-space representation of $\hat{A}$.
Although the basic formulation in the remainder of this section applies to $-1 \leq s \leq 1$,
in the subsequent sections,
we focus on integer $s$ ($=0,\pm1$).

We further introduce the function $A^e_s(\vec{\alpha} + \vec{\zeta},\vec{\alpha}^* + \vec{\xi}^*)$, whose arguments are not in the complex conjugated pairs, as 
\begin{align}
    \label{eq:definition of extended s-ordered phase-space representation of A}
    A^e_{s}(\vec{\alpha} + \vec{\zeta},\vec{\alpha}^* + \vec{\xi}^*) = {\rm exp}\left\{\sum_{m=1}^M\left(\zeta_{m}\frac{\partial}{\partial\alpha_{m}} + \xi^*_{m}\frac{\partial}{\partial\alpha^*_{m}}\right)\right\}A_{s}(\vec{\alpha},\vec{\alpha}^*).
\end{align}
This is equivalent to the one obtained by formally replacing the arguments $\vec{\alpha}$ and $\vec{\alpha}^*$ with $\vec{\alpha} + \vec{\zeta}$ and $\vec{\alpha}^* + \vec{\xi}^*$, respectively, in $A_s(\vec{\alpha},\vec{\alpha}^*)$.
Accordingly, $A^e_s(\vec{\alpha},\vec{\alpha}^*)=A_s(\vec{\alpha},\vec{\alpha}^*)$ holds.
Here, we make two remarks about Eq.~\eqref{eq:definition of extended s-ordered phase-space representation of A}.
First, $A^e_s(\vec{\alpha} + \vec{\zeta},\vec{\alpha}^* + \vec{\xi}^*)$ is not the same as the one defined in the doubled phase-space representation, such as the positive-P representation \cite{Gardiner,Drummond1980,Oliveira,Plimak}.
To avoid confusion, in this paper, we refer to the function $A^e_s(\vec{\alpha} + \vec{\zeta},\vec{\alpha}^* + \vec{\xi}^*)$ as the extended $s$-ordered phase-space representation of $\hat{A}$ \cite{Yoneya2026}.
Second, $[A^{e}_s(\vec{\alpha} + \vec{\zeta},\vec{\alpha}^* + \vec{\xi}^*)]^*$ is not the extended $s$-ordered phase-space representation of $\hat{A}^{\dagger}$, where the latter is obtained by replacing $\vec{\alpha}$ and $\vec{\alpha}^*$ in $A_s^*(\vec{\alpha},\vec{\alpha}^*)$ with $\vec{\alpha}+\vec{\zeta}$ and $\vec{\alpha}^*+\vec{\xi}^*$, respectively, and is defined as
\begin{align}
    \label{eq:definition of extended s-ordered phase-space representation of A dagger}
    \bar{A}^e_{s}(\vec{\alpha} + \vec{\zeta},\vec{\alpha}^* + \vec{\xi}^*) = {\rm exp}\left\{\sum_{m=1}^M\left(\zeta_{m}\frac{\partial}{\partial\alpha_{m}} + \xi^*_{m}\frac{\partial}{\partial\alpha^*_{m}}\right)\right\}A^*_{s}(\vec{\alpha},\vec{\alpha}^*).
\end{align}

We specifically refer to the $(-s)$-ordered phase-space representation of the density operator $\hat{\rho}(t)$ as the $s$-ordered quasiprobability distribution function $W_s(\vec{\alpha},\vec{\alpha}^*,t) \in \mathbb{R}$, which is defined by
\begin{gather}
    \label{eq:definition of the s-ordered quasiprobability distribution function}
    W_s(\vec{\alpha},\vec{\alpha}^*,t) = \int\frac{d^2\vec{\eta}}{\pi^M}\chi_{\rho}(\vec{\eta},-s)e^{\vec{\alpha}^*\cdot\vec{\eta} - \vec{\alpha}\cdot\vec{\eta}^*}, \\
    \label{eq:definition of the characteristic function of the s-ordered quasiprobability distribution function}
    \chi_{\rho}(\vec{\eta},-s) = {\rm Tr}\left[\hat{\rho}(t)\hat{D}^{\dagger}(\vec{\eta},s)\right].
\end{gather}
Here, $W_s(\vec{\alpha},\vec{\alpha}^*,t)$ with $s=1,0,$ and $-1$ corresponds to the Glauber-Sudarshan P function, the Wigner function, and the Husimi Q function, respectively.
By using the relation \cite{Cahill1,Cahill2}
\begin{align}
    \label{eq:s-parametrized representation of TrAB}
    {\rm Tr}[\hat{A}\hat{B}] = \int\frac{d^2\vec{\alpha}}{\pi^M}A_s(\vec{\alpha},\vec{\alpha}^*)B_{-s}(\vec{\alpha},\vec{\alpha}^*)
\end{align}
with $\hat{B} = \hat{\rho}(t)$, we can evaluate the expectation value of a physical quantity $\braket{\hat{A}(t)} = {\rm Tr}[\hat{A}\hat{\rho}(t)]$ as
\begin{align}
    \label{eq:physical quantity in the s-ordered phase space}
    \braket{\hat{A}(t)} = \int\frac{d^2\vec{\alpha}}{\pi^M}A_s(\vec{\alpha},\vec{\alpha}^*)W_s(\vec{\alpha},\vec{\alpha}^*,t).
\end{align}
When we choose $\hat{A}$ as the identity operator $\hat{1}$ and use the normalization property of the density operator ${\rm Tr}[\hat{\rho}(t)] = 1$, we obtain the normalization condition for $W_s(\vec{\alpha},\vec{\alpha}^*,t)$:
\begin{align}
    \label{eq:normalization condition for the quasiprobability distribution function}
    \int\frac{d^2\vec{\alpha}}{\pi^M}W_s(\vec{\alpha},\vec{\alpha}^*,t) = 1.
\end{align}
The quasiprobability distribution function can generally take negative values except for the Husimi Q function $W_{s=-1}(\vec{\alpha},\vec{\alpha}^*,t)$, which can take only non-negative values.


\subsection{\label{subsec:Path-integral representation}Path-integral representation}
The phase-space mapping Eq.~\eqref{sec:Review of functional representation of Markovian open quantum systems in the phase space} transforms the GKSL equation~\eqref{eq:def of GKSL equation} into the partial differential equation for the $s$-ordered quasiprobability distribution function in phase space and, its formal solution is given by \cite{Yoneya2026}
\begin{gather}
    \label{eq:path-integral representaiton discrete}
    W_{s}(\vec{\alpha}_{\rm f},\vec{\alpha}^*_{\rm f},t) = \lim_{\Delta t \to 0}\prod_{j=0}^{N_t-1}\int\frac{d^2\vec{\alpha}_j d^2\vec{\eta}_{j+1}}{\pi^{2M}}e^{i\Delta t \mathcal{L}^{s}_j/\hbar} W_{s}(\vec{\alpha}_0,\vec{\alpha}^*_0,t_0), \\
    \label{eq:action discrete}
    \mathcal{L}^{s}_j = i\hbar\left\{\vec{\eta}_{j+1}\cdot\left(\frac{\vec{\alpha}_{j+1}^* - \vec{\alpha}_j^*}{\Delta t}\right) - \vec{\eta}_{j+1}^*\cdot\left(\frac{\vec{\alpha}_{j+1} - \vec{\alpha}_j}{\Delta t}\right)\right\} + H^e_{s}(\vec{\psi}^+_{s,j},\vec{\psi}^{+*}_{-s,j}) - H^e_{s}(\vec{\psi}^-_{-s,j},\vec{\psi}^{-*}_{s,j}) - i\hbar\mathcal{D}_{s}(\vec{\psi}^+_{s,j},\vec{\psi}^{+*}_{-s,j},\vec{\psi}^-_{-s,j},\vec{\psi}^{-*}_{s,j}),
\end{gather}
where $\mathcal{L}^{s}_j$ is the Lagrangian of the system, and we discretize the time interval $[t_0,t]$ into $N_t$ steps of size $\Delta t$:
\begin{align}
    N_t=\frac{t-t_0}{\Delta t},\quad
    t_j = t_0 + j\Delta t,\quad
    t_{N_t} = t,\quad
    \alpha_{N_t} = \alpha_{\rm f}.
\end{align}
The details of the derivation of Eqs.~\eqref{eq:path-integral representaiton discrete} and \eqref{eq:action discrete} are give in Refs.~\cite{Yoneya2025,Yoneya2026}.
In Eqs.~\eqref{eq:path-integral representaiton discrete} and \eqref{eq:action discrete}, we have introduced the fields $\vec{\psi}^{+}_{s,j}$ and $\vec{\psi}^{-}_{s,j}$ as
\begin{gather}
    \label{eq:+ vectors contains alpha and eta}
    \vec{\psi}^+_{s,j} = \left(\alpha_{1,j} + \frac{1+s}{2}\eta_{1,j+1},\alpha_{2,j} + \frac{1+s}{2}\eta_{2,j+1},\dots,\alpha_{M,j} + \frac{1+s}{2}\eta_{M,j+1}\right), \\
    \label{eq:- vectors contains alpha and eta}
    \vec{\psi}^-_{s,j} = \left(\alpha_{1,j} - \frac{1+s}{2}\eta_{1,j+1},\alpha_{2,j} - \frac{1+s}{2}\eta_{2,j+1},\dots,\alpha_{M,j} - \frac{1+s}{2}\eta_{M,j+1}\right),
\end{gather}
and defined $\mathcal{D}_{s}(\vec{\alpha},\vec{\beta},\vec{\gamma},\vec{\delta})$ as
\begin{align}
    \label{eq:definition of the non-unitary term of the propagator of the s-ordered quasiprobability distributino function}
    \mathcal{D}_{s}(\vec{\alpha},\vec{\beta},\vec{\gamma},\vec{\delta}) = \sum_{k=1}^{k_{\rm max}}\gamma_k\left\{\bar{L}^e_{ks}(\vec{\alpha},\vec{\beta})\star_{s} L^e_{ks}(\vec{\gamma},\vec{\delta}) - \frac{1}{2}\bar{L}^e_{ks}(\vec{\alpha},\vec{\beta})\star_{s} L^e_{ks}(\vec{\alpha},\vec{\beta}) - \frac{1}{2}\bar{L}^e_{ks}(\vec{\gamma},\vec{\delta})\star_{s} L^e_{ks}(\vec{\gamma},\vec{\delta})\right\},
\end{align}
where $H^e_{s}(\vec{\alpha},\vec{\beta})$, $L^e_{ks}(\vec{\alpha},\vec{\beta})$,
and $\bar{L}^e_{ks}(\vec{\alpha},\vec{\beta})$ are, respectively, the extended $s$-ordered phase-space representation of $\hat{H}$, $\hat{L}_k$, and $\hat{L}^{\dagger}_k$ defined by Eqs.~\eqref{eq:definition of extended s-ordered phase-space representation of A} and \eqref{eq:definition of extended s-ordered phase-space representation of A dagger}, and we have introduced the differential operator $\star_{s}$ as
\begin{align}
    \label{eq:definition of the s-ordered Moyal product}
    A^e_{s}(\vec{\alpha},\vec{\gamma})\star_{s} B^e_{s}(\vec{\beta},\vec{\delta}) = A^e_{s}(\vec{\alpha},\vec{\gamma})e^{\hat{\phi}_{s}[\vec{\alpha},\vec{\beta},\vec{\gamma},\vec{\delta}]/2}B^e_{s}(\vec{\beta},\vec{\delta}),
\end{align}
where $\hat{\phi}_{s}[\vec{\alpha},\vec{\beta},\vec{\gamma},\vec{\delta}]$ is the differential operator defined by
\begin{align}
    \label{eq:operator phi}
    \hat{\phi}_{s}[\vec{\alpha},\vec{\beta},\vec{\gamma},\vec{\delta}] = \sum_{m=1}^{M}\left\{(1+s)\frac{\overleftarrow{\partial}}{\partial \alpha_m}\frac{\overrightarrow{\partial}}{\partial \delta_m} - (1-s)\frac{\overleftarrow{\partial}}{\partial \gamma_m}\frac{\overrightarrow{\partial}}{\partial \beta_m}\right\}
\end{align}
When we choose $\vec{\beta} = \vec{\alpha}$, $\vec{\gamma} = \vec{\delta} = \vec{\alpha}^*$, and $s=0$ in Eq.~\eqref{eq:definition of the s-ordered Moyal product}, the differential operator $\star_{s}$ reduces to the Moyal product.

Eqs.~\eqref{eq:path-integral representaiton discrete} and \eqref{eq:action discrete} reduce to the path-integral representation for the $s$-ordered quasiprobability distribution function for an isolated system \cite{plimak2009} when we choose $\gamma_k=0$ for $\forall k$, and the one for the Wigner function \cite{Yoneya2025,Polkovnikov2010,Polkovnikov2009} when we choose $s=0$.
From these correspondences, we can respectively regard the fields $\vec{\alpha}_j$ and $\vec{\eta}_{j+1}$ as classical and quantum fields,
where the classical fields describe the classical motion of the system and the quantum fields characterize quantum fluctuations around the classical motion \cite{Polkovnikov2010,Polkovnikov2003,plimak2009,Polkovnikov2009,Marinov,Dittrich2006,Dittrich2010,Gozzi,Pagani}.

In the continuous-time limit,
Eqs.~\eqref{eq:path-integral representaiton discrete} and \eqref{eq:action discrete} can be formally cast into the following path-integral form:
\begin{gather}
    \label{eq:path-integral representaiton continuous}
    W_{s}(\vec{\alpha},\vec{\alpha}^*,t) = \int\mathscr{D}^2\vec{\alpha}\mathscr{D}^2\vec{\eta} e^{i\mathcal{S}[\vec{\alpha},\vec{\eta}]/\hbar}W_{s}(\vec{\alpha}_0,\vec{\alpha}^*_0,t_0), \\
    \label{eq:action continuous}
    \mathcal{S}[\vec{\alpha},\vec{\eta}] = \int^t_{t_0}d\tau \left\{i\hbar\left(\vec{\eta}\cdot\frac{d\vec{\alpha}^*}{d\tau} - \vec{\eta}^*\cdot\frac{d\vec{\alpha}}{d\tau}\right) + H^e_{s}(\vec{\psi}^+_{s},\vec{\psi}^{+*}_{-s}) - H^e_{s}(\vec{\psi}^-_{-s},\vec{\psi}^{-*}_{s}) - i\hbar\mathcal{D}_{s}(\vec{\psi}^+_{s},\vec{\psi}^{+*}_{-s},\vec{\psi}^-_{-s},\vec{\psi}^{-*}_{s})\right\},
\end{gather}
where $\mathcal{S}[\vec{\alpha},\vec{\eta}]$ is the action of the system.
At the boundaries, while the classical fields take $\vec{\alpha}(t_0) = \vec{\alpha}_0$ and $\vec{\alpha}(t) = \vec{\alpha}$, the quantum fields are unconstrained.
Fig.~\ref{fig:Path integral short summary}(a) displays a schematic illustration for the path-integral representation for a system with a single degree of freedom.
When we choose a point in phase space as an initial state, the point moves along infinite paths in the time evolution.
Eq.~\eqref{eq:path-integral representaiton continuous} says that we need to sum up all of the paths with multiplying the appropriate phase factor $e^{i\mathcal{S}[\vec{\alpha},\vec{\eta}]/\hbar}$.
Then, we can obtain the time-evolved $s$-ordered quasiprobability distribution function $W_{s}(\vec{\alpha}_{\rm f},\vec{\alpha}^*_{\rm f},t)$ by applying the same procedure for all initial points in phase space and taking the ensemble average of the results weighted by $W_{s}(\vec{\alpha}_0,\vec{\alpha}^*_0,t_0)$.


\subsection{\label{subsec:Equations of motion in the phase space}Equations of motion in phase space}
\begin{figure}[t]
	\centering 
	\includegraphics[width = \linewidth]{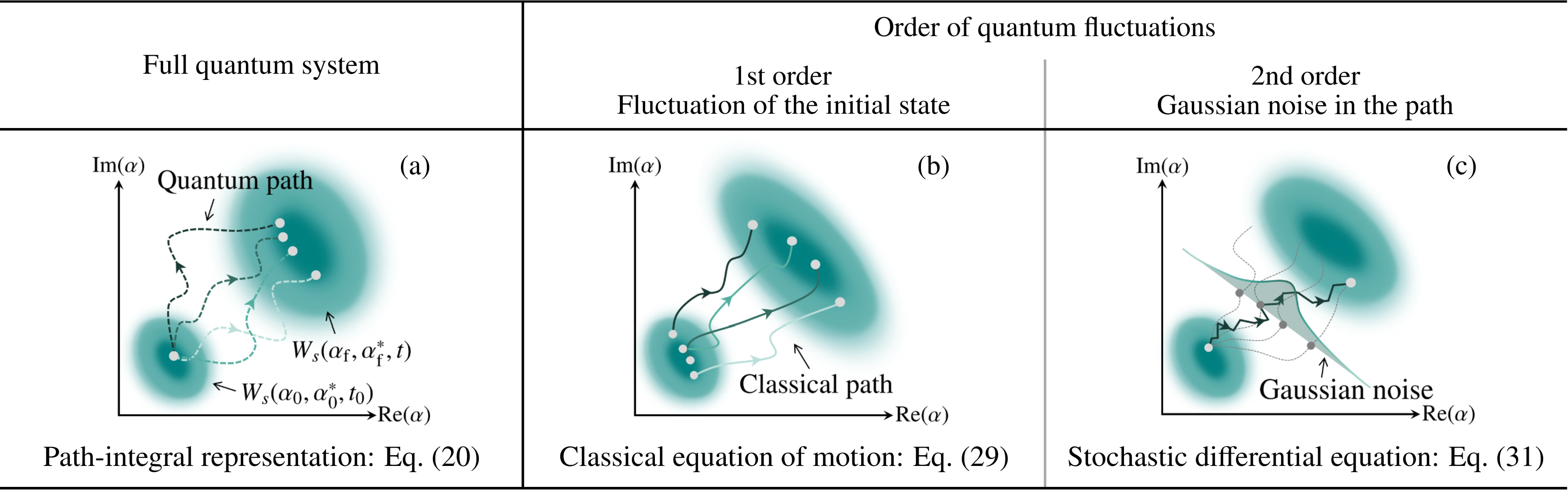}
	\caption{Schematic images of (a) the path-integral representation \eqref{eq:path-integral representaiton continuous} and (b)-(c) the approximations of the GKSL equation in phase space.
(b) Within the first-order approximation,
the GKSL equation is approximated into the generalized Liouville equation \cite{Gerlich,Steeb}, where each point distributed by the initial $s$-ordered quasiprobability distribution function $W_{s}(\vec{\alpha}_0,\vec{\alpha}^*_0,t_0)$ follows the classical equation of motion, the equation of motion of the classical path.
(c) The effects of the second order of the quantum fluctuations are incorporated into the classical path as Gaussian noises,
where each point follows the stochastic differential equation.
Here, the GKSL equation is approximated into the Fokker--Planck
equation.
The details of the generalized Liouville equation and the Fokker--Planck equation are given in Refs.~\cite{Yoneya2025,Yoneya2026}.
This figure is reproduced from Ref.~\cite{Yoneya2026} with modifications.
}
	\label{fig:Path integral short summary}
\end{figure}
By assuming small quantum fluctuations and expanding $\mathcal{L}^{s}_j$ in Eq.~\eqref{eq:action discrete} with respect to the quantum fields $\vec{\eta}_{j+1}$ order by order up to second order, we obtain
\begin{align}
    \label{eq:expansion of the Lagrangian}
    \mathcal{L}^{s}_j = \mathcal{L}^{s(1)}_j + \mathcal{L}^{s(2)}_j + o(\vec{\eta}^2_{j+1}),
\end{align}
where $\mathcal{L}^{s(1)}_j$ and $\mathcal{L}^{s(2)}_j$ respectively denote the first- and second-order contribution of the quantum fields in $\mathcal{L}^{s}_j$:
\begin{gather}
    \label{eq:first order of the action}
    \mathcal{L}^{s(1)}_j = -\sum_{m=1}^{M}\eta^*_{m,j+1}\left\{i\hbar\left(\frac{\alpha_{m,j+1} - \alpha_{m,j}}{\Delta t}\right) - \frac{\partial H_{s}(\vec{\alpha}_j,\vec{\alpha}^*_j)}{\partial\alpha^*_{m,j}}  - \frac{i\hbar}{2}\mathcal{K}^{s}_m(\vec{\alpha}_j,\vec{\alpha}^*_j)\right\} + {\rm c.c.}, \\
    \label{eq:second order of the action}
    \mathcal{L}^{s(2)}_j = \frac{i\hbar}{2}
    \begin{bmatrix}
        \vec{\eta}_{j+1}^{*{\rm T}},\vec{\eta}^{\rm T}_{j+1}
    \end{bmatrix}
    \bm{\mathcal{A}}^{s}(\vec{\alpha}_j,\vec{\alpha}^*_j)
    \begin{bmatrix}
        \vec{\eta}_{j+1} \\
        \vec{\eta}_{j+1}^{*}
    \end{bmatrix}.
\end{gather}
Here,
$\mathcal{K}^s_m$ is the dissipative drift term defined by 
\begin{align}
    \label{eq:def of dissipative drift term}
    \mathcal{K}^{s}_m(\vec{\alpha},\vec{\alpha}^*) = \sum_{k=1}^{k_{\rm max}}\gamma_k\left\{L^*_{ks}(\vec{\alpha},\vec{\alpha}^*)\star_{s}\frac{\partial L_{ks}(\vec{\alpha},\vec{\alpha}^*)}{\partial\alpha^*_{m}} - \frac{\partial L^*_{ks}(\vec{\alpha},\vec{\alpha}^*)}{\partial\alpha^*_{m}}\star_{s}L_{ks}(\vec{\alpha},\vec{\alpha}^*)\right\},
\end{align}
and $\bm{\mathcal{A}}^{s}$ is a $2M\times 2M$ Hermitian matrix given by
\begin{align}
    \label{eq:definitnioa of the diffusion matrix A}
    \bm{\mathcal{A}}^{s} = 2
    \begin{bmatrix}
        \bm{\Lambda}^{s} & \bm{\lambda}^{s} \\
        \bm{\lambda}^{s*} & \bm{\Lambda}^{s*}
    \end{bmatrix},
\end{align}
where $\bm{\lambda}^{s}$ and $\bm{\Lambda}^{s}$ are,
respectively,
$M\times M$ symmetric and Hermitian matrices whose matrix elements are given by
\begin{gather}
    \label{eq:definitnion of lambdamn}
    \lambda^{s}_{mn} = \sum_{k=1}^{k_{\rm max}}\frac{\gamma_k}{4}\left\{\frac{\partial L^*_{ks}}{\partial\alpha^*_{m}}\star_{s} \frac{\partial L_{ks}}{\partial\alpha^*_{n}} + \frac{\partial L^*_{ks}}{\partial\alpha^*_{n}}\star_{s} \frac{\partial L_{ks}}{\partial\alpha^*_{m}}- s\left(L^*_{ks}\star_{s}\frac{\partial^2 L_{ks}}{\partial\alpha^*_{m}\partial\alpha^*_{n}} - \frac{\partial^2 L^*_{ks}}{\partial\alpha^*_{m}\partial\alpha^*_{n}}\star_{s} L_{ks}\right)\right\} + \frac{is}{2\hbar}\frac{\partial^2 H_{s}}{\partial\alpha^*_{m}\partial\alpha^*_{n}}, \\
    \label{eq:definitnion of Lambdamn}
    \Lambda^{s}_{mn} = \sum_{k=1}^{k_{\rm max}}\frac{\gamma_k}{4}\left\{(1-s)\frac{\partial L^*_{ks}}{\partial\alpha^*_{m}}\star_{s} \frac{\partial L_{ks}}{\partial\alpha_{n}} + (1+s)\frac{\partial L^*_{ks}}{\partial\alpha_{n}}\star_{s} \frac{\partial L_{ks}}{\partial\alpha^*_{m}}\right\}.
\end{gather}
We can obtain the equations of motion at each order of quantum fluctuations by approximating $\mathcal{L}^{s}_j$ as $\mathcal{L}^{s(1)}_j$ or $\mathcal{L}^{s(1)}_j + \mathcal{L}^{s(2)}_j$ in Eq.~\eqref{eq:path-integral representaiton discrete} and performing the integration with respect to the quantum fields.
We summarize the schematic images of the dynamics in phase space at first- and second-order approximation in phase space in Figs.~\ref{fig:Path integral short summary}(b) and (c), respectively

Within the first-order approximation [Fig.~\ref{fig:Path integral short summary}(b)], the each point in phase space follows the classical path described by the following classical equation of motion:
\begin{align}
        \label{eq:classical equation of motion}
        i\hbar\frac{d\alpha_m}{dt} = \frac{\partial H_s}{\partial\alpha^*_{m}} + \frac{i\hbar}{2}\sum_{k=1}^{k_{\rm max}}\gamma_k\left(L^*_{ks}\star_s\frac{\partial L_{ks}}{\partial\alpha^*_m} - \frac{\partial L^*_{ks}}{\partial\alpha^*_m}\star_sL_{ks}\right).
\end{align}
Here, $W_s(\vec{\alpha},\vec{\alpha}^*,t)$ follows the generalized Liouville equation whose detailed expression is given in Refs.~\cite{Yoneya2025,Yoneya2026}.

By taking the effects of the second-order contributions of quantum fluctuations [Fig.~\ref{fig:Path integral short summary}(c)], we obtain the stochastic differential equation that governs the motion of points in phase space.
In that case,
$W_s(\vec{\alpha},\vec{\alpha}^*,t)$ follows the Fokker--Planck equation \cite{Yoneya2025,Yoneya2026}.
Here, however, the stochastic differential equation is not always obtainable depending on the details of the Hamiltonian and jump operators and the choice of the quasiprobability distribution function.
When the matrix $\bm{\mathcal{A}}^s$ is positive semidefinite,
i.e.,
\begin{gather}
    \label{eq:condition for obtaining the stochastic differential equation}
    \bm{\mathcal{A}}^s\succeq0,
\end{gather}
we obtain the following stochastic differential equation:
\begin{align}
    \label{eq:stochastic differential equation using B}
    i\hbar d\alpha_m = \left[\frac{\partial H_s}{\partial\alpha^*_{m}} + \frac{i\hbar}{2}\sum_{k=1}^{k_{\rm max}}\gamma_k\left(L^*_{ks}\star_s\frac{\partial L_{ks}}{\partial\alpha^*_m} - \frac{\partial L^*_{ks}}{\partial\alpha^*_m}\star_sL_{ks}\right)\right]dt + i\hbar\left[\bm{\mathcal{B}}^s\cdot d\overrightarrow{\mathcal{W}}(t)\right]_m.
\end{align}
The details of the derivation of Eq.~\eqref{eq:stochastic differential equation using B} is given in \ref{appendix:Derivations of the Stochastic differential equation}.
In Eq.~\eqref{eq:stochastic differential equation using B}, $\cdot$ denotes the Ito product \cite{Risken} and $\overrightarrow{\mathcal{W}}(t) \in \mathbb{R}^{2M}$ is a real stochastic process vector whose components are Wiener processes and independent each other, i.e., the changes of $\Delta\mathcal{W}_{\mu} = \mathcal{W}_{\mu}(t + \Delta t) - \mathcal{W}_{\mu}(t)$ ($\mu = 1,2,\dots,2M$) in the time interval $\Delta t$ obey the following Gaussian distribution function:
\begin{align}
    \label{eq:Wiener process}
    P\left[\Delta \mathcal{W}_{\mu}\right] &= \frac{1}{\sqrt{2\pi\Delta t}}{\rm exp}\left(-\frac{\Delta \mathcal{W}_{\mu}^2}{2\Delta t}\right),
\end{align}
and $\bm{\mathcal{B}}^s$ is a $2M\times 2M$ complex matrix:
\begin{align}
    \label{eq:def of matrix B}
    \bm{\mathcal{B}}^s = i\bm{\mathcal{P}}\bm{\mathcal{C}}^{s},
\end{align}
where $\bm{\mathcal{P}}$ is a $2M\times 2M$ unitary matrix given by
\begin{align}
    \label{eq:def of matrix P}
    \bm{\mathcal{P}} = \frac{1}{\sqrt{2}}
    \begin{bmatrix}
        \bm{1} & i\bm{1} \\
        \bm{1} & -i\bm{1}
    \end{bmatrix}
\end{align}
with $\bm{1}$ being the $M\times M$ identity matrix,
and $\bm{\mathcal{C}}^{s}$ is a $2M\times 2M$ real matrix defined by decomposing the real symmetric matrix $\bm{\mathcal{P}}^{\dagger}\bm{\mathcal{A}}^{s}\bm{\mathcal{P}}$ as 
\begin{align}
    \label{eq:def of matrix C}
    \bm{\mathcal{P}}^{\dagger}\bm{\mathcal{A}}^{s}\bm{\mathcal{P}} = \bm{\mathcal{C}}^{s}\bm{\mathcal{C}}^{s\rm T}.
\end{align}
Here, the matrix decomposition Eq.~\eqref{eq:def of matrix C} is feasible only when the matrix $\bm{\mathcal{A}}^{s}$ is positive semidefinite.
In general, to solve the stochastic differential equation~\eqref{eq:stochastic differential equation using B},
we first confirm at each time step that all eigenvalues of 
$\bm{\mathcal{A}}^{s}$ are non-negative,
and then numerically decompose $\bm{\mathcal{A}}^{s}$ as in Eq.~\eqref{eq:def of matrix C} to obtain $\bm{\mathcal{B}}^{s}$.
However,
as shown in Sec.~\ref{sec:Feasibility of the second-order approximation using the stochastic differential equation},
for systems satisfying the conditions Eqs.~\eqref{eq:condition for lambda_mn two jumps} and \eqref{eq:condition for Lambda_mn two jumps},
by appropriately choosing the value of $s$ depending on the form of $\hat{H}$ and $\hat{L}_k$,
we can analytically determine $\bm{\mathcal{B}}^{s}$,
thereby avoiding the diagonalization and decomposition of $\bm{\mathcal{A}}^s$ and significantly reducing the numerical cost.

From Eq.~\eqref{eq:physical quantity in the s-ordered phase space},
we can write the expectation value of a physical quantity $\braket{\hat{A}(t)}$ at each order of quantum fluctuations as
\begin{align}
    \label{eq:physica quantity at each order of quantum fluctuations}
    \braket{\hat{A}(t)} = \int\frac{d^2\vec{\alpha}_0}{\pi^M}A_s(\vec{\alpha}_{\rm approx}(t),\vec{\alpha}^*_{\rm approx}(t))W_s(\vec{\alpha}_0,\vec{\alpha}^*_0),
\end{align}
where $\vec{\alpha}_{\rm approx}(t)$ obeys the classical equation of motion~\eqref{eq:classical equation of motion} and stochastic differential equation~\eqref{eq:stochastic differential equation using B} within the first- and second-order approximation, respectively.
Eq.~\eqref{eq:physica quantity at each order of quantum fluctuations} means that we can calculate $\braket{\hat{A}(t)}$ by a Monte Carlo simulation:
Within the second-order [first-order] quantum fluctuations, we iteratively solve Eq.~\eqref{eq:stochastic differential equation using B} [Eq.~\eqref{eq:classical equation of motion}] for $\forall m$ with various initial conditions stochastically sampled from $W_{s}(\vec{\alpha}_0,\vec{\alpha}^*_0,t_0)$, calculate $A_{s}(\vec{\alpha}_{\rm approx}(t),\vec{\alpha}^*_{\rm approx}(t))$, and take the ensemble average over the results.


\section{\label{sec:Feasibility of the second-order approximation using the stochastic differential equation}Feasibility of the second-order approximation using the stochastic differential equation}
Within the second-order approximation, the stochastic differential equation~\eqref{eq:stochastic differential equation using B} is obtainable when the matrix $\bm{\mathcal{A}}^s$ is positive semidefinite.
Here,
we discuss sufficient conditions under which $\bm{\mathcal{A}}^s$ is positive semidefinite.
In Sec.~\ref{subsec:Sufficient conditions for A>0},
we present the sufficient conditions Eqs.~\eqref{eq:condition for lambda_mn two jumps} and \eqref{eq:condition for Lambda_mn two jumps} and provide the explicit expression for the matrix $\bm{\mathcal{B}}^s$ in the stochastic differential equation~\eqref{eq:stochastic differential equation using B}.
In Tab.~\ref{tab:conditions for positive-semidefinite},
we summarize examples of the system setups that satisfy the conditions.
We generalize the results in Sec.~\ref{subsec:Generalization}.
In Sec.~\ref{subsec:Additional sufficient condition for the Wigner function},
we show an additional condition on quadratic jump operators satisfying $\bm{\mathcal{A}}^{s=0}\succeq 0$ for the Wigner function ($s=0$),
and in practice,
derive the stochastic differential equations for a one-dimensional lattice model.


\subsection{\label{subsec:Sufficient conditions for A>0}Sufficient conditions for \texorpdfstring{$\bm{\mathcal{A}^{s}}\succeq 0$}{TEXT}}
\begin{table}[t]
    \centering
    \caption{Representative examples of system setups for $\bm{\mathcal{A}}^s\succeq 0$.
    We consider systems with one ($k_{\rm max} = 1$) or two ($k_{\rm max} = 2$) jump operators that involve terms up to quadratic order in $\hat{a}^{\dagger}_m$ and $\hat{a}_m$. The linear and quadratic contributions are denoted by $\hat{L}^{(1)}_k$ and $\hat{L}^{(2)}_k$,
    respectively.
    For systems with two jump operators ($k_{\rm max}=2$),
    the operators are labeled by $k=1,2$,
    whereas the subscript $k$ is omitted for systems with a single jump operator ($k_{\rm max}=1$).
    Here,
    $\land$ denotes logical AND.
    For all cases, the Hamiltonian must satisfy Eq.~\eqref{eq:condition for the Hamiltonian}.}
    \label{tab:conditions for positive-semidefinite}
    \resizebox{\textwidth}{!}{
        \begin{tabular}{ccccccccc}
            \hline\hline
             & $s$ & $k_{\rm max}$ & $\gamma_k$ & $L^{(1)}_{ks}$ & $L^{(2)}_{ks}$ & $H_s$ & $\mathcal{F}^s_m$ & $\mathcal{G}^s_m$\\ \hline
            (1--i) & $0,\pm1$ & $1$ & $\gamma$ & $\displaystyle \frac{\partial L^{(1)}_{s}}{\partial\alpha^*_m}=0$ & $L^{(2)}_s = 0$ & $\displaystyle s\frac{\partial^2 H_s}{\partial\alpha_m\partial\alpha_n} = 0$ & $0$ & $\displaystyle \sqrt{\frac{\gamma}{2}(1-s)}\frac{\partial L_{s}}{\partial\alpha_{m}}$\\
            (1--ii) & $0,\pm1$ & $1$ & $\gamma$ & $\displaystyle \frac{\partial L^{(1)}_{s}}{\partial\alpha_m}=0$ & $L^{(2)}_s = 0$ & $\displaystyle s\frac{\partial^2 H_s}{\partial\alpha_m\partial\alpha_n} = 0$ & $\displaystyle \sqrt{\frac{\gamma}{2}(1+s)}\frac{\partial L_{s}}{\partial\alpha^*_{m}}$ & 0\\
            (1--iii) & $0$ & $1$ & $\gamma$ & No restriction & $L^{(2)}_{s=0} = L^{(2)*}_{s=0}$ & No restriction & $\displaystyle\sqrt{\frac{\gamma}{2}}\frac{\partial L_{s=0}}{\partial\alpha^*_m}$ & $\displaystyle\sqrt{\frac{\gamma}{2}}\frac{\partial L_{s=0}}{\partial\alpha_m}$\\
            (1--iv) & $\pm1$ & $1$ & $\gamma$ & $L^{(1)}_{s=\pm1} = L^{(1)*}_{s=\pm1}$ & $\displaystyle L^{(2)}_{s=\pm1} = L^{(2)*}_{s=\pm1}\land\frac{\partial^2 L^{(2)}_{s=\pm1}}{\partial\alpha^*_m\partial\alpha^*_n}=0$ & $\displaystyle \frac{\partial^2 H_{s=\pm1}}{\partial\alpha_m\partial\alpha_n} = 0$ & $\displaystyle\sqrt{\frac{\gamma}{2}}\frac{\partial L_{s=\pm1}}{\partial\alpha^*_m}$ & $\displaystyle\sqrt{\frac{\gamma}{2}}\frac{\partial L_{s=\pm1}}{\partial\alpha_m}$\\
            (1--v) & $0$ & $2$ & $\gamma_{k=1} = \gamma_{k=2}$ & $L^{(1)}_{k=1,s=0} = L^{(1)*}_{k=2,s=0}$ & $L^{(2)}_{k=1,s=0} = L^{(2)*}_{k=2,s=0}$ & No restriction &$\displaystyle\sqrt{\gamma_{k=1}}\frac{\partial L_{k=1,s=0}}{\partial\alpha^*_m}$ & $\displaystyle\sqrt{\gamma_{k=1}}\frac{\partial L_{k=1,s=0}}{\partial\alpha_m}$\\
            (1--vi) & $\pm1$ & $2$ & $\gamma_{k=1} = \gamma_{k=2}$ & $L^{(1)}_{k=1,s=\pm1} = L^{(1)*}_{k=2,s=\pm1}$ & $\displaystyle L^{(2)}_{k=1,s=\pm1} = L^{(2)*}_{k=2,s=\pm1}\land\frac{\partial^2 L^{(2)}_{k=1,s=\pm1}}{\partial\alpha^*_m\partial\alpha^*_n}=\frac{\partial^2 L^{(2)}_{k=1,s=\pm1}}{\partial\alpha_m\partial\alpha_n}=0$ & $\displaystyle \frac{\partial^2 H_{s=\pm1}}{\partial\alpha_m\partial\alpha_n} = 0$ &$\displaystyle\sqrt{\gamma_{k=1}}\frac{\partial L_{k=1,s=\pm1}}{\partial\alpha^*_m}$ & $\displaystyle\sqrt{\gamma_{k=1}}\frac{\partial L_{k=1,s=\pm1}}{\partial\alpha_m}$\\ \hline\hline
        \end{tabular}
    }
\end{table}

When we can write $\lambda^s_{mn}$ and $\Lambda^s_{mn}$ as
\begin{gather}
    \label{eq:condition for lambda_mn two jumps}
    \lambda^{s}_{mn} = \frac{1}{2}(\mathcal{G}^{s*}_m\mathcal{F}^{s}_n + \mathcal{F}^{s}_m\mathcal{G}^{s*}_n), \\
    \label{eq:condition for Lambda_mn two jumps}
    \Lambda^{s}_{mn} = \frac{1}{2}(\mathcal{G}^{s*}_m\mathcal{G}^{s}_n + \mathcal{F}^{s}_m\mathcal{F}^{s*}_n)
\end{gather}
with $\mathcal{G}^{s}_m(\vec{\alpha},\vec{\alpha}^*)$ and $\mathcal{F}^{s}_m(\vec{\alpha},\vec{\alpha}^*)$ being arbitrary complex functions, the matrix $\bm{\mathcal{A}}^s$ becomes positive semidefinite (see \ref{appendix:Positive-semidefiniteness of A}).
Indeed,
in this case,
we can analytically decompose $\bm{\mathcal{A}}^s$ and obtain $\bm{\mathcal{B}}^s$ as
\begin{align}
    \label{eq:block component of matrix Bs}
    \bm{\mathcal{B}}^{s} =
    \begin{bmatrix}
        \bm{B}^s_{11} & \bm{B}^s_{12} \\
        \bm{B}^s_{21} & \bm{B}^s_{22}
    \end{bmatrix},
\end{align}
where $\bm{B}^s_{11}$, $\bm{B}^s_{12}$, $\bm{B}^s_{21}$, and $\bm{B}^s_{22}$ are $M\times M$ complex matrices given by
\begin{align}
    \label{eq:matrix B using G and F}
    [\bm{B}^{s}_{11}]_{mn} &= \frac{i}{\sqrt{2}}\left(\mathcal{F}^{s}_m + \mathcal{G}^{s*}_m\right)e_n,\quad
    [\bm{B}^{s}_{12}]_{mn} = \frac{1}{\sqrt{2}}\left(\mathcal{F}^{s}_m - \mathcal{G}^{s*}_m\right)e_n,\quad \bm{B}^{s}_{21} = -\bm{B}^{s*}_{11},\quad \bm{B}^{s}_{22} = -\bm{B}^{s*}_{12},
\end{align}
with $e_n$ being the $n$th component of a normalized real vector $\vec{e}\in \mathbb{R}^{M}$.
By substituting $\bm{\mathcal{B}}^{s}$ into the stochastic term of Eq.~\eqref{eq:stochastic differential equation using B} and setting $\vec{e} = (1,0,\dots,0)$, we obtain
\begin{align}
    \label{eq:stochastic term for systems with two jumps}
    \left[\bm{\mathcal{B}}^{s}\cdot d\overrightarrow{\mathcal{W}}(t)\right]_m = \frac{i}{\sqrt{2}}\left\{(\mathcal{F}^{s}_m + \mathcal{G}^{s*}_m)\cdot d\mathcal{W}_{+}(t) - i(\mathcal{F}^{s}_m - \mathcal{G}^{s*}_m)\cdot d\mathcal{W}_{-}(t)\right\},
\end{align}
where $\mathcal{W}_+ = \mathcal{W}_1$ and $\mathcal{W}_- = \mathcal{W}_{M+1}$.
In the general form of Eq.~\eqref{eq:stochastic differential equation using B}, we need at most $2M$ stochastic processes $\mathcal{W}_\mu$ for $\mu = 1,2,\dots,2M$.
However,
the number of stochastic processes can be reduced by appropriately choosing $\bm{\mathcal{C}}^s$,
i.e.,
choosing $\vec{e}$,
since $\bm{\mathcal{C}}^s$ is non-unique and can be multiplied on the right by an arbitrary orthogonal matrix $\bm{\mathcal{Q}}$,
i.e.,
$\bm{\mathcal{C}}^s \to \bm{\mathcal{C}}^s\bm{\mathcal{Q}}$.
Systems satisfying Eqs.~\eqref{eq:condition for lambda_mn two jumps} and \eqref{eq:condition for Lambda_mn two jumps} correspond to a special case and can be described by only two stochastic processes,
$\mathcal{W}_+$ and $\mathcal{W}_-$.
Here, we note that although the derivation of Eq.~\eqref{eq:matrix B using G and F} is heuristic, we have confirmed that $\bm{\mathcal{B}}^s$ in Eqs.~\eqref{eq:block component of matrix Bs} and \eqref{eq:matrix B using G and F} satisfies Eqs.~\eqref{eq:def of matrix B} and \eqref{eq:def of matrix C}.
The conditions~\eqref{eq:condition for lambda_mn two jumps} and \eqref{eq:condition for Lambda_mn two jumps} allow for jump operators that couple different degrees of freedom, and therefore apply to a broader class of systems than those considered in our previous works \cite{Yoneya2025,Yoneya2026}.

In Tab.~\ref{tab:conditions for positive-semidefinite}, we summarize examples for the Hamiltonian, jump operators, and the choice of $s$ for $\lambda^s_{mn}$ and $\Lambda^s_{mn}$ satisfying Eqs.~\eqref{eq:condition for lambda_mn two jumps} and \eqref{eq:condition for Lambda_mn two jumps}.
In all cases considered here,
the Hamiltonian satisfies
\begin{align}
    \label{eq:condition for the Hamiltonian}
    s\frac{\partial^2 H_s}{\partial\alpha_m\partial\alpha_n} = 0
\end{align}
for $\forall m,n$ so that the matrix $\bm{\mathcal{A}}^s$ does not contain any Hamiltonian-dependent terms.
Consequently,
the condition $\bm{\mathcal{A}}^s\succeq0$ depends only on the details of the jump operators.
For $s=\pm1$,
condition~\eqref{eq:condition for the Hamiltonian} is satisfied by the non-interacting Hamiltonian,
whereas for $s=0$ it is satisfied regardless of the details of the Hamiltonian.
The jump operators considered in Tab.~\ref{tab:conditions for positive-semidefinite} consist of linear ($\hat{L}_k^{(1)}$) and quadratic ($\hat{L}_k^{(2)}$) terms in $\hat{a}_m$ and $\hat{a}_m^\dagger$,
and we restrict ourselves to cases with one or two jump operators ($k_{\rm max}=1$ or $2$). For systems with two jump operators ($k_{\rm max}=2$), the operators are labeled by $k=1,2$,
whereas the subscript $k$ is omitted for systems with a single jump operator ($k_{\rm max}=1$).
Although the examples in Tab.~\ref{tab:conditions for positive-semidefinite} are restricted to linear and quadratic jump operators,
Eqs.~\eqref{eq:condition for lambda_mn two jumps} and \eqref{eq:condition for Lambda_mn two jumps} remain valid even when the jump operators include higher-body interaction terms,
thereby allowing for the construction of a broader class of jump operators $\hat{L}_k$.

We note that Eqs.~\eqref{eq:condition for lambda_mn two jumps} and \eqref{eq:condition for Lambda_mn two jumps} are not necessary conditions for $\bm{\mathcal{A}}^s$ to be positive semidefinite and that $\bm{\mathcal{A}}^s$ may remain positive semidefinite even when they are not satisfied.
In such cases,
the stochastic terms differ from those in Eq.~\eqref{eq:stochastic term for systems with two jumps} \cite{Yoneya2026,Huberskin}.


\subsection{\label{subsec:Generalization}Generalization}
We generalize the results in Sec.~\ref{subsec:Sufficient conditions for A>0} to cases where $\lambda^s_{mn}$ and $\Lambda^s_{mn}$ can be decomposed into sums of contributions satisfying Eqs.~\eqref{eq:condition for lambda_mn two jumps} and \eqref{eq:condition for Lambda_mn two jumps},
i.e.,
\begin{gather}
    \label{eq:decomposition of lambda}
    \lambda^{s}_{mn} = \sum_{\ell = 1}^{\ell_{\rm max}}\lambda^{s(\ell)}_{mn}, \\
    \label{eq:decomposition of Lambda}
    \Lambda^{s}_{mn} = \sum_{\ell = 1}^{\ell_{\rm max}}\Lambda^{s(\ell)}_{mn},
\end{gather}
where the subscript $\ell$ labels the different contributions $(\ell = 1,2,\dots,\ell_{\rm max})$,
and $\lambda^{s(\ell)}_{mn}$ and $\Lambda^{s(\ell)}_{mn}$ are given by
\begin{gather}
    \label{eq:condition for lambda_mn multiple jumps}
    \lambda^{s(\ell)}_{mn} = \frac{1}{2}(\mathcal{G}^{s(\ell)*}_m\mathcal{F}^{s(\ell)}_n + \mathcal{F}^{s(\ell)}_m\mathcal{G}^{s(\ell)*}_n), \\
    \label{eq:condition for Lambda_mn multiple jumps}
    \Lambda^{s(\ell)}_{mn} = \frac{1}{2}(\mathcal{G}^{s(\ell)*}_m\mathcal{G}^{s(\ell)}_n + \mathcal{F}^{s(\ell)}_m\mathcal{F}^{s(\ell)*}_n)
\end{gather}
with $\mathcal{G}^{s(\ell)}_m(\vec{\alpha},\vec{\alpha}^*)$ and $\mathcal{F}^{s(\ell)}_m(\vec{\alpha},\vec{\alpha}^*)$ being arbitrary complex functions.
Under this decomposition,
we can also show that the matrix $\bm{\mathcal{A}}^s$ is positive semidefinite:
From Eqs.~\eqref{eq:decomposition of lambda} and \eqref{eq:decomposition of Lambda},
we can rewrite $\bm{\mathcal{A}}^s$ as
\begin{align}
    \label{eq:matrix decomposing of A}
    \bm{\mathcal{A}}^{s} = \sum_{\ell=1}^{\ell_{\rm max}}\bm{\mathcal{A}}^{s(\ell)}
\end{align}
with
\begin{align}
    \label{eq:definitnioa of the diffusion matrix A(ell)}
    \bm{\mathcal{A}}^{s(\ell)} = 2
    \begin{bmatrix}
        \bm{\Lambda}^{s(\ell)} & \bm{\lambda}^{s(\ell)} \\
        \bm{\lambda}^{s(\ell)*} & \bm{\Lambda}^{s(\ell)*}
    \end{bmatrix},
\end{align}
where $\bm{\lambda}^{s(\ell)}$ and $\bm{\Lambda}^{s(\ell)}$ are $M\times M$ symmetric and Hermitian matrices whose matrix elements are given by Eqs.~\eqref{eq:condition for lambda_mn multiple jumps} and \eqref{eq:condition for Lambda_mn multiple jumps},
respectively;
From Sec.~\ref{subsec:Sufficient conditions for A>0},
since $\bm{\mathcal{A}}^{s(\ell)}\succeq 0$ for $\forall \ell$,
we can show that $\bm{\mathcal{A}}^s\succeq 0$.
The corresponding stochastic differential equation is given by
\begin{align}
    \label{eq:stochastic differential equation using B(ell)}
    i\hbar d\alpha_m = \left[\frac{\partial H_s}{\partial\alpha^*_{m}} + \frac{i\hbar}{2}\sum_{k=1}^{k_{\rm max}}\gamma_k\left(L^*_{ks}\star_s\frac{\partial L_{ks}}{\partial\alpha^*_m} - \frac{\partial L^*_{ks}}{\partial\alpha^*_m}\star_sL_{ks}\right)\right]dt + i\hbar\sum_{\ell=1}^{\ell_{\rm max}}\left[\bm{\mathcal{B}^{s(\ell)}}\cdot d\overrightarrow{\mathcal{W}}^{(\ell)}(t)\right]_m,
\end{align}
which we derive in \ref{appendix:Derivations of the Stochastic differential equation}.
In Eq.~\eqref{eq:stochastic differential equation using B(ell)},
$\overrightarrow{\mathcal{W}}^{(\ell)}(t) \in \mathbb{R}^{2M}$ is a real stochastic process vector whose components are Wiener processes satisfying Eq.~\eqref{eq:Wiener process} and independent each other,
and $\bm{\mathcal{B}^{s(\ell)}}$ is a $2M\times 2M$ complex matrix given by
\begin{align}
    \label{eq:def of matrix B(ell)}
    \bm{\mathcal{B}^{s(\ell)}} = i\bm{\mathcal{P}}\bm{\mathcal{C}}^{s(\ell)},
\end{align}
with $\bm{\mathcal{C}}^{s(\ell)}$ being a $2M\times 2M$ real matrix:
\begin{align}
    \label{eq:def of matrix C(ell)}
    \bm{\mathcal{P}}^{\dagger}\bm{\mathcal{A}}^{s(\ell)}\bm{\mathcal{P}} = \bm{\mathcal{C}}^{s(\ell)}\bm{\mathcal{C}}^{s(\ell)\rm T}.
\end{align}
Using the results in Sec.~\ref{subsec:Sufficient conditions for A>0},
we can choose $\bm{\mathcal{C}}^{s(\ell)}$ such that only two Wiener processes are required for each $\ell$.
The stochastic term in Eq.~\eqref{eq:stochastic differential equation using B(ell)} can then be written as
\begin{align}
    \label{eq:stochastic term for systems with multiple jumps}
    \left[\bm{\mathcal{B}}^{s(\ell)}\cdot d\overrightarrow{\mathcal{W}}^{(\ell)}(t)\right]_m = \frac{i}{\sqrt{2}}\left\{(\mathcal{F}^{s(\ell)}_m + \mathcal{G}^{s(\ell)*}_m)\cdot d\mathcal{W}^{(\ell)}_{+}(t) - i(\mathcal{F}^{s(\ell)}_m - \mathcal{G}^{s(\ell)*}_m)\cdot d\mathcal{W}^{(\ell)}_{-}(t)\right\},
\end{align}
where $\mathcal{W}^{(\ell)}_+ = \mathcal{W}^{(\ell)}_1$ and $\mathcal{W}^{(\ell)}_- = \mathcal{W}^{(\ell)}_{M+1}$.
Thus, a total of $2\ell_{\max}$ Wiener processes are sufficient to describe the stochastic evolution.


\subsection{\label{subsec:Additional sufficient condition for the Wigner function}Additional sufficient condition for the Wigner function}
Although Eqs.~\eqref{eq:decomposition of lambda} and \eqref{eq:decomposition of Lambda}  do not provide an explicit condition on the operators, for the Wigner function we heuristically find an explicit condition for the jump operators that satisfy Eqs.~\eqref{eq:decomposition of lambda} and \eqref{eq:decomposition of Lambda} under the assumption that the jump operators are at most quadratic in $\hat{a}^{\dagger}_m$ and $\hat{a}_m$.
The resulting condition is given by
\begin{align}
    \label{eq:cond for A>0 for Wigner}
    \sum_{k=1}^{k_{\rm max}}\gamma_k\left[\hat{L}^{\dagger}_k,\hat{L}_k\right]_- = \sum_{m=1}^M(l_m\hat{a}_m + \bar{l}_m\hat{a}^{\dagger}_m) + {\rm Const.},
\end{align}
where $l_m,\bar{l}_m\in\mathbb{C}$.
For $s=0$,
Eq.~\eqref{eq:condition for the Hamiltonian} imposes no restriction on the Hamiltonian $\hat{H}$,
which may therefore contain higher-body interaction terms.
When the right-hand side of Eq.~\eqref{eq:cond for A>0 for Wigner} vanishes, as in cases (1--iii) and (1--v) of Tab.~\ref{tab:conditions for positive-semidefinite}, Eq.~\eqref{eq:cond for A>0 for Wigner} coincides with the necessary and sufficient for the infinite-temperature state to be a steady-state solution of the GKSL equation~\eqref{eq:def of GKSL equation} \cite{Breuer},
as shown in \ref{appendix:Infinite-temperature steady state}.

When Eq.~\eqref{eq:cond for A>0 for Wigner} is satisfied, the corresponding $\lambda^{s=0}_{mn}$ and $\Lambda^{s=0}_{mn}$ are obtained by replacing $\star_{s=0}$ with the identity operator, namely,
\begin{gather}
    \label{eq:lambda Wigner quadratic}
    \lambda^{s=0}_{mn} = \sum_{k=1}^{k_{\rm max}}\frac{\gamma_k}{4}\left(\frac{\partial L^*_{ks=0}}{\partial\alpha^*_m}\frac{\partial L_{ks=0}}{\partial\alpha^*_n} + \frac{\partial L^*_{ks=0}}{\partial\alpha^*_n}\frac{\partial L_{ks=0}}{\partial\alpha^*_m}\right), \\
    \label{eq:Lambda Wigner quadratic}
    \Lambda^{s=0}_{mn} = \sum_{k=1}^{k_{\rm max}}\frac{\gamma_k}{4}\left(\frac{\partial L^*_{ks=0}}{\partial\alpha^*_m}\frac{\partial L_{ks=0}}{\partial\alpha_n} + \frac{\partial L^*_{ks=0}}{\partial\alpha_n}\frac{\partial L_{ks=0}}{\partial\alpha^*_m}\right).
\end{gather}
The contributions arising from the difference between $\star_{s=0}$ and the identity operator vanish identically.
This can be shown by expressing Eq.~\eqref{eq:cond for A>0 for Wigner} in the phase-space representation and taking the second-order partial derivatives $\partial_{\alpha_m^*}\partial_{\alpha_n^*}$ and $\partial_{\alpha_m^*}\partial_{\alpha_n}$ of both sides.
See \ref{appendix:lambda and Lambda for jump operators satisfying} for the details.
Eqs.~\eqref{eq:lambda Wigner quadratic} and \eqref{eq:Lambda Wigner quadratic} clearly satisfy Eqs.~\eqref{eq:decomposition of lambda} and \eqref{eq:decomposition of Lambda},
respectively,
with $\ell = k$ and $\ell_{\rm max} = k_{\rm max}$.
Here,
$\mathcal{F}^{s=0(k)}_m$ and $\mathcal{G}^{s=0(k)}_m$ are given by
\begin{gather}
    \label{eq:F and G for Wigner positive}
    \mathcal{F}^{s=0(k)}_m = \sqrt{\frac{\gamma_k}{2}}\frac{\partial L_{ks=0}}{\partial\alpha^*_m},\quad \mathcal{G}^{s=0(k)}_m = \sqrt{\frac{\gamma_k}{2}}\frac{\partial L_{ks=0}}{\partial\alpha_m}.
\end{gather}

As an illustrative example satisfying the condition~\eqref{eq:cond for A>0 for Wigner},
we consider a one-dimensional lattice system with unidirectional incoherent hopping under periodic boundary conditions,
described by the following GKSL equation:
\begin{gather}
    \label{eq:GKSL equation discussion}
    \frac{d\hat{\rho}(t)}{dt} = -\frac{i}{\hbar}\left[\hat{H},\hat{\rho}(t)\right]_- + \gamma\sum_{k=1}^M\left(\hat{L}_k\hat{\rho}(t)\hat{L}^{\dagger}_k - \frac{1}{2}\left[\hat{L}^{\dagger}_k\hat{L}_k,\hat{\rho}(t)\right]_+\right),\\
    \label{eq:jump operators discussion}
    \hat{L}_{k} =
    \begin{cases}
        \hat{a}^{\dagger}_{k+1}\hat{a}_k & \text{for}\quad k=1,2,\dots,M-1 \\
        \hat{a}^{\dagger}_1\hat{a}_M & \text{for}\quad k=M
    \end{cases},
\end{gather}
where we assume that the incoherent hopping is unidirectional and that all jump operators have the same strength,
i.e.,
$\gamma_k = \gamma$ for $\forall k$,
and $\hat{H}$ may contain higher-body interaction terms.
In this system,
the jump operators in Eq.~\eqref{eq:jump operators discussion} satisfy the condition~\eqref{eq:cond for A>0 for Wigner},
for which the right-hand side vanishes,
indicating that the infinite-temperature state is a steady-state solution of the GKSL equation~\eqref{eq:GKSL equation discussion}.
Here,
from Eq.~\eqref{eq:F and G for Wigner positive},
$\mathcal{F}^{s=0(k)}_m$ and $\mathcal{G}^{s=0(k)}_m$ are given by
\begin{gather}
    \label{eq:Fm for incoherent hopping}
    \mathcal{F}^{s=0(k)}_m = \sqrt{\frac{\gamma}{2}}\frac{\partial L_{ks=0}}{\partial\alpha^*_m} =
    \begin{cases}
        \displaystyle\sqrt{\frac{\gamma}{2}}\alpha_{m-1} & \text{for}~k=m-1,\\
        0 & \text{for}~\text{others},
    \end{cases} \\
    \label{eq:Gm for incoherent hopping}
    \mathcal{G}^{s=0(k)}_m = \sqrt{\frac{\gamma}{2}}\frac{\partial L_{ks=0}}{\partial\alpha_m} =
    \begin{cases}
        \displaystyle\sqrt{\frac{\gamma}{2}}\alpha^*_{m+1} & \text{for}~ k=m, \\
        0 & \text{for}~\text{others},
    \end{cases}
\end{gather}
where $\alpha_{M+1} = \alpha_1$ and $\alpha_{0} = \alpha_M$ from the periodic boundary condition.
By substituting Eqs.~\eqref{eq:Fm for incoherent hopping} and \eqref{eq:Gm for incoherent hopping} into Eq.~\eqref{eq:stochastic term for systems with multiple jumps} and \eqref{eq:stochastic differential equation using B(ell)},
we obtain the following stochastic differential equation to be solved:
{\small
\begin{align}
    \label{eq:SDEm discussion}
    i\hbar d\alpha_m = \left[\frac{\partial H_s}{\partial\alpha_m^*} + \frac{i\hbar\gamma}{2}\alpha_m(|\alpha_{m-1}|^2 - |\alpha_{m+1}|^2 - 1)\right]dt + i\hbar\sqrt{\frac{\gamma}{4}}\left\{\alpha_{m-1}\cdot\left(id\mathcal{W}^{(m-1)}_+ + d\mathcal{W}^{(m-1)}_-\right) + \alpha_{m+1}\cdot\left(id\mathcal{W}^{(m)}_+ - d\mathcal{W}^{(m)}_-\right)\right\}.
\end{align}}
In Sec.~\ref{subsec:Model2},
we investigate the validity of the stochastic differential equation~\eqref{eq:SDEm discussion} for the three-site ($M=3$) Bose--Hubbard model.

\section{\label{sec:Equation of motion in the high occupancy limit}Equation of motion in the high-occupancy limit}
Generally,
by appropriately scaling the Hamiltonian parameters so that the
thermodynamic limit is well defined,
the mean-field approximation becomes valid in the thermodynamic limit for isolated bosonic systems \cite{Hepp,Spohn,Yaffe}.
In this section,
we consider an analogous limit for open systems,
which we refer to as the high-occupancy limit.
Within our theoretical framework,
the counterpart of the mean-field approximation is given by the classical equation of motion,
Eq.~\eqref{eq:classical equation of motion}.
Interestingly,
the validity of Eq.~\eqref{eq:classical equation of motion} in the high-occupancy limit depends on the form of the jump operators, and Eq.~\eqref{eq:classical equation of motion} can be insufficient for certain classes of jump operators.

In the following,
we first define the high-occupancy limit in Sec.~\ref{subsec:High-occupancy limit}.
In Sec.~\ref{subsec:Dissipative drift and breakdown of the classical description},
we show that the dissipative drift term $\mathcal{K}_m^s$,
which is defined in Eq.~\eqref{eq:def of dissipative drift term},
serves as diagnostic of the breakdown of the classical description.
In particular,
the vanishing of $\mathcal{K}_m^s$ provides a criterion for the breakdown of the classical description,
indicating that the effects of the second-order quantum fluctuations must be taken into account.
In Sec.~\ref{subsec:Positive diffusion approximation for the Wigner function},
we show that,
for the Wigner function,
the positive semidefiniteness of $\bm{\mathcal{A}}^{s=0}$ is always preserved in the high-occupancy limit,
regardless of the details of the jump operators.
This implies that the dynamics of the quantum system can always be described by a stochastic differential equation in this limit.


\subsection{\label{subsec:High-occupancy limit}High-occupancy limit}
We assume that,
on average,
each degree of freedom is occupied by $N$ bosons,
i.e.,
$\braket{\hat{a}_m^\dagger \hat{a}_m} = O(N)$ for $\forall m$.
Consequently,
the relation $\braket{\hat{a}_m^\dagger \hat{a}_m} = |\alpha_m|^2 = O(N)$ implies that $\alpha_m$ scales as $\alpha_m = O(N^{1/2})$.
We define the high-occupancy limit as the limit $N \to \infty$.
We scale the system parameters so that each contribution of $H_s$ and $L_{ks}$ in $\mathcal{L}_j^s$ neither vanishes nor diverges in the high-occupancy limit,
ensuring that both remain physically meaningful.

For clarity, before proceeding to the general discussion, we illustrate the parameter scaling using the Bose--Hubbard Hamiltonian for an isolated system, given by
\begin{align}
    \label{eq:Bose--Hubbard Hamiltonian general}
    \hat{H}_{\rm BH} &= -\mu\sum_{m=1}^M\hat{a}^{\dagger}_m\hat{a}_m - J\sum_{\braket{m,n}}(\hat{a}^{\dagger}_m\hat{a}_n + {\rm h.c.}) + \frac{1}{2}\sum_{m=1}^MU_{mm}\hat{a}^{\dagger}_m\hat{a}^{\dagger}_m\hat{a}_m\hat{a}_m,
\end{align}
where $\mu$ is the chemical potential, $J$ is the hopping amplitude between the nearest-neighboring lattice sites $\braket{m,n}$, and $U_{mm}$ is the on-site interaction energy.
For the one-dimensional case, the corresponding classical equation of motion,
derived from the first-order contribution $\mathcal{L}^{s(1)}_j$,
becomes
\begin{gather}
    \label{eq:classical equation of motion for the Bose Hubbard model}
    i\hbar \frac{d\alpha_m}{dt} = -\mu\alpha_{m} - J(\alpha_{m+1} + \alpha_{m-1}) + U_{mm}\alpha_m(|\alpha_m|^2 - 1 + s).
\end{gather}
For the high-occupation limit to be well defined, the leading-order terms on both sides of Eq.~\eqref{eq:classical equation of motion for the Bose Hubbard model} must have the same scaling order.
Here,
from $\alpha_m=O(N^{1/2})$, the term on the left-hand side of Eq.~\eqref{eq:classical equation of motion for the Bose Hubbard model} and the terms proportional to $\mu$ and $J$ on the right-hand side are $O(N^{1/2})$, whereas the factor multiplying $U_{mm}$ is $O(N^{3/2})$.
Therefore,
$U_{mm}$ must be scaled as $U_{mm} = O(N^{-1})$,
so that $NU_{mm}$ remains constant.
Under this scaling, contributions higher than second order in $\mathcal{L}^s_j$ vanish in the high-occupancy limit, as will be shown in below.

We now perform a scaling analysis of the general form of the Lagrangian in Eq.~\eqref{eq:action discrete}.
From $\alpha_m=O(N^{1/2})$, the first term on the right-hand side of Eq.~\eqref{eq:action discrete} is formally of order $O(\eta N^{1/2})$. Since the dominant contribution to the integral over $\eta_m$ comes from the region $0 < |\eta_m| < N^{-1/2}$,
the relevant values of $\eta_m$ are of order $O(N^{-1/2})$. Consequently, the first term on the right-hand side of Eq.~\eqref{eq:action discrete} is $O(1)$.

We next consider the terms involving $H_s$ [the second and third terms on the right-hand side of Eq.~\eqref{eq:action discrete}
].
Suppose that $H_s = O(N^{n_H})$.
Expanding these terms in powers of $\eta_m$, we find that the $n$-th order term scales as $O(N^{n_H-n})$,
while the zeroth-order term vanishes.
Therefore,
the requirement that the leading non-vanishing contribution to remain $O(1)$ in the high-occupancy limit implies $n_H=1$,
namely,
$\braket{\hat{H}}=O(N)$.
The single-body Hamiltonian naturally satisfies this condition.
For many-body interaction terms,
the coupling constants are scaled so that each interaction contributes at most $O(N)$ to the total energy.
Under this condition,
the contributions from higher-order terms in $\eta_m$ vanish in the high-occupancy limit.
Therefore,
in the case of an isolated system,
$\mathcal{L}_j^s$ is dominated by its lowest-order contribution,
$\mathcal{L}_j^{s(1)}$,
and hence $\mathcal{L}_j^s \to \mathcal{L}_j^{s(1)}$ in the high-occupancy limit.
This corresponds to the well-known fact that the mean-field approximation becomes exact in the thermodynamic limit \cite{Hepp,Spohn,Yaffe}.

\subsection{\label{subsec:Dissipative drift and breakdown of the classical description}Dissipative drift and breakdown of the classical description}
Without loss of generality,
we set $H_s=0$ and estimate the leading-order behavior of $\mathcal{D}_s$ in Eq.~\eqref{eq:action discrete}.
We note that,
from the scaling behavior of $\alpha_m$ and $\eta_m$,
$\star_s$ defined in Eq.~\eqref{eq:definition of the s-ordered Moyal product} scales as
\begin{align}
    \label{eq:star}
    \star_s = 1 + O(1/N).
\end{align}
As in the case of the terms involving $H_s$,
the leading contribution to $\mathcal{D}_s$ in the high-occupancy limit generally arises from the first-order terms in $\eta_m$,
namely,
the term proportional to $\mathcal{K}_m^s$ defined in Eq.~\eqref{eq:def of dissipative drift term}.
However,
depending on the form of the jump operators,
$\mathcal{K}_m^s$ may vanish,
or the contribution proportional to $\eta_m \mathcal{K}_m^s$ may no longer dominate over that arising from $\mathcal{L}_j^{s(2)}$.
Below,
we analyze such cases in detail.

Since the following analysis applies independently to each $k$,
we consider a single jump operator and omit the index $k$.
In general,
$L_s$ is a polynomial in $\alpha_m$ and $\alpha^*_m$ for $m=1,2,\dots,M$.
Let $n_L$ denote the highest scaling exponent of $L_s$ in the high-occupancy limit.
We then define $L_s^{({\rm LO})}$ and $L_s^{({\rm NLO})}$ as the parts of $L_s$ scaling as $O(N^{n_L})$ and $O(N^{n_L-1/2})$,
respectively.
We can expand the dissipative drift term $\mathcal{K}^s_m$ order by order with respect to $N$ as
\begin{align}
    \label{eq:order expansion of the disipative drift term}
    \mathcal{K}^s_m = \mathcal{K}^{s({\rm LO})}_m + \mathcal{K}^{s({\rm NLO})}_m + \gamma O(N^{2n_L - 3/2}),
\end{align}
where $\mathcal{K}^{s({\rm LO})}_m$ and $\mathcal{K}^{s({\rm NLO})}_m$ are respectively given by
\begin{gather}
    \label{eq:K^s_m LO}
    \mathcal{K}^{s(\rm LO)}_m = \gamma\left\{L^{(\rm LO)*}_{s}\frac{\partial L^{(\rm LO)}_{s}}{\partial\alpha^*_{m}} - \frac{\partial L^{(\rm LO)*}_{s}}{\partial\alpha^*_{m}}L^{(\rm LO)}_{s}\right\} = \gamma O(N^{2n_L - 1/2}),\\
    \label{eq:K^s_m NLO}
    \mathcal{K}^{s(\rm NLO)}_m = \gamma\left\{L^{(\rm LO)*}_{s}\frac{\partial L^{(\rm NLO)}_{s}}{\partial\alpha^*_{m}} + L^{(\rm NLO)*}_{s}\frac{\partial L^{(\rm LO)}_{s}}{\partial\alpha^*_{m}} - \frac{\partial L^{(\rm LO)*}_{s}}{\partial\alpha^*_{m}}L^{(\rm NLO)}_{s} - \frac{\partial L^{(\rm NLO)*}_{s}}{\partial\alpha^*_{m}}L^{(\rm LO)}_{s}\right\} = \gamma O(N^{2n_L - 1}).
\end{gather}
In order for $\mathcal{D}_s$ to remain physically meaningful in the high-occupancy limit,
we scale $\gamma$ such that $\eta_m\mathcal{K}^s_m = O(1)$.
Thus, depending on whether $\mathcal{K}^{s({\rm LO})}_m$ and $\mathcal{K}^{s({\rm NLO})}_m$ vanish or remain nonzero,
the scaling of $\gamma$ falls into one of the following three cases:
\begin{align}
    \label{eq:condition for the absence of classical counterpart}
    \def\arraystretch{1.5}
    \begin{array}{cl}
       {\rm (a)}   & \mathcal{K}^{s(\rm LO)}_m \neq 0~\text{for}~\exists m,~\text{yielding}~\gamma = O(N^{1-2n_L}),    \\
       {\rm (b)}  & \mathcal{K}^{s(\rm LO)}_m = 0~\text{for}~\forall m,~\text{but}~\mathcal{K}^{s(\rm NLO)}_m \neq 0~\text{for}~\exists m,~\text{yielding}~\gamma = O(N^{3/2-2n_L}),     \\
       {\rm (c)} &  \mathcal{K}^{s(\rm LO)}_m = \mathcal{K}^{s(\rm NLO)}_m = 0~\text{for}~\forall m,~\text{requiring}~\gamma = O(N^{2-2n_L}).
    \end{array}
\end{align}
In cases (a) and (b),
$\mathcal{L}_j^{s(2)}$ scales as $O(N^{-1})$ and $O(N^{-1/2})$, respectively, in the high-occupancy limit, and hence $\mathcal{L}_j^s$ is asymptotically well approximated by $\mathcal{L}_j^{s(1)}$.
Therefore,
in the high-occupancy limit,
the system dynamics is well described by the classical equation of motion,
Eq.~\eqref{eq:classical equation of motion}.
On the other hand,
in case~(c),
if the first non-vanishing contribution to $\mathcal{L}_j^{s(1)}$ and the leading-order contribution to $\mathcal{L}_j^{s(2)}$ are of the same order in $\alpha$,
both contributions being $O(1)$ requires $\gamma=O(N^{2-2n_L})$ in the high-occupancy limit.
This means that the classical equation of motion no longer provides a valid description of the dynamics,
and that the second-order approximation is necessary for describing the dynamics accurately.
A notable example is provided by a jump operator containing only terms linear in $\hat{a}_m$ and $\hat{a}_m^\dagger$,
for which $\mathcal{K}_m^s$ vanishes identically under condition~(c). Consequently,
the effects of the jump operator originate entirely from second-order quantum fluctuations.
We note that, for case~(c), if the first nonvanishing contribution to $\mathcal{L}_j^{s(1)}$,
namely the term of order $O(N^{2n_{L}-3/2})$,
vanishes,
the scaling of $\gamma$ differs from that in case~(c) of Eq.~\eqref{eq:condition for the absence of classical counterpart}. However, this is beyond the scope of this work.

\begin{table}[t]
    \centering
    \caption{Examples of jump operators for which the second-order quantum-fluctuation contribution becomes dominant in the high-occupancy limit,
    i.e.,
    those satisfying condition~(c) of Eq.~\eqref{eq:condition for the absence of classical counterpart}.
    In cases~(2--i) and (2--ii), we consider linear jump operators, whereas in cases~(2--iii) and (2--iv), we allow jump operators of arbitrary order.
    When a linear jump operator is Hermitian [case (2--i)] or when two linear jump operators form a Hermitian pair [case (2--ii)],
    $\mathcal{K}_m^s=0$ identically and condition~(c) is automatically satisfied. Even when $\mathcal{K}_m^s \neq 0$,
    the second-order contribution cannot be neglected and remains as a leading-order term in the high-occupancy limit if both the leading- and next-to-leading-order contributions of a jump operator are Hermitian [case (2--iii)],
    or if the leading- and next-to-leading-order contributions of two jump operators form Hermitian pairs [case (2--iv)].
    }
    \label{tab:conditions for the absence of classical counterpart}
    \scalebox{0.8}{
        \begin{tabular}{cccccc}
            \hline\hline
             & $s$ & $n_L$ & $k_{\rm max}$ & $\gamma_k$ & $L_{ks}$  \\ \hline
             (2--i) & $0,\pm1$ & $1/2$ & $1$ & $\gamma$ & $L_s = L^*_s$  \\
             (2--ii) & $0,\pm1$ & $1/2$ & $2$ & $\gamma_{k=1} = \gamma_{k=2}$ & $L_{k=1,s} = L^*_{k=2,s}$  \\
            (2--iii) & $0,\pm1$ & $\geq 1$ & $1$ & $\gamma$ & $L^{(\rm LO)}_s = L^{(\rm LO)*}_s\land L^{(\rm NLO)}_s = L^{(\rm NLO)*}_s$  \\
            (2--iv) & $0,\pm1$ & $\geq 1$ & $2$ & $\gamma_{k=1} = \gamma_{k=2}$ & $L^{(\rm LO)}_{k=1,s} = L^{(\rm LO)*}_{k=2,s}\land L^{(\rm NLO)}_{k=1,s} = L^{(\rm NLO)*}_{k=2,s}$  \\ \hline\hline
        \end{tabular}
    }
\end{table}
Two particularly important situations in which condition~(c) is satisfied are Hermitian jump operators and pairs of jump operators related by Hermitian conjugation.
For a Hermitian jump operator,
one can show that $\mathcal{K}_m^s$ vanishes identically.
Likewise, when two jump operators satisfy $\hat{L}_1=\hat{L}_2^\dagger$ and have equal strengths ($\gamma_1=\gamma_2$),
$\mathcal{K}_m^s$ also vanishes identically,
even though neither jump operator individually satisfies condition~(c).
Tab.~\ref{tab:conditions for the absence of classical counterpart} provides additional examples of jump operators satisfying condition~(c),
including cases in which both the leading- and next-to-leading-order terms are either Hermitian or form Hermitian-conjugate pairs.

\subsection{\label{subsec:Positive diffusion approximation for the Wigner function}Positive semidefiniteness of \texorpdfstring{$\bm{\mathcal{A}}^{s=0}$}{TEXT} in the high-occupancy limit}
In the previous section, we examined the conditions under which the second-order approximation is necessary to describe the dynamics in the high-occupancy limit. However, it remained unclear whether such an approximation can always be implemented in practice.
In this section,
we demonstrate that, for the Wigner function ($s=0$), the matrix $\bm{\mathcal{A}}^{s=0}$ is always positive semidefinite in the high-occupancy limit,
regardless of the details of the jump operators. Consequently,
a well-defined stochastic differential equation can always be constructed.

For simplicity,
we first consider a system with a single jump operator and omit the subscript $k$ in $\gamma_k$ and $\hat{L}_k$.
Let $\lambda_{mn}^{s=0({\rm LO})}$ and $\Lambda_{mn}^{s=0({\rm LO})}$ be the leading-order contributions to $\lambda_{mn}^{s=0}$ and $\Lambda_{mn}^{s=0}$,
respectively.
In the high-occupancy limit,
only these leading-order terms survive.
Here,
from Eq.~\eqref{eq:star},
the details of $\lambda_{mn}^{s=0({\rm LO})}$ and $\Lambda_{mn}^{s=0({\rm LO})}$ can be derived by replacing $\star_{s=0}$ and $L_{s=0}$ as $\star_{s=0} = 1$ and $L_{s=0} = L^{(\rm LO)}_{s=0}$,
respectively in Eqs.~\eqref{eq:definitnion of lambdamn} and \eqref{eq:definitnion of Lambdamn},
obtaining
\begin{gather}
    \label{eq:definitnion of lambdamn s=0 star=1}
    \lambda^{s=0({\rm LO})}_{mn} = \frac{\gamma}{4}\left\{\frac{\partial L^{(\rm LO)*}_{s=0}}{\partial\alpha^*_{m}}\frac{\partial L^{(\rm LO)}_{s=0}}{\partial\alpha^*_{n}} + \frac{\partial L^{(\rm LO)*}_{s=0}}{\partial\alpha^*_{n}}\frac{\partial L^{(\rm LO)}_{s=0}}{\partial\alpha^*_{m}}\right\}, \\
    \label{eq:definitnion of Lambdamn s=0 star=1}
    \Lambda^{s=0({\rm LO})}_{mn} = \frac{\gamma}{4}\left\{\frac{\partial L^{(\rm LO)*}_{s=0}}{\partial\alpha^*_{m}}\frac{\partial L^{(\rm LO)}_{s=0}}{\partial\alpha_{n}} + \frac{\partial L^{(\rm LO)*}_{s=0}}{\partial\alpha_{n}}\frac{\partial L^{(\rm LO)}_{s=0}}{\partial\alpha^*_{m}}\right\},
\end{gather}
which clearly satisfy the conditions~\eqref{eq:condition for lambda_mn two jumps} and \eqref{eq:condition for Lambda_mn two jumps},
ensuring that $\bm{\mathcal{A}}^{s=0}$ is positive semidefinite.
The stochastic term of the corresponding stochastic differential equation can be obtained by substituting
\begin{gather}
    \mathcal{F}^{s=0}_m = \sqrt{\frac{\gamma}{2}}\frac{\partial L^{(\rm LO)}_{s=0}}{\partial\alpha^*_m},\quad \mathcal{G}^{s=0}_m = \sqrt{\frac{\gamma}{2}}\frac{\partial L^{(\rm LO)}_{s=0}}{\partial\alpha_m}
\end{gather}
into Eq.~\eqref{eq:stochastic term for systems with two jumps}.
For systems with multiple jump operators,
we can also show that $\bm{\mathcal{A}}^{s=0}$ is positive semidefinite in the high-occupancy limit.
This follows from the fact that $\lambda_{mn}
^{s=0({\rm LO})}$ and $\Lambda_{mn}
^{s=0({\rm LO})}$
 can always be decomposed as in Eqs.~\eqref{eq:decomposition of lambda} and \eqref{eq:decomposition of Lambda},
 respectively,
 with each term taking the form of Eqs.~\eqref{eq:definitnion of lambdamn s=0 star=1} and \eqref{eq:definitnion of Lambdamn s=0 star=1}.
Thus,
for $s=0$,
whenever the jump operators satisfy the condition~(c) of Eq.~\eqref{eq:condition for the absence of classical counterpart},
including the examples listed in Tab.~\ref{tab:conditions for the absence of classical counterpart},
we can conclude that the second-order approximation is necessary and is always feasible in the high-occupancy limit.

\section{\label{sec:Higher order of quantum fluctuations}Higher-order contributions of quantum fluctuations}
So far,
we have described system dynamics within the second-order approximation.
In this section,
we discuss the effects of the higher-order quantum fluctuations,
particularly cases in which their contributions vanish,
i.e.,
$\mathcal{L}^{s(n_{\eta})}_j = 0$ for $n_{\eta}\geq 3$.
Below,
we first consider the higher-order contributions from the Hamiltonian and then these from the jump operators.


\subsection{\label{subsec:Contributions of the Hamiltonian in the higher order of quantum fluctuations}Contributions from the Hamiltonian}
\begin{table}[t]
    \centering
    \caption{Sufficient conditions on the Hamiltonian $\hat{H}$ for all third- and higher-order contributions in the quantum fluctuations to vanish.
    Because $\hat{H}$ is Hermitian,
    the condition for $s=0$ implies that $\hat{H}$ contains at most quadratic terms in $\hat{a}_m$ and $\hat{a}_m^\dagger$.
    For $s=\pm1$,
    however,
    U(1)-symmetric quartic terms are also permitted.
    }
    \label{tab:conditions for zero higher order of quantum fluctuations Hamiltonian}
    \scalebox{0.8}{
        \begin{tabular}{ccc}
            \hline\hline
             & $s$ & $H_{s}$ \\ \hline
            (3--i) & $0$ & $\displaystyle \frac{\partial^3 H_{s=0}}{\partial\alpha_m\partial\alpha_n\partial\alpha_p} = \frac{\partial^3 H_{s=0}}{\partial\alpha_m\partial\alpha_n\partial\alpha^*_p} = 0$ for $\forall m,n,p$  \\
            (3--ii) & $\pm1$ & $\displaystyle \frac{\partial^3 H_{s=\pm1}}{\partial\alpha_m\partial\alpha_n\partial\alpha_p} = 0$ for $\forall m,n,p$  \\ \hline\hline
        \end{tabular}
    }
\end{table}
We expand the Lagrangian $\mathcal{L}_j^s$ in Eq.~\eqref{eq:action discrete} in powers of the quantum fields $\vec{\eta}_{j+1}$.
The expressions for the third- and fourth-order terms are summarized in \ref{appendix:Third and fourth order of quantum fluctuations}.
Among the contributions to $\mathcal{L}_j^{s(3)}$ and $\mathcal{L}_j^{s(4)}$,
those originating from the Hamiltonian is given by Eqs.~\eqref{eq:third order of the action unitary} and \eqref{eq:fourth order of the action unitary},
respectively.
These expressions show that the third- and fourth-order contributions vanish when the Hamiltonian satisfies

\begin{gather}
    \label{eq:condition for H higher order 1}
    \frac{\partial^3 H_s}{\partial\alpha_m\partial\alpha_n\partial\alpha_p} = \frac{\partial^3 H_s}{\partial\alpha_m\partial\alpha_n\partial\alpha^*_p} = 0~(s=0), \\
    \label{eq:condition for H higher order 2}
    \frac{\partial^3 H_s}{\partial\alpha_m\partial\alpha_n\partial\alpha_p} = 0~(s=\pm1),
\end{gather}
for $\forall m, n$,
and $p$.
Any quadratic Hamiltonian,
i.e.,
a single-body Hamiltonian, satisfies Eq.~\eqref{eq:condition for H higher order 1},
whereas Hamiltonian containing up to two-body interaction satisfies Eq.~\eqref{eq:condition for H higher order 2}.
Since the $n$th-order term contains the $n$th derivative of $H_s$ with respect to $\alpha_m$ and $\alpha_m^*$,
Hamiltonian containing at most two-body interactions do not generate any terms beyond fourth order.
For example,
for the Bose-Hubbard model given by Eq.~\eqref{eq:Bose--Hubbard Hamiltonian general},
the quantum dynamics is described exactly up to the second order in the quantum fluctuations for $s=\pm 1$,
and up to the third order for $s=0$.
The conditions on the Hamiltonian are summarized in Tab.~\ref{tab:conditions for zero higher order of quantum fluctuations Hamiltonian}.


\subsection{\label{subsec:Contributions of the jump operators in the higher order of quantum fluctuations}Contributions from the jump operators}
\begin{table}[t]
    \centering
    \caption{Sufficient conditions on jump operators containing terms up to quadratic order in $\hat{a}_m$ and $\hat{a}_m^\dagger$ for all third- and higher-order contributions in the quantum fluctuations to vanish.
    }
    \label{tab:conditions for zero higher order of quantum fluctuations jump operator}
    \scalebox{0.8}{
        \begin{tabular}{cccccc}
            \hline\hline
             & $s$ & $k_{\rm max}$ & $\gamma_k$ & $L^{(1)}_{ks}$ & $L^{(2)}_{ks}$ \\ \hline
            (4--i) & $0,\pm1$ & $1$ & $\gamma$ & No restriction & $L^{(2)}_s = 0$ \\
            (4--ii) & $1$ & $1$ & $\gamma$ & No restriction & $\displaystyle \frac{\partial L^{(2)}_{s=1}}{\partial\alpha^*_m} = 0$ \\
            (4--iii) & $-1$ & $1$ & $\gamma$ & No restriction & $\displaystyle \frac{\partial L^{(2)}_{s=-1}}{\partial\alpha_m} = 0$ \\
            (4--iv) & $0$ & $1$ & $\gamma$ & $L^{(1)}_{s=0} = L^{(1)*}_{s=0}$ & $L^{(2)}_{s=0} = L^{(2)*}_{s=0}$ \\
            (4--v) & $\pm1$ & $1$ & $\gamma$ & $L^{(1)}_{s=\pm1} = L^{(1)*}_{s=\pm1}$ & $\displaystyle L^{(2)}_{s=\pm1} = L^{(2)*}_{s=\pm1}\land\frac{\partial^2 L^{(2)}_{s=\pm1}}{\partial\alpha^*_m\partial\alpha^*_n}=0$ \\
            (4--vi) & $0$ & $2$ & $\gamma_{k=1} = \gamma_{k=2}$ & $L^{(1)}_{k=1,s=0} = L^{(1)*}_{k=2,s=0}$ & $L^{(2)}_{k=1,s=0} = L^{(2)*}_{k=2,s=0}$ \\
            (4--vii) & $\pm1$ & $2$ & $\gamma_{k=1} = \gamma_{k=2}$ & $L^{(1)}_{k=1,s=\pm1} = L^{(1)*}_{k=2,s=\pm1}$ & $\displaystyle L^{(2)}_{k=1,s=\pm1} = L^{(2)*}_{k=2,s=\pm1}\land\frac{\partial^2 L^{(2)}_{k=1,s=\pm1}}{\partial\alpha^*_m\partial\alpha^*_n}=\frac{\partial^2 L^{(2)}_{k=1,s=\pm1}}{\partial\alpha_m\partial\alpha_n}=0$ \\ \hline\hline
        \end{tabular}
    }
\end{table}
We next consider the contributions from the jump operators.
In \ref{appendix:Third and fourth order of quantum fluctuations},
we restrict our analysis to jump operators containing terms up to quadratic order in $\hat{a}_m$ and $\hat{a}_m^\dagger$.
The contributions to $\mathcal{L}_j^s$ arising from third- and fourth-order quantum fluctuations are given by Eqs.~\eqref{eq:third order of the action non-unitary} and \eqref{eq:fourth order of the action non-unitary},
respectively.
Representative examples for which the third- and fourth-order contributions vanish are summarized in Tab.~\ref{tab:conditions for zero higher order of quantum fluctuations jump operator}.
When these conditions are satisfied,
all higher-order contributions also vanish provided that the jump operators contain terms up to quadratic order in $\hat{a}_m$ and $\hat{a}_m^\dagger$.
Consequently,
the second-order approximation becomes exact.

\begin{table}[t]
    \centering
    \caption{Summary of the conditions obtained in this work:
    (i) the diffusion matrix is positive semidefinite ($\bm{\mathcal{A}}^s\succeq 0$),
    (ii) the mean-field approximation breaks down and the second-order approximation is necessary,
    and (iii) all higher-order quantum fluctuations vanish. Representative examples of system setups satisfying these conditions are listed.
    The listed conditions on $H_s$ for $s=0$ and $s=\pm1$ imply that the Hamiltonian must be non-interacting.
    }
    \label{tab:brief summary}
    \resizebox{\textwidth}{!}{
        \begin{tabular}{ccccccccc}
            \hline\hline
             & $s$ & $k_{\rm max}$ & $\gamma_k$ & $L^{(1)}_{ks}$ & $L^{(2)}_{ks}$ & $H_s$ & $\mathcal{F}^s_m$ & $\mathcal{G}^s_m$\\ \hline
            (I) & $0$ & $1$ & $\gamma$ & $L^{(1)}_{s=0} = L^{(1)*}_{s=0}$ & $L^{(2)}_{s=0} = L^{(2)*}_{s=0}$ & $\displaystyle \frac{\partial^3 H_{s=0}}{\partial\alpha_m\partial\alpha_n\partial\alpha_p} = \frac{\partial^3 H_{s=0}}{\partial\alpha_m\partial\alpha_n\partial\alpha^*_p} = 0$ &  $\displaystyle\sqrt{\frac{\gamma}{2}}\frac{\partial L_{s=0}}{\partial\alpha_m}$ & $\displaystyle\sqrt{\frac{\gamma}{2}}\frac{\partial L_{s=0}}{\partial\alpha_m}$\\
            (II) & $\pm1$ & $1$ & $\gamma$ & $L^{(1)}_{s=\pm1} = L^{(1)*}_{s=\pm1}$ & $\displaystyle L^{(2)}_{s=\pm1} = L^{(2)*}_{s=\pm1}\land\frac{\partial^2 L^{(2)}_{s=\pm1}}{\partial\alpha^*_m\partial\alpha^*_n}=0$ & $\displaystyle \frac{\partial^2 H_{s=\pm1}}{\partial\alpha_m\partial\alpha_n} = 0$ & $\displaystyle\sqrt{\frac{\gamma}{2}}\frac{\partial L_{s=\pm1}}{\partial\alpha^*_m}$ & $\displaystyle\sqrt{\frac{\gamma}{2}}\frac{\partial L_{s=\pm1}}{\partial\alpha_m}$\\
            (III) & $0$ & $2$ & $\gamma_{k=1} = \gamma_{k=2}$ & $L^{(1)}_{k=1,s=0} = L^{(1)*}_{k=2,s=0}$ & $L^{(2)}_{k=1,s=0} = L^{(2)*}_{k=2,s=0}$ & $\displaystyle \frac{\partial^3 H_{s=0}}{\partial\alpha_m\partial\alpha_n\partial\alpha_p} = \frac{\partial^3 H_{s=0}}{\partial\alpha_m\partial\alpha_n\partial\alpha^*_p} = 0$ & $\displaystyle\sqrt{\gamma_{k=1}}\frac{\partial L_{k=1,s=0}}{\partial\alpha^*_m}$ & $\displaystyle\sqrt{\gamma_{k=1}}\frac{\partial L_{k=1,s=0}}{\partial\alpha_m}$\\
            (IV) & $\pm1$ & $2$ & $\gamma_{k=1} = \gamma_{k=2}$ & $L^{(1)}_{k=1,s=\pm1} = L^{(1)*}_{k=2,s=\pm1}$ & $\displaystyle L^{(2)}_{k=1,s=\pm1} = L^{(2)*}_{k=2,s=\pm1}\land\frac{\partial^2 L^{(2)}_{k=1,s=\pm1}}{\partial\alpha^*_m\partial\alpha^*_n}=\frac{\partial^2 L^{(2)}_{k=1,s=\pm1}}{\partial\alpha_m\partial\alpha_n}=0$ & $\displaystyle \frac{\partial^2 H_{s=\pm1}}{\partial\alpha_m\partial\alpha_n} = 0$ & $\displaystyle\sqrt{\gamma_{k=1}}\frac{\partial L_{k=1,s=\pm1}}{\partial\alpha^*_m}$ & $\displaystyle\sqrt{\gamma_{k=1}}\frac{\partial L_{k=1,s=\pm1}}{\partial\alpha_m}$\\ \hline\hline
        \end{tabular}
    }
\end{table}
On the other hand,
the existence of an exact second-order description does not necessarily imply that the dynamics can be implemented as a stochastic differential equation.
Based on the results of Sec.~\ref{sec:Feasibility of the second-order approximation using the stochastic differential equation},
Tab.~\ref{tab:brief summary} summarizes representative examples for which the dynamics is both exactly described within the second-order approximation and can be formulated in terms of a stochastic differential equation, i.e., for which $\bm{\mathcal{A}}^s$ is positive semidefinite. Although the conditions listed in Tab.~\ref{tab:brief summary} are not necessary,
they are sufficient.
Therefore,
whenever these conditions are satisfied,
simulations based on the corresponding stochastic differential equation yield exact results.


\section{\label{sec:Benchmark calculations}Benchmark calculations}
We numerically verify the results obtained in the previous sections.
After describing the numerical setup in Sec.~\ref{sec:Error estimation},
we investigate the validity of the classical description in the high-occupancy limit and show that condition~(c) of Eq.~\eqref{eq:condition for the absence of classical counterpart} correctly predicts its validity in Sec.~\ref{subsec:Model1}.
In Sec.~\ref{subsec:Model2},
we demonstrate the applicability of the stochastic differential equation by applying it to the Bose–Hubbard model with unidirectional incoherent hopping, using the analytical expression for the diffusion term derived in Sec.~\ref{subsec:Additional sufficient condition for the Wigner function}.
\begin{table}[t]
    \centering
    \caption{Quasiprobability distribution functions and abbreviations}
    \label{tab:quasiprobability and abbreviation}
    \scalebox{0.8}{
    \begin{tabular}{ccc}
        \hline\hline
        $s$ & Quasiprobability distribution function & Abbreviation \\ \hline
        $1$ & Glauber-Sudarshan P & P \\ 
        $0$ & Wigner & W \\ 
        $-1$ & Husimi Q & Q \\ \hline\hline
    \end{tabular}
    }
\end{table}
In all the figures,
the Glauber--Sudarshan P,
Wigner,
and Husimi Q functions are denoted simply as ``P'',
``W'',
and ``Q'',
respectively,
as summarized in Tab.~\ref{tab:quasiprobability and abbreviation},
and we abbreviate the numerically exact result as ``Exact'' and results of the first- and second-order approximations as ``Prob:$1$st'' and ``Prob:$2$nd'', respectively,
where Prob $=$ P,
W,
and Q.


\subsection{\label{sec:Error estimation}Numerical methods and error estimation}

We summarize the numerical methods used in the benchmark calculations, as well as the error estimation in the Monte Carlo simulations.
In the numerical calculations,
we use the fourth-order Runge-Kutta method to solve the GKSL equation and the classical equations of motion.
To solve the stochastic differential equations, we use the weak second-order Platen scheme \cite{Platen}.
In the Monte Carlo simulations,
we take $N_{\rm sample} = 10^5$ samples of the initial conditions.
In the second-order approximation,
both the initial condition and the stochastic process are independently sampled for each trajectory.
Thus,
both in the first- and second-order approximation,
we obtain total $N_{\rm sample} = 10^5$ trajectories.
For all numerical simulations, we use a time step of $10^{-3}$ in the dimensionless time shown on the horizontal axis of each figure, and have confirmed convergence with respect to the time step.

For the first- and second-order calculations,
we calculate the standard error of the Monte Carlo estimate.
By solving the classical equations of motion or the stochastic differential equations, we obtain $\alpha_{m,i}(t)$ for $m=1,2,\dots,M$, where the subscripts $i=1,2,\dots,N_{\rm sample}$ identifies the label of the trajectories.
The Monte Carlo estimate of the physical quantity $\braket{\hat{A}(t)}$ evaluated within the first- or second-order approximation,
which we denote as $A^s_{\rm approx}(t)$,
is given by
\begin{align}
    A^{s}_{\rm approx}(t) = \frac{1}{N_{\rm sample}}\sum_{i=1}^{N_{\rm sample}}A_s(\vec{\alpha}_i(t)),
\end{align}
where $A_s(\vec{\alpha}_i(t))$ is a $s$-ordered phase-space representation of the physical quantity $\hat{A}$ using $\alpha_{m,i}(t)$ for $m=1,2,\dots,M$.
We then evaluate the standard error defined by \cite{Barlow}
\begin{align}
    \label{eq:standard error}
    \sigma^{s}_{A}(t) = \frac{1}{\sqrt{N_{\rm sample}}}\sqrt{\sum_{i=1}^{N_{\rm sample}}\frac{\left\{A_s(\vec{\alpha}_i(t)) - A^{s}_{\rm approx}(t)\right\}^2}{N_{\rm sample}-1}},
\end{align}
which measures the statistical uncertainly of the sample mean and decreases as $N_{\rm sample}$ increases and vanishes in the  limit $N_{\rm sample}\to\infty$.
Below, we depict the standard error by shading the region between $A^{s}_{\rm approx}\pm\sigma^{s}_A$ or by error bars in the figures of the numerical results.

\subsection{\label{subsec:Model1}Two-site non-interacting atoms}
In this section,
we consider a two-site non-interacting system described by the following Hamiltonian:
\begin{gather}
    \label{eq:Hamiltonian free boson}
    \hat{H}_{\rm FB} = -\mu\sum_{m=1,2}\hat{a}^{\dagger}_m\hat{a}_m - J(\hat{a}^{\dagger}_2\hat{a}_1 + \hat{a}_1^{\dagger}\hat{a}_2),
\end{gather}
where $\hat{a}^{\dagger}_m$ and $\hat{a}_m$ are the creation and annihilation operators,
respectively,
for atoms at site $m=1,2$,
$\mu$ is the chemical potential,
and $J$ is the hopping amplitude.
In the subsequent sections~\ref{subsubsec:Model11} and \ref{subsubsec:Model21},
we consider the dynamics of a two-site system of non-interacting atoms with jump operators describing the non-local losses [Eq.~\eqref{eq:jump operator nonlocal loss}] and symmetric incoherent hopping [Eq.~\eqref{eq:jump operators symmetric incoherent hopping}]. 
The former does not satisfy condition~(c) of Eq.~\eqref{eq:condition for the absence of classical counterpart},
whereas the latter does.
\begin{table}[t]
    \centering
    \caption{The derivability of the stochastic differential equations,
    the validity of the description based on the classical equations of motion,
    and the higher-order contributions of quantum fluctuations for the systems in Secs~\ref{subsubsec:Model11} and \ref{subsubsec:Model21}.
    In the high-occupancy limit,
    the classical description is invalid (valid) when condition~(c) of Eq.~\eqref{eq:condition for the absence of classical counterpart} is satisfied (not satisfied).}
    \label{tab:benchmark calculation}
    \scalebox{0.8}{
        \begin{tabular}{cccc}
            \hline\hline
            \multirow{2}{*}{Model} &\multirow{2}{*}{Stochastic differential equations} & Validity of classical description & \multirow{2}{*}{Higher-order quantum fluctuations} \\
            &  & (High-occupancy limit)\\\hline
            Sec.~\ref{subsubsec:Model11} & Derivable  & Valid & No effect\\ 
            Sec.~\ref{subsubsec:Model21} & Derivable & Invalid & No effect \\ \hline\hline 
        \end{tabular}
    }
\end{table}
Tab.~\ref{tab:benchmark calculation} summarizes the derivability of the stochastic differential equations,
the validity of the classical description in the high-occupancy limit,
and the higher-order contributions of quantum fluctuations for the systems in Secs~\ref{subsubsec:Model11} and \ref{subsubsec:Model21}.
In both systems,
the quantum dynamics can be described by stochastic differential equations for all $s=0,\pm 1$.
These equations are exact because the contributions from higher-order quantum fluctuations vanish identically.

As the initial state, we prepare a pure coherent state $\hat{\rho}(0) = \ket{\alpha_{{\rm I}1},\alpha_{{\rm I}2}}\bra{\alpha_{{\rm I}1},\alpha_{{\rm I}2}}$, where $\hat{a}_m\ket{\alpha_{{\rm I}1},\alpha_{{\rm I}2}} = \alpha_{{\rm I}m}\ket{\alpha_{{\rm I}1},\alpha_{{\rm I}2}}$ for $m = 1$ and $2$, and we choose $\alpha_{{\rm I}1} = \sqrt{N_{{\rm I}1}}e^{i\pi/8}$ and $\alpha_{{\rm I}2} = \sqrt{N_{{\rm I}2}}e^{i\pi/4}$ with $N_{\rm I1}$ and $N_{\rm I2}$ being the number of atoms at each site.
Thus,
the total number of atoms $N_{\rm I}$ is $N_{\rm I} = N_{\rm I1} + N_{\rm I2}$.
Below, 
in the results in the main panels of all the figures,
we take $N_{\rm I1}/N_{\rm I} = 0.8$ and $N_{\rm I2}/N_{\rm I} = 0.2$ with $N_{\rm I} = 10$,
i.e.,
$N_{\rm I1} = 8$ and $N_{\rm I2} = 2$.
In the results in the insets of the figures in Secs.~\ref{subsubsec:Model11} and \ref{subsubsec:Model21},
we change the value of $N_{\rm I}$ with a fixed fractions of atoms at the initial state $N_{\rm I1}/N_{\rm I} = 0.8$ and $N_{\rm I2}/N_{\rm I} = 0.2$.
The corresponding initial quasiprobability distribution function $W_{s}(\vec{\alpha},\vec{\alpha}^*,t_0 = 0)$ is the Gaussian function for $s =0,-1$ and the Dirac delta function for $s=1$, i.e.,
\begin{align}
    \label{eq:quasiprobability for coherent state}
    W_s(\vec{\alpha},\vec{\alpha}^*,0) = 
    \begin{cases}
    \displaystyle\prod_{m=1,2}\dfrac{2}{1-s}e^{-2|\alpha_m-\alpha_{{\rm I}m}|^2/(1-s)} &\text{for}~ s=0,-1, \\
    \displaystyle\prod_{m=1,2}\pi\delta^{(2)}(\alpha_m - \alpha_{{\rm I}m}) &\text{for}~ s=1.
    \end{cases}
\end{align}

Under these setups, we investigate the time evolution of the fraction of the remaining atoms in the site $m=1$, and those of the correlation between atoms at different sites which are respectively defined by
\begin{align}
    \label{eq:physical quantities}
    n_1 = \frac{\braket{\hat{a}_1^{\dagger}\hat{a}_1}}{N_{\rm I}},\quad C_{12} = \frac{\braket{\hat{a}^{\dagger}_1\hat{a}_2} + \braket{\hat{a}^{\dagger}_2\hat{a}_1}}{2N_{\rm I}}.
\end{align}
We also calculate the difference between the results of the first- and second-order approximation ($A_{\rm approx}$) and the numerically exact one ($A_{\rm Exact}$) defined by $\delta A = A_{\rm approx} - A_{\rm Exact}$ with $A$ being one of the physical quantities.


\subsubsection{\label{subsubsec:Model11}Validity of the classical description: Non-local losses}
We first consider the system described by the following GKSL equation:
\begin{gather}
    \label{eq:GKSL equation model1}
    \frac{d\hat{\rho}(t)}{dt} = -\frac{i}{\hbar}\left[\hat{H}_{\rm FB},\hat{\rho}(t)\right]_- + \gamma\left(\hat{L}\hat{\rho}(t)\hat{L}^{\dagger} - \frac{1}{2}\left[\hat{L}^{\dagger}\hat{L},\hat{\rho}(t)\right]_+\right), \\
    \label{eq:jump operator nonlocal loss}
    \hat{L} = \hat{a}_1 + \hat{a}_2,
\end{gather}
where the jump operator describes the non-local losses of atoms and $\gamma$ denotes its strength.
In this system, the jump operator corresponds to case~(a) in Eq.~\eqref{eq:condition for the absence of classical counterpart}, and therefore the classical description is valid in the high-occupation limit. For condition~(a) in Eq.~\eqref{eq:condition for the absence of classical counterpart}, since $n_{L} = 1/2$ in this case, $\gamma$ does not need to be scaled.
Here,
since the GKSL equation~\eqref{eq:GKSL equation model1} is quadratic in $\hat{a}^{\dagger}_m$ and $\hat{a}_m$,
the hierarchy of equations of motion for the correlation function $\braket{\hat{a}^{\dagger}_m\hat{a}_n}$ closes \cite{Eisler,Medvedyeva,Barthel,Zunkovic}, and is given by
\begin{gather}
    \label{eq:equation of motion of the correlation function1 Model11}
    i\hbar\frac{d\braket{\hat{a}^{\dagger}_1\hat{a}_1}}{dt} = -2iJ{\rm Im}[\braket{\hat{a}^{\dagger}_1\hat{a}_2}] - i\hbar\gamma(\braket{\hat{a}^{\dagger}_1\hat{a}_1} + {\rm Re}[\braket{\hat{a}^{\dagger}_1\hat{a}_2}]),\\
    \label{eq:equation of motion of the correlation function2 Model11}
    i\hbar\frac{d\braket{\hat{a}^{\dagger}_1\hat{a}_2}}{dt} = -J(\braket{\hat{a}^{\dagger}_1\hat{a}_1} - \braket{\hat{a}^{\dagger}_2\hat{a}_2}) - i\hbar\gamma\left\{\braket{\hat{a}^{\dagger}_1\hat{a}_2} + \frac{1}{2}(\braket{\hat{a}^{\dagger}_1\hat{a}_1} + \braket{\hat{a}^{\dagger}_2\hat{a}_2})\right\},\\
    \label{eq:equation of motion of the correlation function3 Model11}
    i\hbar\frac{d\braket{\hat{a}^{\dagger}_2\hat{a}_2}}{dt} = 2iJ{\rm Im}[\braket{\hat{a}^{\dagger}_1\hat{a}_2}] - i\hbar\gamma(\braket{\hat{a}^{\dagger}_2\hat{a}_2} + {\rm Re}[\braket{\hat{a}^{\dagger}_1\hat{a}_2}]).
\end{gather}
We obtain the numerically exact dynamics by solving Eqs.~\eqref{eq:equation of motion of the correlation function1 Model11}--\eqref{eq:equation of motion of the correlation function3 Model11}.

We next derive the stochastic differential equation to be solved in the second-order approximation.
This system falls into case~(1--i) of Tab.~\ref{tab:conditions for positive-semidefinite}.
The corresponding stochastic differential equation is derived by substituting $\mathcal{F}^s_1 = \mathcal{F}^s_2 = 0$ and $\mathcal{G}^s_1 = \mathcal{G}^s_2 = \sqrt{\gamma(1-s)/2}$ into Eq.~\eqref{eq:stochastic term for systems with two jumps} and then using Eq.~\eqref{eq:stochastic differential equation using B},
yielding
\begin{gather}
    \label{eq:SDE for site1 Model11}
    i\hbar d\alpha_{1} = \left[-\mu\alpha_{1} - J\alpha_{2} - \frac{i\hbar\gamma}{2}(\alpha_{1} + \alpha_2)\right]dt - \hbar\sqrt{\frac{\gamma}{4}(1-s)}\cdot(d\mathcal{W}_+ + id\mathcal{W}_-), \\
    \label{eq:SDE for site2 Model11}
    i\hbar d\alpha_{2} = \left[-\mu\alpha_{2} - J\alpha_{1} - \frac{i\hbar\gamma}{2}(\alpha_{1} + \alpha_2)\right]dt - \hbar\sqrt{\frac{\gamma}{4}(1-s)}\cdot(d\mathcal{W}_+ + id\mathcal{W}_-),
\end{gather}
where $\mathcal{W}_+$ and $\mathcal{W}_-$ are independent real scalar Wiener processes.
The classical equation of motion in the first-order calculation is given by Eqs.~\eqref{eq:SDE for site1 Model11} and \eqref{eq:SDE for site2 Model11} with neglecting the stochastic terms.
For $s=1$,
the stochastic terms of Eqs.~\eqref{eq:SDE for site1 Model11} and \eqref{eq:SDE for site2 Model11} vanish identically.
Considering that the initial Glauber-Sudarshan P function is a Dirac delta function,
the exact dynamics can then be calculated simply by solving the classical equations of motion with the initial conditions $\alpha_1(0) = \alpha_{{\rm I}1}$ and $\alpha_2(0) = \alpha_{{\rm I}2}$.
On the other hand,
for $s=0$ or $-1$,
we evaluate the dynamics by using the Monte Carlo trajectory sampling based on Eqs.~\eqref{eq:SDE for site1 Model11} and \eqref{eq:SDE for site2 Model11}.

\begin{figure}[t]
	\centering 
	\includegraphics[width = \linewidth]{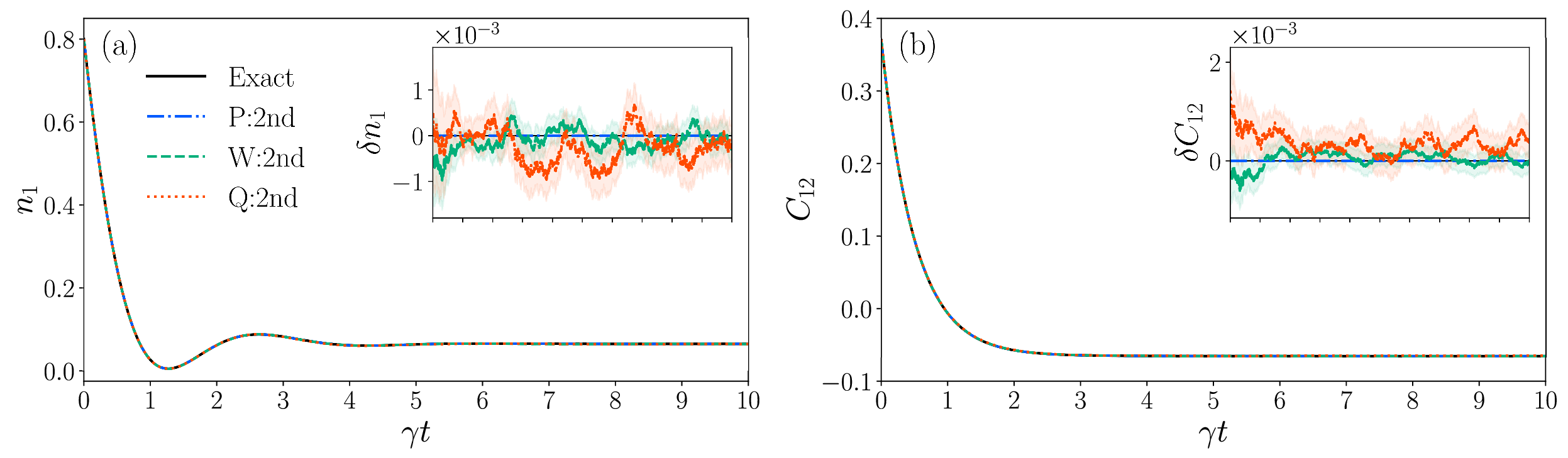}
	\caption{Relaxation dynamics of a two-site system of non-interacting atoms obeying the GKSL equation~\eqref{eq:GKSL equation model1} starting from the pure coherent state $\hat{\rho}(0) = \ket{\alpha_{{\rm I}1},\alpha_{{\rm I}2}}\bra{\alpha_{{\rm I}1},\alpha_{{\rm I}2}}$,
    where $\alpha_{{\rm I}1} = \sqrt{N_{{\rm I}1}}e^{i\pi/8}$ and $\alpha_{{\rm I}2} = \sqrt{N_{{\rm I}2}}e^{i\pi/4}$ with $N_{{\rm I}1} = 8$ and $N_{{\rm I}2} = 2$.
    Shown are (a) the remaining fraction of atoms $n_1$ at site $m=1$ and (b) the correlation of atoms at different sites $C_{12}$,
    which are defined by Eq.~\eqref{eq:physical quantities}.
    In each panel,
    we compare the numerically exact result (Exact) obtained by solving Eqs.~\eqref{eq:equation of motion of the correlation function1 Model11}--\eqref{eq:equation of motion of the correlation function3 Model11} and the ones of the second-order approximation using the Glauber-Sudarshan P (P:$2$nd),
    Wigner (W:$2$nd),
    and Husimi Q function (Q:$2$nd).
    We choose $\mu/(\hbar\gamma) = 1$ and $J/(\hbar\gamma) = 1$,
    and take $10^5$ samples for the first- and second-order approximations.
    The insets depict the difference between the results of each approximation and the numerically exact one.}
	\label{fig:Model11second}
\end{figure}
We first study the validity of the derived stochastic differential equations~\eqref{eq:SDE for site1 Model11} and \eqref{eq:SDE for site2 Model11}.
Fig.~\ref{fig:Model11second} shows the relaxation dynamics of $n_1$ and $C_{12}$,
where we choose the parameters as $\mu/(\hbar\gamma) = 1$ and $J/(\hbar\gamma) = 1$.
The insets depict the difference between the results of the second-order approximation (P:$2$nd, W:$2$nd and Q:$2$nd) and the numerically exact one (Exact).
In all the panels,
the results of the second-order approximations show good agreement with the numerically exact one.

\begin{figure}[t]
	\centering 
	\includegraphics[width = \linewidth]{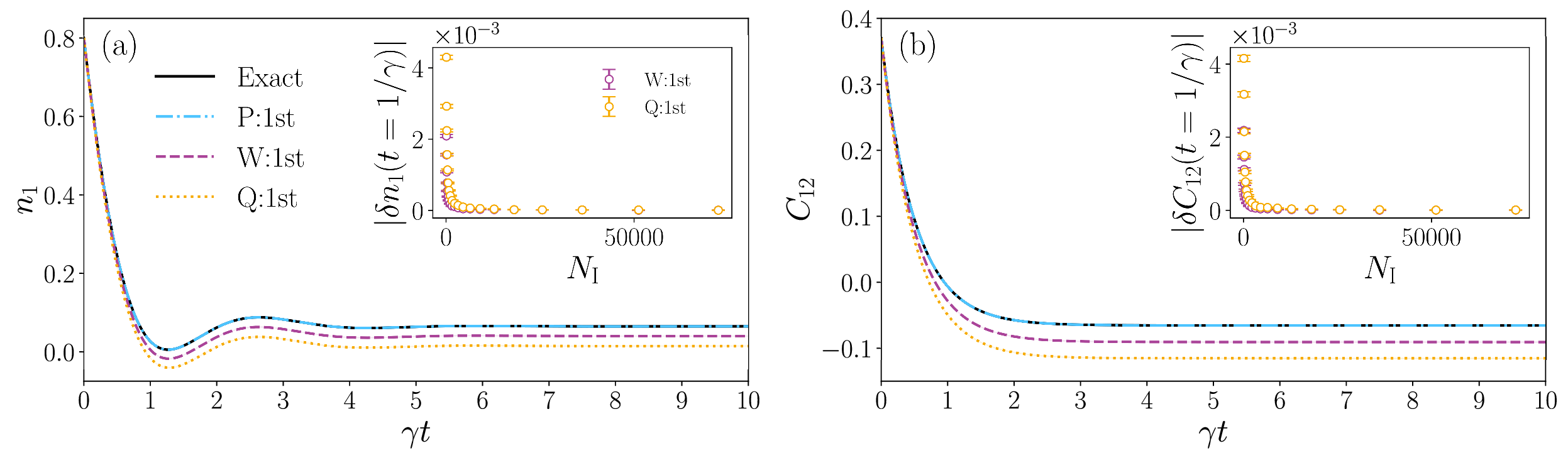}
	\caption{Relaxation dynamics of a two-site system of non-interacting atoms obeying the GKSL equation~\eqref{eq:GKSL equation model1},
    starting from the pure coherent state.
    The quantities,
    parameters and the notations are the same as in Fig.~\ref{fig:Model11second},
    except that the first-order approximation is applied using the Glauber-Sudarshan P (P:$1$st),
    Wigner (W:$1$st),
    and Husimi Q functions (Q:$1$st).
    The insets depict $N_{\rm I}$-dependence of the deviation between each approximation and the numerically exact result at $\gamma t = 1$.
    When varying $N_{\rm I}$,
    the fractions of atoms at each site are fixed as $N_{\rm I1}/N_{\rm I} = 0.8$ and $N_{\rm I2}/N_{\rm I} = 0.2$ in the initial state.}
	\label{fig:Model11first}
\end{figure}
Next,
we investigate the dynamics in the high-occupancy limit.
Fig.~\ref{fig:Model11first} shows the relaxation dynamics of $n_1$ and $C_{12}$ within the first-order approximation.
The parameters are identical to those used in Fig.~\ref{fig:Model11second}.
The main panels show the dynamics with total $N_{\rm I} = 10$ atoms being condensed at the initial state.
In the insets,
we vary the initial total number of atoms and investigate the $N_{\rm I}$-dependence of the deviation between the results of the first-order approximation (W:$1$st and Q:$1$st) and the numerically exact one at $\gamma t = 1$.
Here,
since the first-order approximation for $s=1$ yields the exact dynamics as discussed in the previous paragraph, 
we omit this case in the inset calculation.
As we can see in the main panels,
the first-order approximation exhibits finite deviations from the numerically exact dynamics.
However,
the insets show that these deviations decrease with increasing $N_{\rm I}$ and eventually converge to zero.
This demonstrates that the first-order approximation becomes asymptotically exact in the high-occupancy limit.

\subsubsection{\label{subsubsec:Model21}Breakdown of the classical description: Symmetric incoherent hopping}
Next,
we investigate the dynamics described by the following GKSL equation:
\begin{gather}
    \label{eq:GKSL equation model2}
    \frac{d\hat{\rho}(t)}{dt} = -\frac{i}{\hbar}\left[\hat{H}_{\rm FB},\hat{\rho}(t)\right]_- + \gamma\sum_{k=1,2}\left(\hat{L}_k\hat{\rho}(t)\hat{L}^{\dagger}_k - \frac{1}{2}\left[\hat{L}^{\dagger}_k\hat{L}_k,\hat{\rho}(t)\right]_+\right), \\
    \label{eq:jump operators symmetric incoherent hopping}
    \hat{L}_1 = \hat{a}^{\dagger}_1\hat{a}_2,~\hat{L}_2 = \hat{a}^{\dagger}_2\hat{a}_1,
\end{gather}
where the jump operators $\hat{L}_1$ and $\hat{L}_2$ describe incoherent transmissions of atoms from site 1 to site 2,
and \textit{vice versa},
respectively.
In this system,
we assume that these transmissions of atoms occur with equal strength,
so that the pair of the jump operators ($\hat{L}_1,\hat{L}_2$) describes the symmetric incoherent hopping of atoms \cite{Temme},
and the jump operators correspond to case~(c) in Eq.~\eqref{eq:condition for the absence of classical counterpart}.
Therefore, the classical description is not valid, and the second-order approximation is required in the high-occupation limit.
For condition~(c) in Eq.~\eqref{eq:condition for the absence of classical counterpart}, since $n_{L} = 1$ in this case, $\gamma$ does not need to be scaled.
As in the case of Sec.~\ref{subsubsec:Model11},
the hierarchy of equations of motion for the correlation function closes \cite{Eisler,Medvedyeva,Barthel,Zunkovic},
which is given by
\begin{gather}
    \label{eq:equation of motion of the correlation function1 Model21}
    i\hbar\frac{d\braket{\hat{a}^{\dagger}_1\hat{a}_1}}{dt} = -2iJ{\rm Im}[\braket{\hat{a}^{\dagger}_1\hat{a}_2}] - i\hbar\gamma(\braket{\hat{a}^{\dagger}_1\hat{a}_1} - \braket{\hat{a}^{\dagger}_2\hat{a}_2}),\\
    \label{eq:equation of motion of the correlation function2 Model21}
    i\hbar\frac{d\braket{\hat{a}^{\dagger}_1\hat{a}_2}}{dt} = -J(\braket{\hat{a}^{\dagger}_1\hat{a}_1} - \braket{\hat{a}^{\dagger}_2\hat{a}_2}) - i\hbar\gamma\braket{\hat{a}^{\dagger}_1\hat{a}_2},\\
    \label{eq:equation of motion of the correlation function3 Model21}
    i\hbar\frac{d\braket{\hat{a}^{\dagger}_2\hat{a}_2}}{dt} = 2iJ{\rm Im}[\braket{\hat{a}^{\dagger}_1\hat{a}_2}] + i\hbar\gamma(\braket{\hat{a}^{\dagger}_1\hat{a}_1} - \braket{\hat{a}^{\dagger}_2\hat{a}_2}).
\end{gather}
By solving Eqs.~\eqref{eq:equation of motion of the correlation function1 Model21}--\eqref{eq:equation of motion of the correlation function3 Model21},
we obtain the numerically exact result.

We next derive the corresponding stochastic differential equations to be solved.
This system falls into case~(1--v) for $s=0$ and case~(1--vi) for $s=\pm1$ of Tab.~\ref{tab:conditions for positive-semidefinite}.
Then,
by substituting $\mathcal{F}^s_1 = \sqrt{\gamma}\alpha_2$, $\mathcal{F}^s_2 = 0$, $\mathcal{G}^s_1 = 0$,
and $\mathcal{G}^s_2 = \sqrt{\gamma}\alpha^*_1$ into Eq.~\eqref{eq:stochastic term for systems with two jumps} and using Eq.~\eqref{eq:stochastic differential equation using B},
we obtain
\begin{gather}
    \label{eq:SDE for site1 Model21}
    i\hbar d\alpha_{1} = \left[-\mu\alpha_{1} - J\alpha_{2} - \frac{i\hbar\gamma}{2}\alpha_{1}\right]dt - \hbar\sqrt{\frac{\gamma}{2}}\alpha_2\cdot(d\mathcal{W}_+ - id\mathcal{W}_-), \\
    \label{eq:SDE for site2 Model21}
    i\hbar d\alpha_{2} = \left[-\mu\alpha_{2} - J\alpha_{1} - \frac{i\hbar\gamma}{2}\alpha_{2}\right]dt - \hbar\sqrt{\frac{\gamma}{2}}\alpha_1\cdot(d\mathcal{W}_+ + id\mathcal{W}_-),
\end{gather}
where $\mathcal{W}_+$ and $\mathcal{W}_-$ are independent real scalar Wiener processes.

\begin{figure}[t]
	\centering 
	\includegraphics[width = \linewidth]{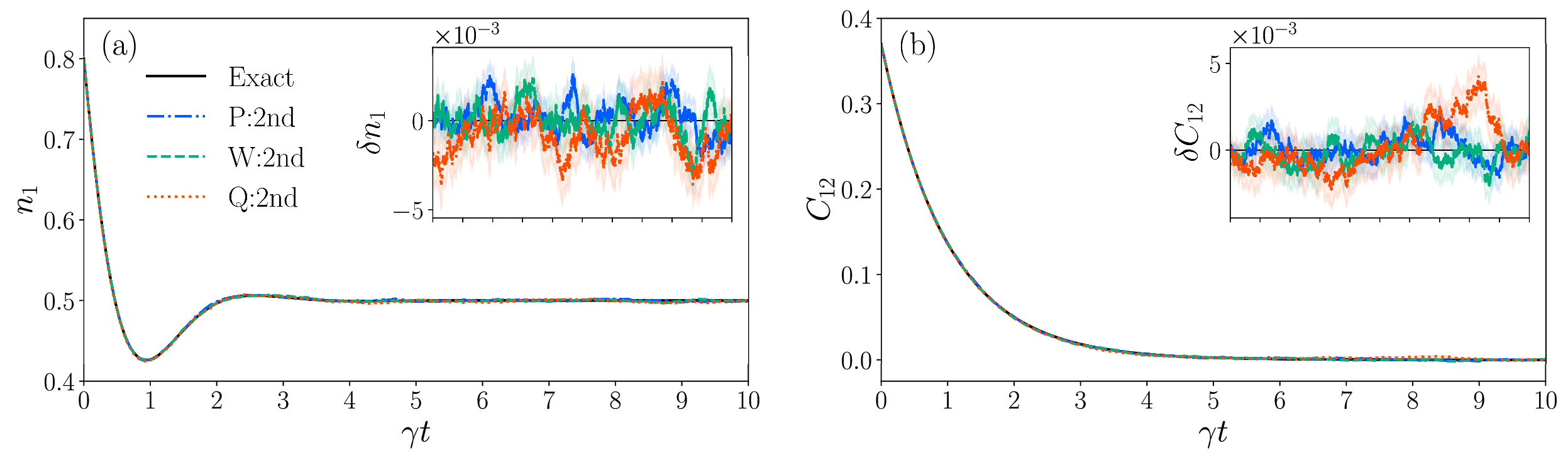}
	\caption{Relaxation dynamics of a two-site system of non-interacting system with the symmetric incoherent hopping of atoms obeying the GKSL equation~\eqref{eq:GKSL equation model2} starting from the pure coherent state.
    The quantities, parameters and the notations are same with the ones in Fig.~\ref{fig:Model11second}.
    The insets depict the difference between the results of the second-order approximation and the numerically exact one.}
	\label{fig:Model21second}
\end{figure}
We first study the validity of Eqs.~\eqref{eq:SDE for site1 Model21} and \eqref{eq:SDE for site2 Model21}.
Fig.~\ref{fig:Model21second} shows the relaxation dynamics of $n_1$ and $C_{12}$ for the parameters $\mu/(\hbar\gamma) = 1$ and $J/(\hbar\gamma) = 1$.
In all the panels,
the results within the second-order approximations show good agreement with the numerically exact result.

\begin{figure}[t]
	\centering 
	\includegraphics[width = \linewidth]{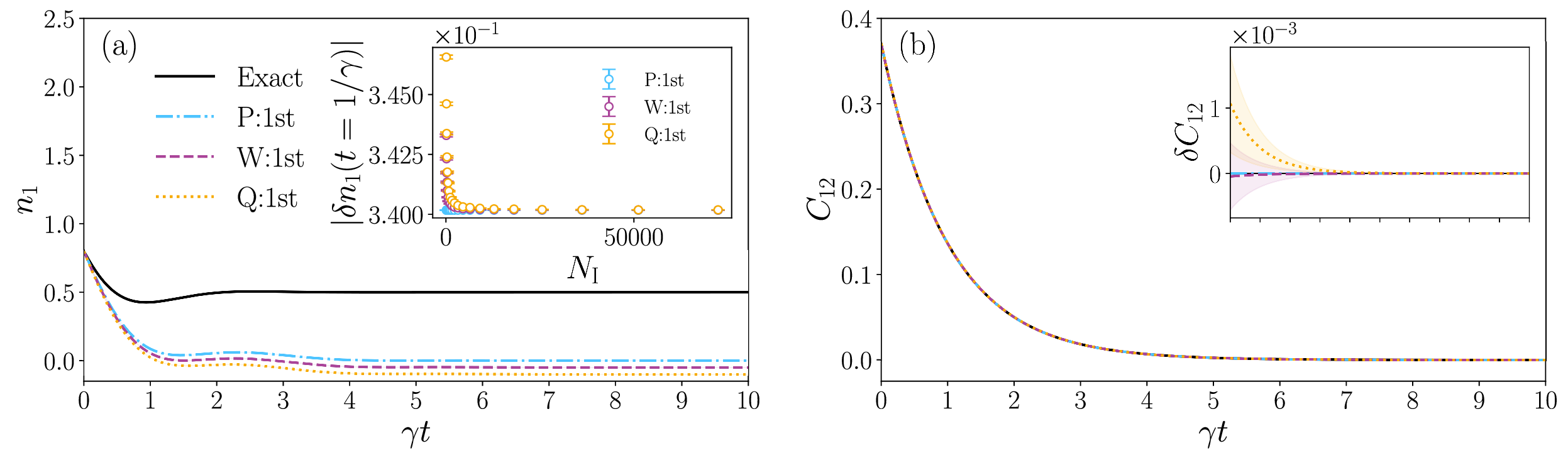}
	\caption{Relaxation dynamics of a two-site system of non-interacting system with the symmetric incoherent hopping of atoms obeying the GKSL equation~\eqref{eq:GKSL equation model2} starting from the pure coherent state.
    The quantities, parameters and the notations are same with the ones in Fig.~\ref{fig:Model11first}.
    The inset of (a) depicts $N_{\rm I}$-dependence of the difference between the results of each approximation and the numerically exact one at time $\gamma t = 1$, and the inset of (b) depicts the difference between the results of the first-order approximation and the numerically exact one.}
	\label{fig:Model21first}
\end{figure}
We next investigate the dynamics in the high-occupancy limit.
Fig.~\ref{fig:Model21first} shows the relaxation dynamics of $n_1$ and $C_{12}$ within the first-order approximation.
The parameters are identical to those in Fig.~\ref{fig:Model21second}.
In the inset of Fig.~\ref{fig:Model21first}(a),
we vary the initial total number of atoms and investigate the $N_{\rm I}$-dependence of the deviation between the dynamics of the first-order approximation and the numerically exact result at $\gamma t = 1$.
The inset shows that the deviation decreases with increasing $N_{\rm I}$, but converge to a finite nonzero value.
This indicates that the first-order approximation fails to provide an accurate description of the dynamics in the high-occupancy limit.
On the other hand,
as shown in Fig.~\ref{fig:Model21first}(b),
the first-order approximation accurately reproduces the relaxation dynamics $C_{12}$.
This is because the contributions of the second-order quantum fluctuations to the dynamics of $C_{12}$ vanish identically,
which we show in \ref{appendix:Absence of the effect of the second order of quantum fluctuations}.

\subsection{\label{subsec:Model2}Three-site Bose--Hubbard model with unidirectional incoherent hopping}
Finally,
we study the validity of the stochastic differential equation~\eqref{eq:SDEm discussion}.
To this end,
we consider a three-site Bose--Hubbard model with the periodic boundary condition obeying the following GKSL equation:
\begin{gather}
    \label{eq:GKSL equation model3}
    \frac{d\hat{\rho}(t)}{dt} = -\frac{i}{\hbar}\left[\hat{H}_{\rm BH},\hat{\rho}(t)\right]_- + \gamma\sum_{k=1,2,3}\left(\hat{L}_k\hat{\rho}(t)\hat{L}^{\dagger}_k - \frac{1}{2}\left[\hat{L}^{\dagger}_k\hat{L}_k,\hat{\rho}(t)\right]_+\right), \\
    \label{eq:Hamiltonian Bose--Hubbard model three sites}
    \hat{H}_{\rm BH} = -\mu\sum_{m=1,2,3}\hat{a}^{\dagger}_m\hat{a}_m - J(\hat{a}^{\dagger}_2\hat{a}_1 + \hat{a}_3^{\dagger}\hat{a}_2 + \hat{a}^{\dagger}_1\hat{a}_3 + {\rm h.c.}) + \frac{U}{2}\sum_{m=1,2,3}\hat{a}^{\dagger}_m\hat{a}^{\dagger}_m\hat{a}_m\hat{a}_m, \\
    \label{eq:jump operators model3}
    \hat{L}_1 = \hat{a}^{\dagger}_2\hat{a}_1,~\hat{L}_2 = \hat{a}^{\dagger}_3\hat{a}_2,~\hat{L}_3 = \hat{a}^{\dagger}_1\hat{a}_3,
\end{gather}
where $\hat{H}_{\rm BH}$ describes the three-site Bose--Hubbard model with $U$ being the on-site interaction energy,
and the jump operators describe the unidirectional incoherent hopping of atoms,
which is identical to Eq.~\eqref{eq:jump operators discussion} with $M=3$.
We note that,
at finite occupancies,
the contributions from higher-order quantum fluctuations do not vanish,
and hence the second-order approximation is not exact.
On the other hand,
since the jump operators do not satisfy condition~(c) of Eq.~\eqref{eq:condition for the absence of classical counterpart},
the classical description becomes valid in the high-occupancy limit.

In this system,
the hierarchy of equations of motion for the correlation function does not close,
and the calculation of the numerically exact dynamics requires a high numerical cost.
However,
for a particular choice of initial state,
the numerical cost can be significantly reduced.
The corresponding initial state is given by
\begin{align}
    \label{eq:initial state for benchmark model3}
    \hat{\rho}(0) = \sum_{n_1,n_2,n_3=0}^{\infty}e^{-(|\alpha_{{\rm I}1}|^2 + |\alpha_{{\rm I}2}|^2 + |\alpha_{{\rm I}3}|^2)}\frac{|\alpha_{{\rm I}1}|^{2n_1}|\alpha_{{\rm I}2}|^{2n_2}|\alpha_{{\rm I}3}|^{2n_3}}{n_1!n_2!n_3!}\ket{n_1,n_2,n_3}\bra{n_1,n_2,n_3},
\end{align}
where $\ket{n_1,n_2,n_3}$ is the Fock state.
As explained in \ref{appendix:Steady state of the system under the dephasing},
Eq.~\eqref{eq:initial state for benchmark model3} corresponds to the diagonal elements of the density matrix of a pure coherent state in the Fock basis.
Since the Hamiltonian and the jump operators conserve the total number of atoms, the GKSL dynamics preserves each fixed-particle-number sector.
Therefore,
starting from the initial state in Eq.~\eqref{eq:initial state for benchmark model3},
we can simulate the dynamics independently within each sector of fixed total number of atoms,
thereby significantly reducing the numerical cost of the exact calculation.
However,
in phase space,
the quasiprobability distribution corresponding to the initial state in Eq.~\eqref{eq:initial state for benchmark model3} is non-Gaussian,
making direct sampling of the initial conditions difficult.
Instead, we exploit the analytically known long-time limit of an auxiliary stochastic differential equation whose stationary distribution corresponds to Eq.~\eqref{eq:initial state for benchmark model3}.
Specifically, we draw samples from the Gaussian quasiprobability distribution of a pure coherent state and independently randomize their phases uniformly over $[0,2\pi)$, while preserving their magnitudes. This procedure directly generates the required initial samples, as detailed in \ref{appendix:Sampling in the phase space}.
In this benchmark calculation,
we set $\alpha_{{\rm I}1} = \sqrt{N_{{\rm I}1}}e^{i\pi/4}$ with $N_{{\rm I}1} = 10$ and $\alpha_{{\rm I}2} = \alpha_{{\rm I}3} = 0$ in Eq.~\eqref{eq:initial state for benchmark model3}.

Under these setups, 
we investigate the time evolution of the population imbalance of the remaining atoms among the three sites and atomic current,
which are respectively defined by
\begin{gather}
    \label{eq:physical quantities model3 atomic number}
    n_{123} = \frac{2\braket{\hat{a}_1^{\dagger}\hat{a}_1} - \braket{\hat{a}_2^{\dagger}\hat{a}_2} - \braket{\hat{a}_3^{\dagger}\hat{a}_3}}{2N_{\rm I}},\\
    \label{eq:physical quantities model3 atomic current}
    I_{123} = \frac{1}{3}\sum_{m=1,2,3}\left\{\frac{iJ}{\hbar N_{\rm I}\gamma}(\braket{\hat{a}^{\dagger}_{m+1}\hat{a}_m} - \braket{\hat{a}^{\dagger}_m\hat{a}_{m+1}}) + \frac{1}{N_{\rm I}}\braket{\hat{a}^{\dagger}_m\hat{a}_m\hat{a}^{\dagger}_{m+1}\hat{a}_{m+1}}\right\},
\end{gather}
where $\hat{a}_4 = \hat{a}_1$ follows from the periodic boundary condition,
and $N_{\rm I}$ is the total mean atomic number in the initial state ($N_{\rm I} = N_{{\rm I}1} = 10$).
For the atomic current $I_{123}$,
the term proportional to $J$ denotes the coherent current originating from the Hamiltonian,
whereas the remaining term describes the incoherent current induced by the jump operators.
The derivation of the atomic current is provided in \ref{appendix:Coherent and incoherent current}.

In the second-order approximation,
the stochastic differential equation for $s=0$ is obtained by substituting by substituting $\hat{H}=\hat{H}_{\rm BH}$ and $M=3$ into Eq.~\eqref{eq:SDEm discussion}:
{\small
\begin{gather}
    i\hbar d\alpha_1 = \left[-\mu\alpha_{1} - J(\alpha_{2} + \alpha_{3}) + U\alpha_1(|\alpha_1|^2 - 1) + \frac{i\hbar\gamma}{2}\alpha_1(|\alpha_3|^2 - |\alpha_2|^2 - 1)\right]dt +  i\hbar\sqrt{\frac{\gamma}{4}}\left\{\alpha_3\cdot\left(id\mathcal{W}^{(3)}_+ + d\mathcal{W}^{(3)}_-\right) + \alpha_2\cdot\left(id\mathcal{W}^{(1)}_+ - d\mathcal{W}^{(1)}_-\right)\right\}, \\
    i\hbar d\alpha_2 = \left[-\mu\alpha_{2} - J(\alpha_{1} + \alpha_{3}) + U\alpha_2(|\alpha_2|^2 - 1) + \frac{i\hbar\gamma}{2}\alpha_2(|\alpha_1|^2 - |\alpha_3|^2 - 1)\right]dt + i\hbar\sqrt{\frac{\gamma}{4}}\left\{\alpha_1\cdot\left(id\mathcal{W}^{(1)}_+ + d\mathcal{W}^{(1)}_-\right) + \alpha_3\cdot\left(id\mathcal{W}^{(2)}_+ - d\mathcal{W}^{(2)}_-\right)\right\}, \\
    i\hbar d\alpha_3 = \left[-\mu\alpha_{3} - J(\alpha_{1} + \alpha_{2}) + U\alpha_3(|\alpha_3|^2 - 1) + \frac{i\hbar\gamma}{2}\alpha_3(|\alpha_2|^2 - |\alpha_1|^2 - 1)\right]dt +  i\hbar\sqrt{\frac{\gamma}{4}}\left\{\alpha_2\cdot\left(id\mathcal{W}^{(2)}_+ + d\mathcal{W}^{(2)}_-\right) + \alpha_1\cdot\left(id\mathcal{W}^{(3)}_+ - d\mathcal{W}^{(3)}_-\right)\right\},
\end{gather}
}
where $\mathcal{W}^{(1)}_{\pm}$,
$\mathcal{W}^{(2)}_{\pm}$,
and $\mathcal{W}^{(3)}_{\pm}$ are mutually independent real scalar Wiener processes.
We also perform the first-order approximation,
which is governed by the following classical equations of motion:
\begin{gather}
    i\hbar \frac{d\alpha_1}{dt} = -\mu\alpha_{1} - J(\alpha_{2} + \alpha_{3}) + U\alpha_1(|\alpha_1|^2 - 1 + s) + \frac{i\hbar\gamma}{2}\alpha_1(|\alpha_3|^2 - |\alpha_2|^2 - 1), \\
    i\hbar \frac{d\alpha_2}{dt} = -\mu\alpha_{2} - J(\alpha_{1} + \alpha_{3}) + U\alpha_2(|\alpha_2|^2 - 1 + s) + \frac{i\hbar\gamma}{2}\alpha_2(|\alpha_1|^2 - |\alpha_3|^2 - 1), \\
    i\hbar \frac{d\alpha_3}{dt} = -\mu\alpha_{3} - J(\alpha_{1} + \alpha_{2}) + U\alpha_3(|\alpha_3|^2 - 1 + s) + \frac{i\hbar\gamma}{2}\alpha_3(|\alpha_2|^2 - |\alpha_1|^2 - 1).
\end{gather}

\begin{figure}[t]
	\centering 
	\includegraphics[width = \linewidth]{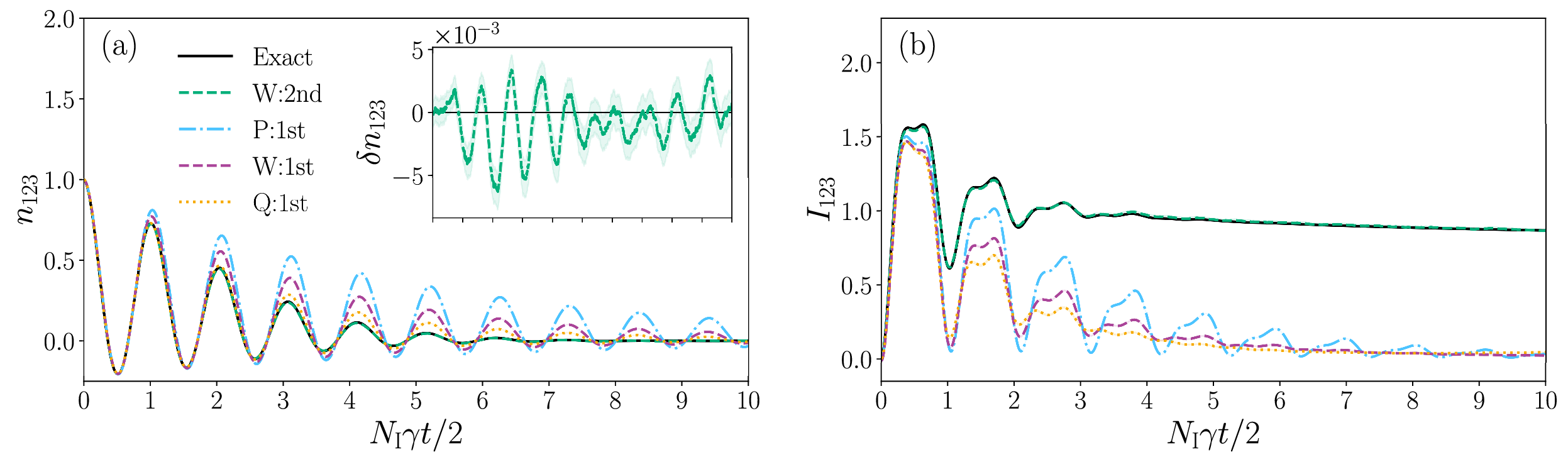}
	\caption{Relaxation dynamics of three-site non-interacting atoms obeying the GKSL equation~\eqref{eq:GKSL equation model3} starting from the state given by Eq.~\eqref{eq:initial state for benchmark model3}.
    Shown are (a) the remaining fraction difference of atoms $n_{123}$ and (b) the atomic current $I_{123}$, which are respectively defined by Eqs.~\eqref{eq:physical quantities model3 atomic number} and \eqref{eq:physical quantities model3 atomic current}.
    In each panel, we compare the numerically exact result (Exact) obtained by directly solving the GKSL equation~\eqref{eq:GKSL equation model3} and the ones of the second-order approximation using the Wigner function (W:$2$nd), and of the first-order approximation using the Glauber-Sudarshan P (P:$1$st) and Husimi Q function (Q:$1$st).
    We choose $\mu/(\hbar N_{\rm I}\gamma) = 1$, $J/(\hbar N_{\rm I}\gamma) = 1$, and $U/(\hbar\gamma)$ and take $10^5$ samples for the first- and second-order approximations.
    The insets depict the difference between the results of each approximation and the numerically exact one.}
	\label{fig:Model3}
\end{figure}
Fig.~\ref{fig:Model3} shows the relaxation dynamics of $n_{123}$ and $I_{123}$.
The parameters are set to $\mu/(\hbar N_{\rm I}\gamma) = 1$, $J/(\hbar N_{\rm I}\gamma) = 1$, and $U/(\hbar\gamma) = 1$.
In all the panels,
the first-order approximation (P:$1$st, W:$1$st and Q:$1$st) exhibits significant deviations from the numerically exact result (Exact),
whereas the second-order approximation (W:$2$nd) well reproduce the exact dynamics.
In particular,
as shown in Fig.~\ref{fig:Model3}(b),
the second-order approximation is necessary for reproducing the non-zero steady-state current.
Recalling that the incoherent atomic current $I_{123}$ is characterized by the fourth-order correlation function $\braket{\hat{a}^{\dagger}_m\hat{a}_m\hat{a}^{\dagger}_{m+1}\hat{a}_{m+1}}$,
this result indicates the importance of the second-order correction in accurately describing the dynamics of higher-order correlations.

\section{\label{sec:Summary and conclusions}Summary and conclusions}
The phase-space formulation of quantum mechanics provides a clear physical interpretation of quantum many-body states and phenomena and enables the study of bosonic quantum many-body dynamics while taking into account quantum fluctuations.
In the phase-space method,
bosonic operators are mapped to $c$-number functions,
and the density operator is represented by a quasiprobability distribution function,
such as the Glauber-Sudarshan P,
Wigner,
and Husimi Q functions.
The GKSL equation is approximated by the Fokker--Planck equation for the quasiprobability distribution function in phase space.
We usually investigate the Fokker--Planck dynamics by deriving the corresponding stochastic differential equations and performing the Monte Carlo simulation.
However,
the Fokker--Planck equation does not always reduce to the stochastic differential equations because the diffusion matrix is not necessarily positive semidefinite and may have negative eigenvalues depending on the details of the Hamiltonian, 
jump operators and choice of quasiprobability distribution function.

In this work,
in Sec.~\ref{sec:Feasibility of the second-order approximation using the stochastic differential equation},
we have first derived the sufficient conditions [Eqs.~\eqref{eq:condition for lambda_mn two jumps} and \eqref{eq:condition for Lambda_mn two jumps}] under which the diffusion matrix is positive semidefinite,
together with the corresponding stochastic differential equations to be solved.
Tab.~\ref{tab:conditions for positive-semidefinite} shows the representative examples satisfying these conditions.
We have also shown a systematic method to derive the stochastic differential equations for more general cases in Sec.~\ref{subsec:Generalization}.
In Sec.~\ref{subsec:Additional sufficient condition for the Wigner function},
we have shown an additional condition on quadratic jump operators satisfying $\bm{\mathcal{A}}^{s=0}\succeq 0$ for the Wigner function ($s=0$),
which indicates that when an infinite-temperature state is one of the steady states,
the diffusion matrix becomes positive semidefinite.
As a specific example,
we have analytically derived the stochastic differential equations for a one-dimensional system with a unidirectional incoherent flow.

Next,
in Sec.~\ref{sec:Equation of motion in the high occupancy limit}
we have investigated the dynamics of the GKSL equation in the high-occupancy limit,
where each degree of freedom,
corresponding to spatial coordinates and/or internal degrees of freedom,
is occupied by a large number of bosons.
The first-order approximation is valid for isolated quantum systems in this limit.
However we have shown that,
in open quantum systems,
it breaks down when the leading-order term of the dissipative drift term in the classical equation of motion vanishes [i.e., the jump operator satisfies condition~(c) in Eq.~\eqref{eq:condition for the absence of classical counterpart}],
the first-order approximation breaks down.
In such cases,
the second-order approximation is required to describe the dynamics accurately.
As summarized in Tab.~\ref{tab:conditions for the absence of classical counterpart},
typical examples include Hermitian jump operators and pairs of Hermitian-conjugate jump operators.
In the high-occupancy limit,
although the feasibility of the second-order approximation is nontrivial,
we have shown,
based on Eqs.~\eqref{eq:condition for lambda_mn two jumps} and \eqref{eq:condition for Lambda_mn two jumps},
that for the Wigner function the stochastic differential equation is always obtainable,
regardless of the details of the jump operators.

In Sec.~\ref{sec:Higher order of quantum fluctuations},
we have investigated the effects of higher-order quantum fluctuations beyond the Fokker--Planck equation.
For quadratic jump operators,
the GKSL equation is generally mapped onto the partial differential equation involving derivative terms up to fourth order, which corresponds to fourth-order quantum fluctuations.
However,
in Sec.~\ref{sec:Higher order of quantum fluctuations},
we have shown that,
when the system setup satisfies one of the conditions in Tab.~\ref{tab:conditions for zero higher order of quantum fluctuations jump operator},
the third- and fourth-order derivative terms vanish,
so that the GKSL equation exactly reduces to the Fokker--Planck equation in phase space.
Hence,
if the corresponding diffusion matrix is positive semidefinite,
the second-order approximation enables exact simulations of the dynamics.
Finally,
Tab.~\ref{tab:brief summary} summarizes the conditions under which the diffusion matrix is positive semidefinite,
the first-order approximation breaks down,
and the contributions from higher-order quantum fluctuations vanish.

In the benchmark calculations in Sec.~\ref{sec:Benchmark calculations},
we have first investigated the relaxation dynamics of a two-site non-interacting system in the high-occupancy limit.
We have considered two representative cases in which the condition~(c) of Eq.~\eqref{eq:condition for the absence of classical counterpart} is either satisfied or not.
By numerically approaching the high-occupancy limit,
we have confirmed that the first-order approximation accurately describes the dynamics when the condition~(c) is not satisfied.
In contrast,
when the condition~(c) is satisfied,
the first-order approximation fails to reproduce the exact dynamics,
whereas the second-order approximation remains in good agreement with the exact result.
Next,
we have investigated the dynamics of the three-site Bose--Hubbard model with the jump operators that induce a unidirectional incoherent flow.
In this system, the incoherent atomic current is characterized by a fourth-order correlation function.
We found that the second-order approximation is necessary to accurately reproduces its dynamics,
demonstrating the importance of the second-order correction for observables characterized by higher-order correlation functions.

Investigating systems in which the mean-field approximation breaks down is an important direction for future work.
In such systems,
our results suggest that second-order quantum fluctuations significantly affect the dynamics,
and nontrivial quantum states and phenomena are expected to emerge.
Accordingly,
it is also important to identify jump operators not listed in Tab.~\ref{tab:conditions for the absence of classical counterpart} for which the mean-field approximation breaks down. In Tab.~\ref{tab:conditions for the absence of classical counterpart},
we have identified cases satisfying condition~(c) of Eq.~\eqref{eq:condition for the absence of classical counterpart} for systems with one or two jump operators. However,
condition~(c) may also be satisfied by three or more jump operators.

\section*{Acknowledgements}
The authors are grateful to Koichiro Furutani and Shunta Yamamoto for helpful discussions. 
This work was supported by JSPS, Japan KAKENHI (Grant Numbers JP26H00385,
JP24K00557 and JP23K13029) and Grant-in-Aid for JSPS Fellows (Grant Number JP25KJ1414).

\appendix


\section{\label{appendix:Derivations of the Stochastic differential equation}Stochastic differential equation: Derivation of Eqs.~\texorpdfstring{\eqref{eq:stochastic differential equation using B} and \eqref{eq:stochastic differential equation using B(ell)}}{TEXT}}
We derive the stochastic differential equations~\eqref{eq:stochastic differential equation using B} and \eqref{eq:stochastic differential equation using B(ell)} from the path-integral representation in Eq.~\eqref{eq:path-integral representaiton discrete}.
We first derive the more general stochastic differential equation~\eqref{eq:stochastic differential equation using B(ell)},
from which Eq.~\eqref{eq:stochastic differential equation using B} follows directly as a special case.
To this end,
we expand the Lagrangian with respect to the quantum fields up to second order.
We then approximate $\mathcal{L}^{s}_j$ in Eq.~\eqref{eq:path-integral representaiton discrete} by $\mathcal{L}^{s(1)}_j + \mathcal{L}^{s(2)}_j$,
obtaining
\begin{align}
    W_{s}(\vec{\alpha}_{\rm f},\vec{\alpha}^*_{\rm f},t) &\approx \lim_{\Delta t \to 0}\prod_{j=0}^{N_t-1}\int\frac{d^2\vec{\alpha}_j d^2\vec{\eta}_{j+1}}{\pi^{2M}}e^{i\Delta t \mathcal{L}^{s(1)}_j/\hbar} e^{i\Delta t \mathcal{L}^{s(2)}_j/\hbar}W_{s}(\vec{\alpha}_0,\vec{\alpha}^*_0,t_0) \\
    \label{eq:calculation in the second order of quantum fluctuations 1}
    &= \lim_{\Delta t \to 0}\prod_{j=0}^{N_t-1}\int\frac{d^2\vec{\alpha}_j d^2\vec{\eta}_{j+1}}{\pi^{2M}}e^{i\Delta t \mathcal{L}^{s(1)}_j/\hbar}{\rm exp}\left\{-\frac{\Delta t}{2}
    \begin{bmatrix}
        \vec{\eta}_{j+1}^{*{\rm T}},\vec{\eta}^{\rm T}_{j+1}
    \end{bmatrix}
    \bm{\mathcal{A}}^{s}(\vec{\alpha}_j,\vec{\alpha}^*_j)
    \begin{bmatrix}
        \vec{\eta}_{j+1} \\
        \vec{\eta}_{j+1}^{*}
    \end{bmatrix}
    \right\}W_{s}(\vec{\alpha}_0,\vec{\alpha}^*_0,t_0),
\end{align}
where we have substituted the explicit form of $\mathcal{L}^{s(2)}_j$ given in Eq.~\eqref{eq:second order of the action}.
By decomposing the matrix $\bm{\mathcal{A}}^s$ into a sum of matrices $\bm{\mathcal{A}}^{s(\ell)}$ as in Eq.~\eqref{eq:matrix decomposing of A},
we can rewrite Eq.~\eqref{eq:calculation in the second order of quantum fluctuations 1} as
\begin{align}
    \label{eq:calculation in the second order of quantum fluctuations 1.5}
    W_{s}(\vec{\alpha}_{\rm f},\vec{\alpha}^*_{\rm f},t) = \lim_{\Delta t \to 0}\prod_{j=0}^{N_t-1}\int\frac{d^2\vec{\alpha}_j d^2\vec{\eta}_{j+1}}{\pi^{2M}}e^{i\Delta t \mathcal{L}^{s(1)}_j/\hbar}\prod_{\ell=1}^{\ell_{\rm max}}{\rm exp}\left\{-\frac{\Delta t}{2}
    \begin{bmatrix}
        \vec{\eta}_{j+1}^{*{\rm T}},\vec{\eta}^{\rm T}_{j+1}
    \end{bmatrix}
    \bm{\mathcal{A}}^{s(\ell)}(\vec{\alpha}_j,\vec{\alpha}^*_j)
    \begin{bmatrix}
        \vec{\eta}_{j+1} \\
        \vec{\eta}_{j+1}^{*}
    \end{bmatrix}
    \right\}W_{s}(\vec{\alpha}_0,\vec{\alpha}^*_0,t_0).
\end{align}

Here,
in order to perform the integration over the quantum fields $\vec{\eta}_{j+1}$,
we perform the following Hubbard--Stratonovich transformation \cite{Yoneya2026},
introducing the auxiliary fields $\Delta\overrightarrow{\mathcal{W}}^{(\ell)} \in \mathbb{R}^{2M}$:
\begin{align}
    \label{eq:Hubbard--Stratonovich transformation}
    {\rm exp}\left\{-\frac{\Delta t}{2} 
    \begin{bmatrix}
        \vec{\eta}_{j+1}^{*{\rm T}},\vec{\eta}^{\rm T}_{j+1}
    \end{bmatrix}
    \bm{\mathcal{A}}^{s(\ell)}(\vec{\alpha}_j,\vec{\alpha}^*_j)
    \begin{bmatrix}
        \vec{\eta}_{j+1} \\
        \vec{\eta}_{j+1}^*
    \end{bmatrix}
    \right\}
    =
    \prod_{\mu =1}^{2M}\int_{-\infty}^{\infty} d\Delta\mathcal{W}^{(\ell)}_\mu\frac{e^{-\Delta\mathcal{W}^{(\ell)2}_\mu/(2\Delta t)}}{\sqrt{2\pi\Delta t}}\prod_{m=1}^{M}{\rm exp}\left(\eta^*_{m,j+1}\left[i\bm{\mathcal{P}}\bm{\mathcal{V}}^{s(\ell)}\sqrt{\bm{\mathcal{A}}^{s(\ell)}_{\rm diag}}\bm{\mathcal{Q}^{(\ell)}}\Delta \overrightarrow{\mathcal{W}}^{(\ell)}\right]_m - {\rm c.c.}\right),
\end{align}
which is feasible when the matrix $\bm{\mathcal{A}}^{s(\ell)}$ is positive semidefinite.
In Eq.~\eqref{eq:Hubbard--Stratonovich transformation}, the unitary matrix $\bm{\mathcal{P}}$ is given by Eq.~\eqref{eq:def of matrix P}, $\bm{\mathcal{Q}}^{s(\ell)}$ is an arbitrary $2M\times 2M$ orthogonal matrix \cite{Gardiner},
and $\bm{\mathcal{V}}^{s(\ell)}$ is the $2M\times 2M$ real matrix satisfying
\begin{align}
    \label{eq:def of matrix V}
    \bm{\mathcal{V}}^{s(\ell)\rm T}\bm{\mathcal{P}}^{\dagger}\bm{\mathcal{A}}^{s(\ell)}\bm{\mathcal{P}}\bm{\mathcal{V}}^{s(\ell)} = \bm{\mathcal{A}}^{s(\ell)}_{\rm diag},
\end{align}
where $\bm{\mathcal{A}}^{s(\ell)}_{\rm diag}$ is a $2M \times 2M$ diagonal matrix whose diagonal entries are the eigenvalues of $\bm{\mathcal{A}}^{s(\ell)}$.
When $\bm{\mathcal{A}}^{s(\ell)}$ is positive semidefinite,
the matrix $\bm{\mathcal{P}}^{\dagger}\bm{\mathcal{A}}^{s(\ell)}\bm{\mathcal{P}}$ can be decomposed as a product of a $2M\times 2M$ real matrix $\bm{\mathcal{C}}^{s(\ell)}$ and its transpose as in Eq.~\eqref{eq:def of matrix C(ell)}.
By substituting Eq.~\eqref{eq:def of matrix C(ell)} into Eq.~\eqref{eq:def of matrix V},
we obtain
\begin{align}
    \label{eq:definitnion of matrix V(ell) with CCT}
    \bm{\mathcal{V}}^{s(\ell)\rm T}\bm{\mathcal{C}}^{s(\ell)}\bm{\mathcal{C}}^{s(\ell)\rm T}\bm{\mathcal{V}}^{s(\ell)} = \bm{\mathcal{A}}^{s(\ell)}_{\rm diag} = \sqrt{\bm{\mathcal{A}}^{s(\ell)}_{\rm diag}}\bm{\mathcal{Q}}^{(\ell)}\bm{\mathcal{Q}}^{(\ell)\rm T}\sqrt{\bm{\mathcal{A}}^{s(\ell)}_{\rm diag}},
\end{align}
yielding
\begin{align}
    \label{eq:rewrite sqrt As(ell)_diag}
    \sqrt{\bm{\mathcal{A}}^{s(\ell)}_{\rm diag}} = \bm{\mathcal{V}}^{s(\ell)\rm T}\bm{\mathcal{C}}^{s(\ell)}\bm{\mathcal{Q}}^{(\ell)\rm T}.
\end{align}
In the right-hand side of Eq.~\eqref{eq:definitnion of matrix V(ell) with CCT},
we have performed the decomposition $\bm{\mathcal{A}}^{s(\ell)}_{\rm diag} = \sqrt{\bm{\mathcal{A}}^{s(\ell)}_{\rm diag}}\sqrt{\bm{\mathcal{A}}^{s(\ell)}_{\rm diag}}$ and substituted the identity $\bm{\mathcal{Q}}^{(\ell)}\bm{\mathcal{Q}}^{(\ell)\rm T}$ in between.
Substituting Eq.~\eqref{eq:rewrite sqrt As(ell)_diag} into Eq.~\eqref{eq:Hubbard--Stratonovich transformation},
we can rewrite the Hubbard--Stratonovich transformation as
\begin{align}
    \label{eq:Hubbard--Stratonovich transformation using B(ell)}
    {\rm exp}\left\{-\frac{\Delta t}{2} 
    \begin{bmatrix}
        \vec{\eta}_{j+1}^{*{\rm T}},\vec{\eta}^{\rm T}_{j+1}
    \end{bmatrix}
    \bm{\mathcal{A}}^{s(\ell)}
    \begin{bmatrix}
        \vec{\eta}_{j+1} \\
        \vec{\eta}_{j+1}^*
    \end{bmatrix}
    \right\}
    =
    \prod_{\mu =1}^{2M}\int_{-\infty}^{\infty} d\Delta\mathcal{W}^{(\ell)}_\mu\frac{e^{-\Delta\mathcal{W}^{(\ell)2}_\mu/(2\Delta t)}}{\sqrt{2\pi\Delta t}}\prod_{m=1}^{M}{\rm exp}\left(\eta^*_{m,j+1}\left[\bm{\mathcal{B}}^{s(\ell)}\Delta \overrightarrow{\mathcal{W}}^{(\ell)}\right]_m - {\rm c.c.}\right),
\end{align}
where $\bm{\mathcal{B}}^{s(\ell)}$ is defined by Eq.~\eqref{eq:def of matrix B(ell)}.

When the matrices $\bm{\mathcal{A}}^{s(\ell)}$ are positive semidefinite for $\forall \ell$,
we can perform the Hubbard--Stratonovich transformation in Eq.~\eqref{eq:Hubbard--Stratonovich transformation using B(ell)} for $\forall \ell$,
obtaining
\begin{align}
    \label{eq:calculation in the second order of quantum fluctuations 2}
    W_{s}(\vec{\alpha}_{\rm f},\vec{\alpha}^*_{\rm f},t) =& \lim_{\Delta t \to 0}\prod_{j=0}^{N_t-1}\int d^2\vec{\alpha}_j
    \left[\prod_{\ell=1}^{\ell_{\rm max}}\prod_{\mu=1}^{2M}\int_{-\infty}^{\infty} d\Delta\mathcal{W}^{(\ell)}_\mu\frac{e^{-\Delta\mathcal{W}^{(\ell)2}_\mu/(2\Delta t)}}{\sqrt{2\pi\Delta t}}\right]\nonumber \\
    &\times\prod_{m=1}^{M}\int\frac{ d^2\eta_{m,j+1}}{\pi^{2}}{\rm exp}\left\{\eta^*_{m,j+1}\left(\alpha_{m,j+1} - \alpha_{m,j} - \frac{\Delta t}{i\hbar}\frac{\partial H_{s}}{\partial\alpha^*_{m,j}} - \frac{\Delta t}{2} \mathcal{K}^{s}_m(\vec{\alpha}_j,\vec{\alpha}^*_j) + \sum_{\ell'=1}^{\ell_{\rm max}}\left[\bm{\mathcal{B}}^{s(\ell')}\Delta \overrightarrow{\mathcal{W}}^{(\ell')}\right]_m\right) - {\rm c.c.}\right\}W_{s}(\vec{\alpha}_0,\vec{\alpha}^*_0,t_0),
\end{align}
where we have substituted Eq.~\eqref{eq:Hubbard--Stratonovich transformation using B(ell)} and the explicit form of $\mathcal{L}^{s(1)}_j$ in Eq.~\eqref{eq:first order of the action} into Eq.~\eqref{eq:calculation in the second order of quantum fluctuations 1.5}.
Here,
$\mathcal{K}^{s}_m$ is the dissipative drift term
given by Eq.~\eqref{eq:def of dissipative drift term}.
Considering that the integration over the quantum fields $\vec{\eta}_{j+1}$ in Eq.~\eqref{eq:calculation in the second order of quantum fluctuations 2} leads to the Dirac delta function:
\begin{align}
    \label{eq:definitnio of the Dirac delta function}
    \int \frac{d^2\vec{\eta}}{\pi^{2M}}e^{\vec{\eta}^*\cdot\vec{\alpha}-\vec{\eta}\cdot\vec{\alpha}^*} = \prod_{m=1}^M\int \frac{d^2\eta_m}{\pi^{2}}e^{\eta_m^*\alpha_m-{\rm c.c.}} =\prod_{m=1}^M\delta^{(2)}(\alpha_m)=\prod_{m=1}^M\delta(\alpha_m^{\rm re})\delta(\alpha_m^{\rm im}),
\end{align}
we can rewrite Eq.~\eqref{eq:calculation in the second order of quantum fluctuations 2} as
\begin{align}
    \label{eq:formal solution of the second-order approximation}
    W_{s}(\vec{\alpha}_{\rm f},\vec{\alpha}^*_{\rm f},t) = \lim_{\Delta t\to 0}\prod_{j=0}^{N_t - 1}\int\frac{d^2\vec{\alpha}_j}{\pi^M}\varUpsilon^{(2)}_{s}(\vec{\alpha}_{j+1},t_{j+1};\vec{\alpha}_j,t_j)W_{s}(\vec{\alpha}_0,\vec{\alpha}^*_0,t_0),
\end{align}
where $\varUpsilon^{(2)}_{s}(\vec{\alpha}_{j+1},t_{j+1};\vec{\alpha}_j,t_j)$ is the second-order propagator given by
\begin{align}
    \label{eq:def of second-order propagator}
    \varUpsilon^{(2)}_{s}(\vec{\alpha}_{j+1},t_{j+1};\vec{\alpha}_j,t_j) = \left[\prod_{\ell=1}^{\ell_{\rm max}}\prod_{\mu=1}^{2M}\int_{-\infty}^{\infty} d\Delta\mathcal{W}^{(\ell)}_\mu\frac{e^{-\Delta\mathcal{W}^{(\ell)2}_\mu/(2\Delta t)}}{\sqrt{2\pi\Delta t}}\right]\prod_{m=1}^{M}\pi\delta^{(2)}\left(\alpha_{m,j+1} - \alpha_{m,j} - \frac{\Delta t}{i\hbar}\frac{\partial H_{s}}{\partial\alpha^*_{m,j}} - \frac{\Delta t}{2} \mathcal{K}^{s}_m + \sum_{\ell'=1}^{\ell_{\rm max}}\left[\bm{\mathcal{B}}^{s(\ell')}\Delta \overrightarrow{\mathcal{W}}^{(\ell')}\right]_m - {\rm c.c.}\right).
\end{align}
As depicted in Fig.~\ref{fig:Path integral short summary}(c),
the sample points drawn from the initial $s$-ordered quasiprobability distribution function $W_s(\vec{\alpha}_0,\vec{\alpha}^*_0)$ evolve according to the stochastic differential equation obtained from the argument of the Dirac delta function on the right-hand side of Eq.~\eqref{eq:def of second-order propagator}:
\begin{align}
    \label{eq:stochastic differential equation discrete}
    \alpha_{m,j+1} - \alpha_{m,j} = \frac{\Delta t}{i\hbar}\frac{\partial H_{s}}{\partial\alpha^*_{m,j}} + \frac{\Delta t}{2}\sum_{k=1}^{k_{\rm max}}\gamma_k\left(L^*_{ks}\star_{s}\frac{\partial L_{ks}}{\partial\alpha^*_{m,j}} - \frac{\partial L^*_{ks}}{\partial\alpha^*_{m,j}}\star_{s}L_{ks}\right) + \sum_{\ell=1}^{\ell_{\rm max}}\left[\bm{\mathcal{B}}^{s(\ell)}\Delta \overrightarrow{\mathcal{W}}\right]_m,
\end{align}
where we have substituted the explicit form of the dissipative drift term Eq.~\eqref{eq:def of dissipative drift term}.
Taking the continuous limit of Eq.~\eqref{eq:stochastic differential equation discrete},
we obtain
\begin{align}
    i\hbar d\alpha_m = \left[\frac{\partial H_s}{\partial\alpha^*_{m}} + \frac{i\hbar}{2}\sum_{k=1}^{k_{\rm max}}\gamma_k\left(L^*_{ks}\star_s\frac{\partial L_{ks}}{\partial\alpha^*_m} - \frac{\partial L^*_{ks}}{\partial\alpha^*_m}\star_sL_{ks}\right)\right]dt + i\hbar\sum_{\ell=1}^{\ell_{\rm max}}\left[\bm{\mathcal{B}}^{s(\ell)}\cdot d\overrightarrow{\mathcal{W}}^{(\ell)}(t)\right]_m.
\end{align}
This completes the derivation of the stochastic differential equation~\eqref{eq:stochastic differential equation using B(ell)}.
Equation~\eqref{eq:stochastic differential equation using B} follows by setting $\ell_{\rm max} = 1$.


\section{\label{appendix:Positive-semidefiniteness of A}Positive semidefiniteness of \texorpdfstring{$\bm{\mathcal{A}}^s$}{Text} satisfying Eqs.~\texorpdfstring{\eqref{eq:condition for lambda_mn two jumps} and \eqref{eq:condition for Lambda_mn two jumps}}{Text}}
We show that the matrix $\bm{\mathcal{A}}^s$ becomes positive semidefinite when $\lambda^s_{mn}$ and $\Lambda^s_{mn}$ satisfy Eqs.~\eqref{eq:condition for lambda_mn two jumps} and \eqref{eq:condition for Lambda_mn two jumps},
respectively.
By definition,
$\bm{\mathcal{A}}^s$ is positive semidefinite if
\begin{align}
    \label{eq:necessary and sufficient condition for positive-semidefiniteness}
    \vec{w}^{*\rm T}\bm{\mathcal{A}}^s\vec{w}\geq 0
\end{align}
for any $\vec{w}\in\mathbb{C}^{2M}$.
We therefore prove this inequality under Eqs.~\eqref{eq:condition for lambda_mn two jumps} and \eqref{eq:condition for Lambda_mn two jumps}.
To this end,
without loss of generality,
we first rewrite the vector $\vec{w}$ as
\begin{align}
    \label{eq:rewriting w}
    \vec{w}=
    \begin{bmatrix}
        \vec{u}^{\rm T} \\
        \vec{v}^{\rm T}
    \end{bmatrix},
\end{align}
where $\vec{u}\in\mathbb{C}^M$ and $\vec{v}\in\mathbb{C}^M$ are arbitrary complex vectors.
Here,
for clarity,
we recall the explicit form of the matrix $\bm{\mathcal{A}}^s$:
\begin{align}
    \label{eq:definitnioa of the diffusion matrix A appendix}
    \bm{\mathcal{A}}^{s} = 2
    \begin{bmatrix}
        \bm{\Lambda}^{s} & \bm{\lambda}^{s} \\
        \bm{\lambda}^{s*} & \bm{\Lambda}^{s*}
    \end{bmatrix}.
\end{align}
Substituting Eqs.~\eqref{eq:rewriting w} and \eqref{eq:definitnioa of the diffusion matrix A appendix} into the left-hand side of Eq.~\eqref{eq:necessary and sufficient condition for positive-semidefiniteness}, we obtain
\begin{align}
    \label{eq:wAw Lambda lambda}
    \vec{w}^{*\rm T}\bm{\mathcal{A}}^s\vec{w} = 2\sum_{m,n=1}^M\left(u_m^*\Lambda^s_{mn}u_n + u_m^*\lambda^s_{mn}v_n + v^*_m\lambda^{s*}_{mn}u_n + v^*_m\Lambda^{s*}_{mn}v_n\right).
\end{align}
We note that Eq.~\eqref{eq:wAw Lambda lambda} does not always satisfy Eq.~\eqref{eq:necessary and sufficient condition for positive-semidefiniteness} depending on the details of $\lambda^s_{mn}$ and $\Lambda^s_{mn}$.
However,
when $\lambda^s_{mn}$ and $\Lambda^s_{mn}$ respectively satisfy Eqs.~\eqref{eq:condition for lambda_mn two jumps} and \eqref{eq:condition for Lambda_mn two jumps}, i.e.,
\begin{gather}
    \label{eq:condition for lambda_mn two jumps appendix}
    \lambda^{s}_{mn} = \frac{1}{2}(\mathcal{G}^{s*}_m\mathcal{F}^{s}_n + \mathcal{F}^{s}_m\mathcal{G}^{s*}_n), \\
    \label{eq:condition for Lambda_mn two jumps appendix}
    \Lambda^{s}_{mn} = \frac{1}{2}(\mathcal{G}^{s*}_m\mathcal{G}^{s}_n + \mathcal{F}^{s}_m\mathcal{F}^{s*}_n),
\end{gather}
we can show that $\vec{w}^{*\rm T}\bm{\mathcal{A}}^s\vec{w}$ takes non-negative value as
\begin{align}
    \vec{w}^{*\rm T}\bm{\mathcal{A}}^s\vec{w} =& \sum_{m,n=1}^M\left(u^*_m\mathcal{G}^{s*}_mu_n\mathcal{G}^s_n + u^*_m\mathcal{G}^{s*}_mv_n\mathcal{F}^s_n + v^*_m\mathcal{F}^{s*}_mu_n\mathcal{G}^s_n + v^*_m\mathcal{F}^{s*}_mv_n\mathcal{F}^s_n\right) \\
    &+ \sum_{m,n=1}^M\left(u^*_m\mathcal{F}^{s}_mu_n\mathcal{F}^{s*}_n + u^*_m\mathcal{F}^{s}_mv_n\mathcal{G}^{s*}_n + v^*_m\mathcal{G}^{s}_mu_n\mathcal{F}^{s*}_n + v^*_m\mathcal{G}^{s}_mv_n\mathcal{G}^{s*}_n\right) \\
    =& \left|\sum_{m=1}^M\left(u_m\mathcal{G}^s_m + v_m\mathcal{F}^s_m\right)\right|^2 + \left|\sum_{m=1}^M\left(u_m\mathcal{F}^{s*}_m + v_m\mathcal{G}^{s*}_m\right)\right|^2 \geq 0.
\end{align}
This complete the proof that Eqs.~\eqref{eq:condition for lambda_mn two jumps} and \eqref{eq:condition for Lambda_mn two jumps} are the sufficient conditions for the positive semidefiniteness of $\bm{\mathcal{A}}^s$.

\section{\label{appendix:Infinite-temperature steady state}Infinite-temperature steady state}
We derive a necessary and sufficient condition under which the infinite-temperature state is a steady-state solution of the GKSL equation~\eqref{eq:def of GKSL equation}. Since the infinite-temperature state $\hat{\rho}_{\infty}$ is proportional to the identity operator, i.e.,
$\hat{\rho}_{\infty}\propto\hat{1}$,
it follows that $\hat{\rho}_{\infty}$ is a steady-state solution of the GKSL equation if and only if the right-hand side of Eq.~\eqref{eq:def of GKSL equation} vanishes when $\hat{\rho}(t)=\hat{1}$ \cite{Breuer}:
\begin{align}
    \label{eq:infinite temperature state1}
     -\frac{i}{\hbar}\left[\hat{H},\hat{1}\right]_- + \sum_{k=1}^{k_{\rm max}}\gamma_k\left(\hat{L}_k\hat{1}\hat{L}^{\dagger}_k - \frac{1}{2}\left[\hat{L}^{\dagger}_k\hat{L}_k,\hat{1}\right]_+\right) = 0.
\end{align}
For the Hamiltonian term, since $\hat{H}$ commutes with $\hat{1}$, we have $[\hat{H},\hat{1}]_-=0$.
For the dissipative term, using $\hat{L}_k\hat{1}=\hat{1}\hat{L}_k=\hat{L}_k$, we obtain
\begin{align}
    \label{eq:infinite temperature state2}
    \sum_{k=1}^{k_{\rm max}}\gamma_k\left(\hat{L}_k\hat{1}\hat{L}^{\dagger}_k - \frac{1}{2}\left[\hat{L}^{\dagger}_k\hat{L}_k,\hat{1}\right]_+\right) = \sum_{k=1}^{k_{\rm max}}\gamma_k\left[\hat{L}_k,\hat{L}^{\dagger}_k\right]_-.
\end{align}
Substituting Eq.~\eqref{eq:infinite temperature state2} into Eq.~\eqref{eq:infinite temperature state1},
we obtain
\begin{align}
    \label{eq:infinite temperature state3}
    \sum_{k=1}^{k_{\rm max}}\gamma_k\left[\hat{L}^{\dagger}_k,\hat{L}_k\right]_-  = 0,
\end{align}
where we have swapped the two terms in the commutator.
Eq.~\eqref{eq:infinite temperature state3} provides the necessary and sufficient condition for the infinite-temperature state to be a steady-state solution of the GKSL equation, and is equivalent to Eq.~\eqref{eq:cond for A>0 for Wigner} with its right-hand side set to zero.

\section{\label{appendix:lambda and Lambda for jump operators satisfying}\texorpdfstring{$\lambda_{mn}^{s=0}$ and $\Lambda_{mn}^{s=0}$}{TEXT} for jump operators satisfying Eq.~\texorpdfstring{\eqref{eq:cond for A>0 for Wigner}}{TEXT}: Derivations of Eqs.~\texorpdfstring{\eqref{eq:lambda Wigner quadratic} and \eqref{eq:Lambda Wigner quadratic}}{TEXT}}
We derive Eqs.~\eqref{eq:lambda Wigner quadratic} and \eqref{eq:Lambda Wigner quadratic}.
To this end,
we first derive an intermediate relation from Eq.~\eqref{eq:cond for A>0 for Wigner}, which will be used in the subsequent derivations of Eqs.~\eqref{eq:lambda Wigner quadratic} and \eqref{eq:Lambda Wigner quadratic}.
Mapping both sides of Eq.~\eqref{eq:cond for A>0 for Wigner} to phase space,
we obtain
\begin{align}
    \label{eq:cond for A>0 for Wigner in phase space1}
    \sum_{k=1}^{k_{\rm max}}\gamma_k \left[[\hat{L}^{\dagger}_k,\hat{L}_k]_-\right]_{s=0}(\vec{\alpha},\vec{\alpha}^*) = \sum_{m=1}^M(l_m\alpha_m + \bar{l}_m\alpha^*_m) + {\rm Const.}
\end{align}
The phase-space representation of the commutation relation for $s=0$ on the left-hand side of Eq.~\eqref{eq:cond for A>0 for Wigner in phase space1} can be calculated as \cite{Polkovnikov2010}
\begin{align}
    \left[[\hat{L}^{\dagger}_k,\hat{L}_k]_-\right]_{s=0}(\vec{\alpha},\vec{\alpha}^*) &= L^*_{ks=0}(\vec{\alpha},\vec{\alpha}^*)\star_{s=0}L_{ks=0}(\vec{\alpha},\vec{\alpha}^*) - L_{ks=0}(\vec{\alpha},\vec{\alpha}^*)\star_{s=0}L^*_{ks=0}(\vec{\alpha},\vec{\alpha}^*)\\
    &= L^*_{ks=0}(\vec{\alpha},\vec{\alpha}^*)e^{\hat{\phi}_{s=0}/2}L_{ks=0}(\vec{\alpha},\vec{\alpha}^*) - L_{ks=0}(\vec{\alpha},\vec{\alpha}^*)e^{\hat{\phi}_{s=0}/2}L^*_{ks=0}(\vec{\alpha},\vec{\alpha}^*)\\
    &= L^*_{ks=0}(\vec{\alpha},\vec{\alpha}^*)e^{\hat{\phi}_{s=0}/2}L_{ks=0}(\vec{\alpha},\vec{\alpha}^*) - L^*_{ks=0}(\vec{\alpha},\vec{\alpha}^*)e^{-\hat{\phi}_{s=0}/2}L_{ks=0}(\vec{\alpha},\vec{\alpha}^*)\\
    \label{eq:cond for A>0 for Wigner in phase space2}
    &=2 L^*_{ks=0}(\vec{\alpha},\vec{\alpha}^*) {\rm sinh}\frac{\hat{\phi}_{s=0}}{2} L_{ks=0}(\vec{\alpha},\vec{\alpha}^*),
\end{align}
where we have used the relation $A_{s=0}(\vec{\alpha},\vec{\alpha}^*)e^{\hat{\phi}_{s=0}/2}B_{s=0}(\vec{\alpha},\vec{\alpha}^*) = B_{s=0}(\vec{\alpha},\vec{\alpha}^*)e^{-\hat{\phi}_{s=0}/2}A_{s=0}(\vec{\alpha},\vec{\alpha}^*)$
with $\hat{\phi}_{s=0}$ being the differential operator defined by Eq.~\eqref{eq:operator phi}.
Assuming that the jump operators are at most quadratic in $\hat{a}^{\dagger}_m$ and $\hat{a}_m$, 
$L_{ks=0}$ contains terms no higher than quadratic order in $\alpha_m$ and $\alpha^*_m$.
Since $\hat{\phi}_{s=0}$ is the differential operator,
it then follows that the right-hand side of Eq.~\eqref{eq:cond for A>0 for Wigner in phase space2}, evaluated for $\hat{\phi}_{s=0}$, remains finite up to second order, yielding
\begin{align}
    \label{eq:cond for A>0 for Wigner in phase space3}
    \left[[\hat{L}^{\dagger}_k,\hat{L}_k]_-\right]_{s=0}(\vec{\alpha},\vec{\alpha}^*) = L^*_{ks=0}(\vec{\alpha},\vec{\alpha}^*)\hat{\phi}_{s=0}L_{ks=0}(\vec{\alpha},\vec{\alpha}^*).
\end{align}
Substituting Eq.~\eqref{eq:cond for A>0 for Wigner in phase space3} into Eq.~\eqref{eq:cond for A>0 for Wigner in phase space1},
we obtain
\begin{align}
    \label{eq:cond for A>0 for Wigner in phase space}
    \sum_{k=1}^{k_{\rm max}}\gamma_k L^*_{ks=0}(\vec{\alpha},\vec{\alpha}^*)\hat{\phi}_{s=0}L_{ks=0}(\vec{\alpha},\vec{\alpha}^*) = \sum_{m=1}^M(l_m\alpha_m + \bar{l}_m\alpha^*_m) + {\rm Const.},
\end{align}
Subsequently, 
we differentiate both sides of Eq.~\eqref{eq:cond for A>0 for Wigner in phase space} twice,
obtaining
\begin{gather}
    \label{eq:condition for vanishing lambda dummy}
    \sum_{k=1}^{k_{\rm max}}\gamma_k\left(\frac{\partial L^*_{ks=0}}{\partial\alpha^*_m}\hat{\phi}_{s=0}\frac{\partial L_{ks=0}}{\partial\alpha^*_n} + \frac{\partial L^*_{ks=0}}{\partial\alpha^*_n}\hat{\phi}_{s=0}\frac{\partial L_{ks=0}}{\partial\alpha^*_m}\right) + \sum_{k=1}^{k_{\rm max}}\gamma_k\left(\frac{\partial^2 L^*_{ks=0}}{\partial\alpha^*_m\partial\alpha^*_n}\hat{\phi}_{s=0}L_{ks=0} + L^*_{ks=0}\hat{\phi}_{s=0}\frac{\partial L_{ks=0}}{\partial\alpha^*_m\partial\alpha^*_n}\right) = 0, \\
    \label{eq:condition for vanishing Lambda dummy}
    \sum_{k=1}^{k_{\rm max}}\gamma_k\left(\frac{\partial L^*_{ks=0}}{\partial\alpha^*_m}\hat{\phi}_{s=0}\frac{\partial L_{ks=0}}{\partial\alpha_n} + \frac{\partial L^*_{ks=0}}{\partial\alpha_n}\hat{\phi}_{s=0}\frac{\partial L_{ks=0}}{\partial\alpha^*_m}\right) + \sum_{k=1}^{k_{\rm max}}\gamma_k\left(\frac{\partial^2 L^*_{ks=0}}{\partial\alpha^*_m\partial\alpha_n}\hat{\phi}_{s=0}L_{ks=0} + L^*_{ks=0}\hat{\phi}_{s=0}\frac{\partial L_{ks=0}}{\partial\alpha^*_m\partial\alpha_n}\right)= 0.
\end{gather}
Here,
since $L_{ks=0}$ contains terms no higher than quadratic order in $\alpha_m$ and $\alpha^*_m$, and $\hat{\phi}_{s=0}$ is the differential operator,
every term in the second summations on the left-hand sides of Eqs.~\eqref{eq:condition for vanishing lambda dummy} and \eqref{eq:condition for vanishing Lambda dummy} vanish,
i.e.,
\begin{align}
    \label{eq:cond for A>0 for Wigner in phase space4}
    \frac{\partial^2 L^*_{ks=0}}{\partial\alpha^*_m\partial\alpha^*_n}\hat{\phi}_{s=0}L_{ks=0} = L^*_{ks=0}\hat{\phi}_{s=0}\frac{\partial L_{ks=0}}{\partial\alpha^*_m\partial\alpha^*_n} = \frac{\partial^2 L^*_{ks=0}}{\partial\alpha^*_m\partial\alpha_n}\hat{\phi}_{s=0}L_{ks=0} = L^*_{ks=0}\hat{\phi}_{s=0}\frac{\partial L_{ks=0}}{\partial\alpha^*_m\partial\alpha_n} = 0,
\end{align}
for $\forall k$.
Finally,
substituting Eq.~\eqref{eq:cond for A>0 for Wigner in phase space4} into Eqs.~\eqref{eq:condition for vanishing lambda dummy} and \eqref{eq:condition for vanishing Lambda dummy},
we obtain
\begin{gather}
    \label{eq:condition for vanishing lambda}
    \sum_{k=1}^{k_{\rm max}}\gamma_k\left(\frac{\partial L^*_{ks=0}}{\partial\alpha^*_m}\hat{\phi}_{s=0}\frac{\partial L_{ks=0}}{\partial\alpha^*_n} + \frac{\partial L^*_{ks=0}}{\partial\alpha^*_n}\hat{\phi}_{s=0}\frac{\partial L_{ks=0}}{\partial\alpha^*_m}\right) = 0, \\
    \label{eq:condition for vanishing Lambda}
    \sum_{k=1}^{k_{\rm max}}\gamma_k\left(\frac{\partial L^*_{ks=0}}{\partial\alpha^*_m}\hat{\phi}_{s=0}\frac{\partial L_{ks=0}}{\partial\alpha_n} + \frac{\partial L^*_{ks=0}}{\partial\alpha_n}\hat{\phi}_{s=0}\frac{\partial L_{ks=0}}{\partial\alpha^*_m}\right)= 0.
\end{gather}

We then derive Eqs.~\eqref{eq:lambda Wigner quadratic} and \eqref{eq:Lambda Wigner quadratic}.
To this end,
we expand $\star_{s=0}$ in $\lambda^{s=0}_{mn}$ and $\Lambda^{s=0}_{mn}$ for $\hat{\phi}_{s=0}$.
Since $\hat{L}_k$ is assumed to be at most quadratic in $\hat{a}^{\dagger}_m$ and $\hat{a}_m$,
only terms up to first order in $\hat{\phi}_{s=0}$ survive in $\lambda^{s=0}_{mn}$ and $\Lambda^{s=0}_{mn}$,
yielding
\begin{gather}
    \label{eq:lambda quadratic}
    \lambda^{s=0}_{mn} = \sum_{k=1}^{k_{\rm max}}\frac{\gamma_k}{4}\left(\frac{\partial L^*_{ks=0}}{\partial\alpha^*_m}\frac{\partial L_{ks=0}}{\partial\alpha^*_n} + \frac{\partial L^*_{ks=0}}{\partial\alpha^*_n}\frac{\partial L_{ks=0}}{\partial\alpha^*_m}\right) + \sum_{k=1}^{k_{\rm max}}\frac{\gamma_k}{8}\left(\frac{\partial L^*_{ks=0}}{\partial\alpha^*_m}\hat{\phi}_{s=0}\frac{\partial L_{ks=0}}{\partial\alpha^*_n} + \frac{\partial L^*_{ks=0}}{\partial\alpha^*_n}\hat{\phi}_{s=0}\frac{\partial L_{ks=0}}{\partial\alpha^*_m}\right), \\
    \label{eq:Lambda quadratic}
    \Lambda^{s=0}_{mn} = \sum_{k=1}^{k_{\rm max}}\frac{\gamma_k}{4}\left(\frac{\partial L^*_{ks=0}}{\partial\alpha^*_m}\frac{\partial L_{ks=0}}{\partial\alpha_n} + \frac{\partial L^*_{ks=0}}{\partial\alpha_n}\frac{\partial L_{ks=0}}{\partial\alpha^*_m}\right) + \sum_{k=1}^{k_{\rm max}}\frac{\gamma_k}{8}\left(\frac{\partial L^*_{ks=0}}{\partial\alpha^*_m}\hat{\phi}_{s=0}\frac{\partial L_{ks=0}}{\partial\alpha_n} + \frac{\partial L^*_{ks=0}}{\partial\alpha_n}\hat{\phi}_{s=0}\frac{\partial L_{ks=0}}{\partial\alpha^*_m}\right).
\end{gather}
When the jump operators satisfy Eq.~\eqref{eq:cond for A>0 for Wigner},
from Eqs.~\eqref{eq:condition for vanishing lambda} and \eqref{eq:condition for vanishing Lambda}, the second summations appearing on the right-hand sides of Eqs.~\eqref{eq:lambda quadratic} and \eqref{eq:Lambda quadratic} vanish,
we obtain
\begin{gather}
    \label{eq:lambda Wigner quadratic appendix}
    \lambda^{s=0}_{mn} = \sum_{k=1}^{k_{\rm max}}\frac{\gamma_k}{4}\left(\frac{\partial L^*_{ks=0}}{\partial\alpha^*_m}\frac{\partial L_{ks=0}}{\partial\alpha^*_n} + \frac{\partial L^*_{ks=0}}{\partial\alpha^*_n}\frac{\partial L_{ks=0}}{\partial\alpha^*_m}\right), \\
    \label{eq:Lambda Wigner quadratic appendix}
    \Lambda^{s=0}_{mn} = \sum_{k=1}^{k_{\rm max}}\frac{\gamma_k}{4}\left(\frac{\partial L^*_{ks=0}}{\partial\alpha^*_m}\frac{\partial L_{ks=0}}{\partial\alpha_n} + \frac{\partial L^*_{ks=0}}{\partial\alpha_n}\frac{\partial L_{ks=0}}{\partial\alpha^*_m}\right).
\end{gather}
This completes the derivations of Eqs.~\eqref{eq:lambda Wigner quadratic} and \eqref{eq:Lambda Wigner quadratic}.


\section{\label{appendix:Third and fourth order of quantum fluctuations}Third- and fourth-order quantum fluctuations}
We show the third- and fourth-order contributions from quantum fluctuations in the Lagrangian $\mathcal{L}^{s}_j$.
Here,
we consider a general Hamiltonian,
while assuming that the jump operators involve at most quadratic terms in $\hat{a}^{\dagger}_m$ and $\hat{a}_m$.
Under these assumptions,
the third-order contribution $\mathcal{L}^{s(3)}_j$ is given by
\begin{gather}
    \label{eq:third order of the action}
    \mathcal{L}_j^{s(3)}  = \mathcal{L}^{s(3:\hat{H})}_j +  \mathcal{L}^{s(3:\{\hat{L}_k\})}_j, \\
    \label{eq:third order of the action unitary}
    \mathcal{L}^{s(3:\hat{H})}_j = \sum_{m,n,p}\left\{\frac{\eta_{m,j+1}\eta_{n,j+1}\eta_{p,j+1}}{24}(1 + 3s^2)\frac{\partial^3 H_s}{\partial\alpha_{m,j}\partial\alpha_{n,j}\partial\alpha_{p,j}} + \frac{\eta_{m,j+1}\eta_{n,j+1}\eta^*_{p,j+1}}{8}(1-s^2)\frac{\partial^3 H_s}{\partial\alpha_{m,j}\partial\alpha_{n,j}\partial\alpha^*_{p,j}}\right\} + {\rm c.c.}\\
    \begin{aligned}
        \label{eq:third order of the action non-unitary}
        \mathcal{L}^{s(3:\{\hat{L}_k\})}_j = -i\hbar\sum_{k=1}^{k_{\rm max}}\gamma_k\sum_{m,n,p}&\left[\frac{\eta_{m,j+1}\eta_{n,j+1}\eta_{p,j+1}}{16}\left\{-(s^2 + 4s - 1)\frac{\partial L^*_{ks}}{\partial\alpha_{m,j}}\frac{\partial^2L_{ks}}{\partial\alpha_{n,j}\partial\alpha_{p,j}} + (s^2 - 4s - 1)\frac{\partial^2 L^*_{ks}}{\partial\alpha_{m,j}\partial\alpha_{n,j}}\frac{\partial L_{ks}}{\partial\alpha_{p,j}}\right\}\right. \\
        &+\frac{\eta_{m,j+1}\eta_{n,j+1}\eta^*_{p,j+1}}{16}\left\{(1-s)(1-3s)\frac{\partial L^*_{ks}}{\partial\alpha^*_{p,j}} \frac{\partial^2 L_{ks}}{\partial\alpha_{m,j}\partial\alpha_{n,j}} - 2(1-s^2)\frac{\partial^2 L^*_{ks}}{\partial\alpha_{m,j}\partial\alpha^*_{p,j}}\frac{\partial L_{ks}}{\partial\alpha_{n,j}}\right. \\ 
        &\hphantom{+\frac{\eta_{m,j+1}\eta_{n,j+1}\eta^*_{p,j+1}}{16}\{}+ 2(1-s^2)\frac{\partial L^*_{ks}}{\partial\alpha_{m,j}}\frac{\partial^2 L_{ks}}{\partial\alpha_{n,j}\partial\alpha^*_{p,j}} - \left.\left.(1+s)(1+3s)\frac{\partial^2L^*_{ks}}{\partial\alpha_{m,j}\partial\alpha_{n,j}}\frac{\partial L_{ks}}{\partial\alpha^*_{p,j}}\right\}\right] + {\rm c.c.}
    \end{aligned}
\end{gather}
and the fourth-order contribution $\mathcal{L}^{s(4)}_j$ is as follows:
\begin{gather}
    \label{eq:fourth order of the action}
    \mathcal{L}_j^{s(4)}  = \mathcal{L}^{s(4:\hat{H})}_j +  \mathcal{L}^{s(4:\{\hat{L}_k\})}_j, \\
    \label{eq:fourth order of the action unitary}
    \mathcal{L}^{s(4:\hat{H})}_j = \sum_{m,n,p,q}\left\{\frac{\eta_{m,j+1}\eta_{n,j+1}\eta_{p,j+1}\eta_{q,j+1}}{48}s(1 + s^2)\frac{\partial^4 H_s}{\partial\alpha_{m,j}\partial\alpha_{n,j}\partial\alpha_{p,j}\partial\alpha_{q,j}} + \frac{\eta_{m,j+1}\eta_{n,j+1}\eta_{p,j+1}\eta^*_{q,j+1}}{24}s(1-s^2)\frac{\partial^4 H_s}{\partial\alpha_{m,j}\partial\alpha_{n,j}\partial\alpha_{p,j}\partial\alpha^*_{q,j}}\right\} - {\rm c.c.,} \\
    \begin{aligned}
        \label{eq:fourth order of the action non-unitary}
        \mathcal{L}_j^{s(4:\{\hat{L}_k\})} =  -i\hbar\sum_{k=1}^{k_{\rm max}}\gamma_k\sum_{m,n,p,q}&\left[-\frac{\eta_{m,j+1}\eta_{n,j+1}\eta_{p,j+1}\eta_{q,j+1}}{8}s^2\frac{\partial^2 L^*_{ks}}{\partial\alpha_{m,j}\partial\alpha_{n,j}}\frac{\partial^2 L_{ks}}{\partial\alpha_{p,j}\partial\alpha_{q,j}}\right.\\
        &+ \frac{\eta_{m,j+1}\eta_{n,j+1}\eta_{p,j+1}\eta^*_{q,j+1}}{16}s(1-s^2)\left(\frac{\partial^2 L^*_{ks}}{\partial\alpha_{m,j}\partial\alpha_{n,j}}\frac{\partial^2 L_{ks}}{\partial\alpha_{p,j}\partial\alpha^*_{q,j}} - \frac{\partial^2 L^*_{ks}}{\partial\alpha_{m,j}\partial\alpha^*_{q,j}}\frac{\partial^2 L_{ks}}{\partial\alpha_{p,j}\partial\alpha_{n,j}}\right)\\
        &+\frac{\eta_{m,j+1}\eta_{n,j+1}\eta^*_{p,j+1}\eta^*_{q,j+1}}{32}s\left\{(1+s)^2\frac{\partial^2 L^*_{ks}}{\partial\alpha_{m,j}\partial\alpha_{n,j}}\frac{\partial^2 L_{ks}}{\partial\alpha_{p,j}^{*}\partial\alpha_{q,j}^{*}} - \left.(1-s)^2\frac{\partial^2 L^*_{ks}}{\partial\alpha_{p,j}^{*}\partial\alpha_{q,j}^{*}}\frac{\partial^2 L_{ks}}{\partial\alpha_{m,j}\partial\alpha_{n,j}}\right\} \right] - {\rm c.c.},
    \end{aligned}
\end{gather}
where $\mathcal{L}^{s(3/4:\hat{H})}_j$ and $\mathcal{L}^{s(3/4:\{\hat{L}_k\})}_j$ are the contributions from the Hamiltonian and jump operators,
respectively.


\section{\label{appendix:Absence of the effect of the second order of quantum fluctuations}Absence of the effect of second-order quantum fluctuations}
We show that the second-order contributions of quantum fluctuations do not affect the dynamics of $C_{12}$ of the system described by the GKSL equation~\eqref{eq:GKSL equation model2},
as discussed in Sec.~\ref{subsubsec:Model21}.
In this system,
the stochastic differential equations can be obtained for $s=0,\pm1$, 
and are given by Eqs.~\eqref{eq:SDE for site1 Model21} and \eqref{eq:SDE for site2 Model21}:
\begin{gather}
    \label{eq:SDE for site1 Model21 appendix}
    i\hbar d\alpha_{1} = \left[-\mu\alpha_{1} - J\alpha_{2} - \frac{i\hbar\gamma}{2}\alpha_{1}\right]dt - \hbar\sqrt{\frac{\gamma}{2}}\alpha_2\cdot(d\mathcal{W}_1 - id\mathcal{W}_2), \\
    \label{eq:SDE for site2 Model21 appendix}
    i\hbar d\alpha_{2} = \left[-\mu\alpha_{2} - J\alpha_{1} - \frac{i\hbar\gamma}{2}\alpha_{2}\right]dt - \hbar\sqrt{\frac{\gamma}{2}}\alpha_1\cdot(d\mathcal{W}_1 + id\mathcal{W}_2).
\end{gather}
Below,
we derive the equation of motion of $C_{12}$ from Eqs.~\eqref{eq:SDE for site1 Model21 appendix} and \eqref{eq:SDE for site2 Model21 appendix},
and show that it does not involve the terms originating from the second-order fluctuations,
i.e.,
contributions from the stochastic terms in Eqs.~\eqref{eq:SDE for site1 Model21 appendix} and \eqref{eq:SDE for site2 Model21 appendix}.
To this end,
we introduce $f_1\in\mathbb{C}$ and $g_1\in\mathbb{C}$ [$f_2\in\mathbb{C}$ and $g_2\in\mathbb{C}$] to denote the contributions from the classical dynamics and second-order fluctuations in Eq.~\eqref{eq:SDE for site1 Model21 appendix} [Eq.~\eqref{eq:SDE for site2 Model21 appendix}],
respectively,
as follows:
\begin{gather}
    \label{eq:classical contribution1 appendix}
    f_1 = \left[-\mu\alpha_{1} - J\alpha_{2} - \frac{i\hbar\gamma}{2}\alpha_{1}\right]dt, \\
    f_2 = \left[-\mu\alpha_{2} - J\alpha_{1} - \frac{i\hbar\gamma}{2}\alpha_{2}\right]dt, \\
    \label{eq:quantum contribution1 appendix}
    g_1 = - \hbar\sqrt{\frac{\gamma}{2}}\alpha_2\cdot(d\mathcal{W}_1 - id\mathcal{W}_2), \\
    \label{eq:quantum contribution2 appendix}
    g_2 = - \hbar\sqrt{\frac{\gamma}{2}}\alpha_1\cdot(d\mathcal{W}_1 + id\mathcal{W}_2).
\end{gather}
Using Eqs.~\eqref{eq:classical contribution1 appendix}--\eqref{eq:quantum contribution2 appendix}, we can rewrite Eqs.~\eqref{eq:SDE for site1 Model21 appendix} and \eqref{eq:SDE for site2 Model21 appendix} as
\begin{align}
    \label{eq:SDE for site1 Model21 appendix fg}
    i\hbar d\alpha_{1} = f_1 + g_1,\\
    \label{eq:SDE for site2 Model21 appendix fg}
    i\hbar d\alpha_{2} = f_2 + g_2.
\end{align}

In order to derive the equation of motion for $C_{12} =  (\braket{\hat{a}^{\dagger}_1\hat{a}_2} + \braket{\hat{a}^{\dagger}_2\hat{a}_1})/(2N_{\rm I}) = {\rm Re}[\braket{\hat{a}^{\dagger}_1\hat{a}_2}]/N_{\rm I}$,
we need to derive the equation of motion for $\braket{\hat{a}^{\dagger}_1\hat{a}_2}$ and then take its real part.
In the Monte Carlo simulation of the stochastic differential equations,
the two-point correlation function $\braket{\hat{a}^{\dagger}_1\hat{a}_2}$ is evaluated as $\braket{\hat{a}^{\dagger}_1\hat{a}_2} = {\rm E}[\alpha^*_1\alpha_2]$,
where ${\rm E}[\dots]$ denotes the ensemble average over both the initial quasiprobability distribution function and stochastic processes.
Noting
\begin{align}
    \label{eq:propertie of the Wiener proccesses}
    {\rm E}[d\mathcal{W}_\mu] = 0,\quad{\rm E}[d\mathcal{W}_\mu d\mathcal{W}_{\mu'}] = \Delta t\delta_{\mu\mu'},
\end{align}
for $\mu,\mu'=1,2$ \cite{Risken},
from Eqs.~\eqref{eq:SDE for site1 Model21 appendix fg} and \eqref{eq:SDE for site2 Model21 appendix fg},
we obtain the equation of motion for ${\rm E}[\alpha^*_1\alpha_2]$ as
\begin{align}
    \label{eq:equatino of motion of correlation E appendix}
    i\hbar d{\rm E}[\alpha_1^*\alpha_2] = {\rm E}[f_2\alpha^*_1 - f^*_1\alpha_2] - {\rm E}[g^*_1g_2] + o(\Delta t),
\end{align}
where ${\rm E}[f_2\alpha^*_1 - f^*_1\alpha_2]$ is the contribution from the classical dynamics given by
\begin{align}
    \label{eq:classical contribution to correlation appendix}
    {\rm E}[f_2\alpha^*_1 - f^*_1\alpha_2] = -J(\braket{\hat{a}^{\dagger}_1\hat{a}_1} - \braket{\hat{a}^{\dagger}_2\hat{a}_2})\Delta t - i\hbar\gamma\braket{\hat{a}^{\dagger}_1\hat{a}_2}\Delta t,
\end{align}
and ${\rm E}[g^*_1g_2]$ represents the contribution from the second-order fluctuations.
However,
from Eqs.~\eqref{eq:quantum contribution1 appendix} and \eqref{eq:quantum contribution2 appendix},
we can show that ${\rm E}[g^*_1g_2]$ vanishes identically,
i.e.,
\begin{align}
    \label{eq:quantum contribution to correlation appendix}
    {\rm E}[g^*_1g_2] = 0.
\end{align}
Thus,
the second-order fluctuations do not affect the dynamics of ${\rm E}[\alpha^*_1\alpha_2]$,
and hence the dynamics of $C_{12}$.
The resulting equation of motion for $C_{12}$ is given by
\begin{align}
    \label{eq:equation of motion of C12 appendix}
    \frac{dC_{12}}{dt} = -\gamma C_{12},
\end{align}
which we can derive by substituting Eqs.~\eqref{eq:classical contribution to correlation appendix} and \eqref{eq:quantum contribution to correlation appendix} into Eq.~\eqref{eq:equatino of motion of correlation E appendix},
taking the limit of $\Delta t\to 0$,
and subsequently taking the real part.


\section{\label{appendix:Sampling for the initial state}Initial-state sampling for the benchmark calculations in \texorpdfstring{Sec.~\ref{subsec:Model2}}{TEXT}}


\subsection{\label{appendix:Steady state of the system under the dephasing}Steady state of the system under the dephasing}
We first show that the quantum state given in Eq.~\eqref{eq:initial state for benchmark model3} corresponds to the diagonal part of the density matrix of a pure coherent state in the Fock basis,
and that it is the steady state of the following GKSL equation:
\begin{gather}
    \label{eq:GKSL equation dephasing}
    \frac{d\hat{\rho}(t)}{dt} = \gamma\sum_{k=1,2,3}\left(\hat{L}_k\hat{\rho}(t)\hat{L}^{\dagger}_k - \frac{1}{2}\left[\hat{L}^{\dagger}_k\hat{L}_k,\hat{\rho}(t)\right]_+\right), \\
    \label{eq:jump operators dephasing}
    \hat{L}_1 = \hat{a}^{\dagger}_1\hat{a}_1,~\hat{L}_2 = \hat{a}^{\dagger}_2\hat{a}_2,~\hat{L}_3 = \hat{a}^{\dagger}_3\hat{a}_3,
\end{gather}
where bosons at each degree of freedom is subjected to dephasing with equal strength.
In order to obtain the steady state of Eq.~\eqref{eq:GKSL equation dephasing},
we expand $\rho(t)$ in the Fock basis as
\begin{align}
    \label{eq:Fock state expansion}
    \hat{\rho}(t) = \sum_{n_1,n_2,n_3=0}^{\infty}\sum_{n'_1,n'_2,n'_3=0}^{\infty}\rho^{n'_1,n'_2,n'_3}_{n_1,n_2,n_3}(t)\ket{n_1,n_2,n_3}\bra{n'_1,n'_2,n'_3},
\end{align}
where $\rho^{n'_1,n'_2,n'_3}_{n_1,n_2,n_3}(t) = \braket{n_1,n_2,n_3|\hat{\rho}(t)|n'_1,n'_2,n'_3}\in\mathbb{C}$.
Substituting Eq.~\eqref{eq:Fock state expansion} into Eq.~\eqref{eq:GKSL equation dephasing}, we obtain the equation of motion for $\rho^{n'_1,n'_2,n'_3}_{n_1,n_2,n_3}(t)$ as
\begin{align}
    \frac{d\rho^{n'_1,n'_2,n'_3}_{n_1,n_2,n_3}(t)}{dt} = - \frac{\gamma}{2}\sum_{m=1,2,3}(n_m - n'_m)^2\rho^{n'_1,n'_2,n'_3}_{n_1,n_2,n_3}(t),
\end{align}
whose solution is
\begin{align}
    \rho^{n'_1,n'_2,n'_3}_{n_1,n_2,n_3}(t) = \rho^{n'_1,n'_2,n'_3}_{n_1,n_2,n_3}(0){\rm exp}\left\{-\frac{\gamma}{2}\sum_{m=1,2,3}(n_m - n'_m)^2t\right\}.
\end{align}
Here,
taking the limit $t\to\infty$ leads to
\begin{align}
    \label{eq:Fock state expansion steady state}
    \rho^{n'_1,n'_2,n'_3}_{n_1,n_2,n_3}(t\to\infty) =
    \begin{cases}
        \rho^{n_1,n_2,n_3}_{n_1,n_2,n_3}(0) & \text{for}~n_m = n'_m~\text{for}~\forall m \\
        0 & \text{for}~\text{others}
    \end{cases}.
\end{align}
Thus,
only the diagonal components of the density matrix survive in the steady state of the GKSL equation~\eqref{eq:GKSL equation dephasing},
i.e.,
\begin{align}
    \label{steady state under dephasing}
    \hat{\rho}(t\to\infty) = \sum_{n_1,n_2,n_3=0}^{\infty}\rho^{n_1,n_2,n_3}_{n_1,n_2,n_3}(0)\ket{n_1,n_2,n_3}\bra{n_1,n_2,n_3}.
\end{align}

When we choose the initial state as a pure coherent state
\begin{align}
    \label{eq:coherent state in Fock}
    \ket{\alpha_{{\rm I}1},\alpha_{{\rm I}2},\alpha_{{\rm I}3}} = e^{-(|\alpha_{{\rm I}1}|^2 + |\alpha_{{\rm I}2}|^2 + |\alpha_{{\rm I}3}|^2)/2}\sum_{n_1,n_2,n_3=0}^{\infty}\frac{\alpha_{{\rm I}1}^{n_1}\alpha_{{\rm I}2}^{n_2}\alpha_{{\rm I}3}^{n_3}}{\sqrt{n_1!n_2!n_3!}}\ket{n_1,n_2,n_3},
\end{align}
the diagonal part of the density matrix $\hat{\rho}(0)=\ket{\alpha_{{\rm I}1},\alpha_{{\rm I}2},\alpha_{{\rm I}3}}\bra{\alpha_{{\rm I}1},\alpha_{{\rm I}2},\alpha_{{\rm I}3}}$ is given by
\begin{align}
    \label{Fock state expansion steady state initial coherent state}
    \rho^{n_1,n_2,n_3}_{n_1,n_2,n_3}(0) = e^{-(|\alpha_{{\rm I}1}|^2 + |\alpha_{{\rm I}2}|^2 + |\alpha_{{\rm I}3}|^2)}\frac{|\alpha_{{\rm I}1}|^{2n_1}|\alpha_{{\rm I}2}|^{2n_2}|\alpha_{{\rm I}3}|^{2n_3}}{n_1!n_2!n_3!}.
\end{align}
By substituting Eq.~\eqref{Fock state expansion steady state initial coherent state} into Eq.~\eqref{steady state under dephasing}, we obtain the steady state of Eq.~\eqref{eq:GKSL equation dephasing} as
\begin{align}
    \hat{\rho}(t\to\infty) = \sum_{n_1,n_2,n_3=0}^{\infty}e^{-(|\alpha_{{\rm I}1}|^2 + |\alpha_{{\rm I}2}|^2 + |\alpha_{{\rm I}3}|^2)}\frac{|\alpha_{{\rm I}1}|^{2n_1}|\alpha_{{\rm I}2}|^{2n_2}|\alpha_{{\rm I}3}|^{2n_3}}{n_1!n_2!n_3!}\ket{n_1,n_2,n_3}\bra{n_1,n_2,n_3},
\end{align}
which is identical to the quantum state in Eq.~\eqref{eq:initial state for benchmark model3}.
In other words,
Eq.~\eqref{eq:initial state for benchmark model3} is the steady state obtained by evolving an initial pure coherent state according to the GKSL equations~\eqref{eq:GKSL equation dephasing} and \eqref{eq:jump operators dephasing}.


\subsection{\label{appendix:Sampling in the phase space}Sampling in phase space}
We describe the procedure for sampling the initial conditions according to Eq.~\eqref{eq:initial state for benchmark model3}.
As explained in \ref{appendix:Steady state of the system under the dephasing},
Eq.~\eqref{eq:initial state for benchmark model3} is the steady state of the GKSL equation~\eqref{eq:GKSL equation dephasing} starting from a pure coherent state,
$\hat{\rho}(0)=\ket{\alpha_{{\rm I}1},\alpha_{{\rm I}2},\alpha_{{\rm I}3}}\bra{\alpha_{{\rm I}1},\alpha_{{\rm I}2},\alpha_{{\rm I}3}}$.
Using this fact,
in Sec.~\ref{subsec:Model2},
we prepare the initial samples corresponding to Eq.~\eqref{eq:initial state for benchmark model3} by calculating the steady state of the GKSL equation~\eqref{eq:GKSL equation dephasing} in phase space using the Wigner function.
Here,
since the GKSL equation~\eqref{eq:GKSL equation dephasing} falls into case~(I) in Tab.~\ref{tab:brief summary},
the second-order approximation becomes exact,
and the stochastic differential equations can be obtained.
The corresponding stochastic differential equations for the Wigner function are given by
\begin{align}
    \label{eq:SDE dephasing}
    i\hbar d\alpha_m = -\frac{i\hbar\gamma}{2}\alpha_mdt - \hbar\sqrt{\gamma}\alpha_m\cdot d\mathcal{W}_m,
\end{align}
where $m=1,2,3$,
and $\mathcal{W}_m$ are the independent Wiener processes.
The dynamics of the different degrees of freedom are decoupled.
Since the stochastic differential equation in Eq.~\eqref{eq:SDE dephasing} is linear in $\alpha_m$,
it can be solved analytically as
\begin{align}
    \label{eq:analytical solution of SDE for dephasing}
    \alpha_m(t) = \alpha_m(0)e^{i\sqrt{\gamma}\mathcal{W}_m(t)}.
\end{align}
Here, $\alpha_m(0)$ denotes the initial condition sampled from the corresponding Wigner function for the coherent state,
$\hat{\rho}(0)=\ket{\alpha_{{\rm I}1},\alpha_{{\rm I}2},\alpha_{{\rm I}3}}\bra{\alpha_{{\rm I}1},\alpha_{{\rm I}2},\alpha_{{\rm I}3}}$,
which is given by the following Gaussian function
\begin{align}
    \label{eq:three Gaussian}
    W_0(\vec{\alpha},\vec{\alpha}^*,0) = 
    \prod_{m=1,2,3}2e^{-2|\alpha_m-\alpha_{{\rm I}m}|^2}.
\end{align}
Thus, the magnitude $|\alpha_m|$ is conserved, and each trajectory stochastically moves along a circle in the complex $\alpha_m$ plane whose radius is fixed by the initial condition $|\alpha_m(0)|$. In the long-time limit $t\to\infty$, the trajectories become uniformly distributed along the circle, and the steady-state samples are given by
\begin{align}
    \label{eq:samples for the initial non-Gaussin state}
    \alpha_m(t\to\infty) = |\alpha_m(0)|e^{i\theta_m},
\end{align}
where $\theta_m$ is uniformly distributed over the interval $0\leq \theta < 2\pi$.
From the discussion in \ref{appendix:Steady state of the system under the dephasing}, the samples in Eq.~\eqref{eq:samples for the initial non-Gaussin state} are equivalent to those obtained by sampling from the quasiprobability distribution function associated with Eq.~\eqref{eq:initial state for benchmark model3}.

\section{\label{appendix:Coherent and incoherent current}Coherent and incoherent current: Derivation of Eq.~\texorpdfstring{\eqref{eq:physical quantities model3 atomic current}}{TEXT}}
We derive the atomic current given in Eq.~\eqref{eq:physical quantities model3 atomic current}.
Below,
we consider the GKSL equation~\eqref{eq:GKSL equation discussion}.
We first introduce the local atomic current $I_{m,m+1}$,
which describes the flow of atoms from site $m$ to site $m+1$ and is defined from the following continuity equation:
\begin{align}
    \label{eq:continuous equation}
    \frac{d\braket{\hat{a}^{\dagger}_m\hat{a}_m}}{dt} = -(I_{m,m+1} - I_{m-1,m}).
\end{align}
By deriving the equation of motion for $\braket{\hat{a}^{\dagger}_m\hat{a}_m}$ from the GKSL equation~\eqref{eq:GKSL equation discussion},
we obtain the explicit form of $I_{m,m+1}$ as
\begin{align}
    I_{m,m+1} &=\frac{iJ}{\hbar}(\braket{\hat{a}^{\dagger}_{m+1}\hat{a}_m} - \braket{\hat{a}^{\dagger}_m\hat{a}_{m+1}}) + \gamma\braket{\hat{a}^{\dagger}_m\hat{a}_m(\hat{a}^{\dagger}_{m+1}\hat{a}_{m+1} + 1)} - \frac{\gamma}{M}\sum_{m=1}^M\braket{\hat{a}^{\dagger}_m\hat{a}_m}. 
\end{align}
We define the atomic current $I_{1,2,\dots,M}$ as the average of the local atomic current over all sites:
\begin{align}
    \label{eq:def of the atomic current M degrees of freedom}
    I_{1,2,\dots,M} &= \frac{1}{N_{\rm I}\gamma M}\sum_{m=1}^MI_{m,m+1}, \\
    &= \frac{1}{M}\sum_{m=1}^{M}\left\{\frac{iJ}{\hbar N_{\rm I}\gamma}(\braket{\hat{a}^{\dagger}_{m+1}\hat{a}_m} - \braket{\hat{a}^{\dagger}_m\hat{a}_{m+1}}) + \frac{1}{N_{\rm I}}\braket{\hat{a}^{\dagger}_m\hat{a}_m\hat{a}^{\dagger}_{m+1}\hat{a}_{m+1}}\right\},
\end{align}
where $\hat{a}_{M+1} = \hat{a}_1$ from the periodic boundary condition,
and $1/(N_{\rm I}\gamma)$ is a normalization constant.
By substituting $M=3$ into Eq.~\eqref{eq:def of the atomic current M degrees of freedom},
we obtain the atomic current Eq.~\eqref{eq:physical quantities model3 atomic current}.

\bibliographystyle{elsarticle-num}
\bibliography{elsarticle-num}






\end{document}